\documentclass[useAMS,usegraphicx]{mn2e}

\def\simlt{\mathrel{\hbox{\rlap{\hbox{\lower4pt\hbox{$\sim$}}}\hbox{$<$}}}}
\def\simgt{\mathrel{\hbox{\rlap{\hbox{\lower4pt\hbox{$\sim$}}}\hbox{$>$}}}}

\def\ale{\mathrel{\hbox{\rlap{\hbox{\lower4pt\hbox{$\sim$}}}\hbox{$<$}}}}
\def\age{\mathrel{\hbox{\rlap{\hbox{\lower4pt\hbox{$\sim$}}}\hbox{$>$}}}}

\def\nodata{---}

\title[Nearby Supernova Rates from LOSS: II. SN LFs and Fractions]{
  Nearby Supernova Rates from the Lick Observatory Supernova Search.
  II. The Observed Luminosity Functions and Fractions of Supernovae in
  a Complete Sample }

\author[Li et al.]
{Weidong Li$^{1}$\thanks{Email: wli@astro.berkeley.edu}, Jesse
Leaman$^{1,2}$, Ryan Chornock$^{1,3}$, Alexei
V. Filippenko$^{1}$, 
\newauthor 
Dovi Poznanski$^{1,4,10}$, Mohan
Ganeshalingam$^{1}$, Xiaofeng Wang$^{1,5,6}$,
\newauthor
Maryam Modjaz$^{1,7,11,12}$, Saurabh Jha$^{1,8}$, Ryan
J. Foley$^{1,3,13}$, and Nathan Smith$^{1,9}$\\ 
$^{1}${Department of Astronomy, University of California,
Berkeley, CA 94720-3411, USA} \\
$^{2}${NASA Ames Research Center, Mountain View, CA 94043, USA} \\
$^{3}${Harvard-Smithsonian Center for Astrophysics, 60
Garden Street, Cambridge, MA 02138, USA} \\
$^{4}$Computational Cosmology Center, Lawrence Berkeley National
Laboratory, 1 Cyclotron Road, Berkeley, CA 94720, USA \\
$^{5}${Department of Physics, Texas A\&M University,
College Station, TX 77843-4242, USA} \\
$^{6}$Physics Department and Tsinghua Center for Astrophysics (THCA),
Tsinghua University, Beijing, 100084, China \\
$^{7}$Columbia Astrophysics Laboratory, Columbia University, New York, NY 10027, USA\\ 
$^{8}${Department of Physics and Astronomy, Rutgers, The
State University of New Jersey, Piscataway, NJ 08854, USA} \\
$^{9}$Steward Observatory, University of Arizona, 933 North Cherry  
Avenue, Tucson, AZ 85721, USA \\
$^{10}$Einstein Fellow \\
$^{11}$Miller Fellow \\
$^{12}$Hubble Fellow \\
$^{13}${Clay Fellow} \\
}

\begin{document}
\maketitle 

\begin{abstract}

This is the second paper of a series in which we present new
measurements of the observed rates of supernovae (SNe) in the local
Universe, determined from the Lick Observatory Supernova Search
(LOSS).  In this paper, a complete SN sample is constructed, and the
observed (uncorrected for host-galaxy extinction) luminosity functions
(LFs) of SNe are derived. These LFs solve two issues that have plagued
previous rate calculations for nearby SNe: the luminosity distribution
of SNe and the host-galaxy extinction.  We select a volume-limited
sample of 175 SNe, collect photometry for every object, and fit a
family of light curves to constrain the peak magnitudes and
light-curve shapes.  The volume-limited LFs show that they are not
well represented by a Gaussian distribution. There are notable
differences in the LFs for galaxies of different Hubble types
(especially for SNe~Ia).  We derive the observed fractions for the
different subclasses in a complete SN sample, and find significant
fractions of SNe~II-L (10\%), IIb (12\%), and IIn (9\%) in the SN~II
sample. Furthermore, we derive the LFs and the observed fractions of
different SN subclasses in a magnitude-limited survey with different
observation intervals, and find that the LFs are enhanced at the
high-luminosity end and appear more ``standard" with smaller scatter,
and that the LFs and fractions of SNe do not change significantly when
the observation interval is shorter than 10~d.  We also discuss the
LFs in different galaxy sizes and inclinations, and for different SN
subclasses. Some notable results are that there is not a strong
correlation between the SN LFs and the host-galaxy size, but there
might be a preference for SNe~IIn to occur in small, late-type spiral
galaxies. The LFs in different inclination bins do not provide strong
evidence for extreme extinction in highly inclined galaxies, though
the sample is still small. The LFs of different SN subclasses show
significant differences. We also find that SNe~Ibc and IIb come from
more luminous galaxies than SNe~II-P, while SNe~IIn come from less
luminous galaxies, suggesting a possible metallicity effect. The
limitations and applications of our LFs are also discussed.

\end{abstract}


\section{Introduction}

The luminosity function (LF) is used to describe the distribution of
intrinsic brightness for a particular type of celestial object, and it
is always intimately connected to the physical processes leading to
the formation of the object of interest.  Specifically, the LF of
supernovae (SNe), among the most luminous and exciting transients,
will provide important information on their progenitor systems and
their evolutionary paths. The intrinsic LF of core-collapse SNe
(CC~SNe, hereafter) can constrain the distribution of ways that
massive stars die at different initial masses (Smith et al. 2011a),
and that of SNe~Ia can illuminate how accreting white dwarfs in the
various binary systems result in a thermonuclear explosion. The
observed LF of SNe will provide information on the extinction they
experienced in their host galaxies and their immediate environments,
thus giving further clues to their physical origins.

From an observational point of view, the LF of SNe is an important
tool for calculating the completeness of a survey or a follow-up
campaign in order to understand the involved selection biases, and for
deriving meaningful SN rates. Knowledge of the SN LF will also provide
guidance on the expected number and brightness distribution of SNe in
several new large surveys (e.g., Pan-STARRS, Kaiser et al. 2002;
Palomar Transient Factory, Law et al. 2009), which can be used to
estimate and coordinate the necessary resources for the follow-up
efforts.

Until now, however, we have had only limited knowledge on the LF of
SNe. Many factors contribute to the difficulties in measuring the
observed SN LF, with the most important being the completeness of
finding all SNe in a survey and gathering follow-up photometry and
spectroscopy.  To study the intrinsic LF of SNe, we need further
knowledge on how the SNe are extinguished in their host
galaxies. There is some theoretical work on this (e.g., Hatano,
Branch, \& Deaton 1998; Riello \& Patat 2005), but there are still
considerable uncertainties in these models.

Many previous measurements of SN rates have adopted different
strategies to derive the survey completeness and control time,
highlighting the uncertainties caused by limited knowledge of the SN
LF.  Some adopted an average luminosity plus a Gaussian scatter for
the SNe (e.g., Cappellaro et al. 1999 [C99, hereafter]; Hardin et
al. 2000; Barris \& Tonry 2006; Botticella et al. 2008), while others
used information from a follow-up sample with unknown completeness and
biases (e.g., Pain et al. 2002; Blanc et al. 2004; Sullivan et
al. 2006; Dilday et al. 2008). Some treat the LFs as observed, while
others consider them as intrinsic and apply additional extinction
corrections.  The host-galaxy extinction correction toward SNe,
however, is another poorly known quantity. Some studies adopted an
arbitrary functional form, such as the positive side of a Gaussian
distribution (Neill et al. 2006; Poznanski et al. 2007), or an
exponential function (Dilday et al. 2008), while others followed the
aforementioned theoretical guidance by Hatano et al. (1998) and Riello
\& Patat (2005) (e.g., Barris \& Tonry 2006; Botticella et al. 2008;
Horesh et al. 2008).

In theory, the observed LF of SNe can be derived from either a volume-
or magnitude-limited search. For a volume-limited survey, the key
factor is to have information (type, luminosity, and light curve) for
all of the SNe in the sample. For a magnitude-limited survey, it is
important to have this information for all of the SNe and then correct
for the different survey volumes of SNe having different brightnesses
(e.g., Bazin et al. 2009).  It is also important for a
magnitude-limited survey to go fairly deep in order to sample the
faint end of the LF.  As discussed in detail by Leaman et al. (2011;
hereafter, Paper I), there are nearly complete spectroscopic
classifications for the SNe discovered in our Lick Observatory SN
Search (LOSS) galaxies. This search goes fairly deep, with a small
observation interval for many nearby galaxies, so a significant
fraction of our survey is in the volume-limited regime.  In
particular, we identified that our survey may have almost full control
for galaxies within 60~Mpc and 80~Mpc for CC~SNe and SNe~Ia,
respectively.  Here we attempt to construct a complete SN sample to
derive the observed LF.

This paper is organised as follows. Section 2 describes the
construction of the complete SN sample, including the adopted light
curves, the collection and fitting of the photometry, and the
completeness study for every SN. In \S 3 we present the observed LFs
and fractions of SNe in a volume-limited survey, while \S 4 gives the
results for a magnitude-limited survey. Section 5 discusses
correlations of the LFs with the SN subtypes and host-galaxy
properties, and possible limitations and caveats in our LFs. Our
conclusions are summarised in \S 6. Throughout the study, we adopt the
WMAP5 Hubble constant of $H_0 = 73$ km s$^{-1}$ Mpc$^{-1}$ (Spergel et
al. 2007), consistent with the recent direct determination based on
Cepheid variables and SNe~Ia by Riess et al. (2009).

\section{The Construction of a Complete SN Sample}

\subsection{The SNe in the Luminosity Function Sample}

Paper I discussed the different subsamples of SNe in our analysis.  We
elect to construct a complete SN sample in the ``season-nosmall"
sample of SNe, consisting of SNe that were discovered ``in season''
but were not in small (major axis $< 1'$) early-type (E/S0) galaxies.
There are considerable advantages of using in-season SNe to construct
the LF; they were discovered young, so there are premaximum data to
help constrain the peak magnitudes. We also limit the sample to the
SNe discovered by the end of year 2006, in accordance with the
reduction of our follow-up photometry database. The reason for the
exclusion of SNe in small early-type galaxies is due to the uncertain
detection efficiency (as discussed in Paper I) which results in an
uncertain completeness correction (\S 2.5). As discussed in \S 5.5,
only two SNe were excluded from the LF study because their host
galaxies are small early-type galaxies, and their inclusion would have
negligible effect on the LFs.

We use a cutoff distance of 80~Mpc for the SN~Ia sample and 60~Mpc for
SNe~Ibc\footnote{We use ``Ibc'' to generically denote the Ib, Ic, and
  hybrid Ib/c objects whose specific Ib or Ic classification is
  uncertain.} and II (see Paper I). In \S 2.5, we will compute the
completeness of our survey for each SN selected in our LF sample. In
total, we select 74 SNe~Ia, 25 SNe~Ibc, and 81 SNe~II, for a grand
total of 180 SNe. Table 1 lists some basic information on the SNe and
their host galaxies (more details can be found in the galaxy and SN
sample tables in Paper I). Five of the SNe (SNe 1999bw, 2000ch,
2001ac, 2002kg, and 2003gm) are so-called ``SN impostors" ---
low-luminosity SNe~IIn that are likely to be superoutbursts of
luminous blue variable stars rather than genuine SNe (e.g., Van Dyk et
al. 2000; Smith et al. 2011b); they are not considered further in this
analysis, but will be discussed in a future paper.

We note that since our survey is conducted without using a filter, the
images are most closely matched to the $R$ band (Li et al.  2003a). In
the following several sections, we therefore focus our effort on
deriving an $R$-band luminosity for the SNe. Some discussion of LFs in
other passbands can be found in \S 5.5, and a full analysis of
multi-colour LFs for SNe~Ia will be presented elsewhere (Li et al. 2011b).

We also note that for all of the LF analysis, our photometry is
corrected for the Galactic extinction adopted from Schlegel,
Finkbeiner, \& Davis (1998) to avoid additional scatter in the LF
caused by the random Galactic extinction that the SNe suffered. Because
of this, our LF is ``pseudo-observed" and only the SN host-galaxy
extinctions are not corrected.  When applying our LF to a known
direction in the Milky Way, the corresponding Galactic extinction
should be applied to the luminosity of the LF SNe.

\subsection{Light-Curve Families for the SNe}

Different types of SNe exhibit a great degree of heterogeneity in
their photometric behaviour (e.g., Barbon, Ciatti, \& Rosino 1979;
Leibundgut et al. 1991). Within a specific SN type, some homogeneity
and correlations are observed, but no type can be well represented by
a single light curve. Ideally, it would be good to have a
well-observed light curve for every SN in the LF sample, but
unfortunately this is not the case (see more details in \S 2.3).  To
quantify the light-curve shape distribution for our LF SNe, we
construct a family of light curves for each type of SN from the
literature and/or our own database of optical photometry.

\subsubsection{Type Ia Supernovae}

With a few exceptions (e.g., Li et al. 2001a, 2003b; Howell et
al. 2006; Foley et al. 2010b), SNe~Ia are generally thought to form a
one-parameter family, with the fast-declining SNe also being
subluminous, and slow decliners being luminous (e.g., Phillips
1993). In the left panel of Figure 1, we plot the $R$-band light
curves of a sample of 83 well-observed SNe~Ia in the LOSS photometry
database (solid lines; Ganeshalingam et al. 2010). The time axis show
the number of days since $R$-band maximum, and the light curves are
plotted on an absolute magnitude scale, after the SNe were corrected
for Galactic extinction. The distances toward the SNe are calculated
from the recession velocities corrected for infall of the Local Group
toward the Virgo cluster. Also overplotted are the light curves of SN
1991T (dash-dotted line, from Lira et al. 1998) and the well-observed
SN 1991bg-like object SN 1999by (dashed line, from Garnavich et
al. 2004), arbitrarily shifted to absolute magnitudes of $-19.5$ and
$-17.5$, respectively.\footnote{The reasons for arbitrarily shifting
  the light curves of SNe 1991T and 1991bg are extinction and
  intrinsic luminosity scatter. Since in general SN 1991T is
  considered to be one of the slowest decliners while SN 1991bg is one
  of the fastest, it is reasonable to shift and place their light
  curves at the two extreme ends of the light-curve distribution.} The
published light curves of SNe 1991T and 1999by have been smoothed with
a spline function (as are all of the other template SN light curves
shown in Figures 1--3). As can be seen, the light curves of SNe 1991T
and 1999by nearly encompass all of the observed SNe~Ia in our
photometry database. We interpolate between the two curves to create
21 light curves (so each curve has a different shoulder prominence and
peak absolute magnitude), and use them as the light-curve family for
SNe~Ia. While our construction of the light-curve family for SNe~Ia is
not drastically different from previous approaches (e.g., the
application of a stretch factor to a template light curve), we need to
use interpolation (rather than stretch) during the construction to
deal with the presence or absence of the shoulder feature in the
$R$-band light curves.

A few SNe in the SN~Ia LF sample belong to the so-called ``SN
2002cx-like objects'' (Filippenko 2003; Li et al. 2003b; Jha et
al. 2006b; Phillips et al. 2007), which show distinct differences from
the rest of the SN~Ia family\footnote{Although we put SNe 2002es and
  1999bh in the ``SN 2002cx-like object'' category because they have
  certain characteristics of this subclass, the two objects also show
  apparent differences from other known members of this subclass,
  perhaps indicating that the subclass is intrinsically heterogeneous
  (e.g., Foley et al. 2009b, 2010c; McClelland et al. 2010; Narayan et
  al. 2011). See Ganeshalingam et al. (2011) for further
  discussion.} Recently, their SN~Ia nature has been questioned
(Valenti et al. 2009; but see Foley et al. 2009b, 2010a).  We
constructed a template light curve from SN 2005hk (Phillips et
al. 2007), a well-observed SN 2002cx-like object.

\subsubsection{Type Ibc Supernovae} 

Compared to the wealth of published photometry for SNe~Ia, the Type
Ibc SNe are not well observed. It is unclear whether they can be
described by a one-parameter family. Some studies suggest that they
can be broadly classified into two bins (e.g., Clocchiatti \& Wheeler
1997): the fast-evolving and the slow-evolving subclasses. In the left
panel of Figure 2, we plot the $R$-band light curves of 8 SNe~Ibc from
our unpublished photometry database. There is no fast-evolving object
among these 8 SNe, so we adopted the photometry of SN 1994I (dashed
line, Richmond et al. 1996), a well-observed object in this
subclass. We have an excellent light curve for the slow-evolving
SN~Ibc 2004dk (dash-dotted line) in our own photometry database which
we use as a template. For the rest of the SNe~Ibc, we construct an
average light curve (solid line). The late-time behaviour of the
average SN~Ibc is not well constrained by our sample, so we utilised
an additional sample of SNe~Ibc from Modjaz (2007).  This family of
three light curves is used to fit the majority of the SNe~Ibc in our
LF sample (without any stretching or interpolating).

For the so-called ``Ca-rich" subclass of peculiar SNe~Ibc, we chose
the light curve of SN 2005E (Perets et al. 2010). Unfortunately, there
are no premaximum data for SN 2005E, so we adopted that portion from
the average SN~Ibc light curve. This light curve is not shown in
Figure 2. The ``SN~Ibc-pec" subclass also contains SN 2003id (Singer
et al. 2003; Hamuy \& Roth 2003), the broad-lined SN Ic 2002ap (Foley
et al. 2003; see more discussion in \S 3.2), and SN 2004bm (\S 3.2).
The photometric behaviours of these SNe~Ibc-pec are all reasonably
represented by the average SN~Ibc light curve.

\subsubsection{Type II Supernovae}

The photometric behaviour of SNe~II is the most heterogeneous among
all SN types, and they can be divided into a few main photometric and
spectroscopic subclasses. SNe~II-P have a prominent ``plateau'' phase
in their light curves, while SNe~II-L decline linearly (in magnitudes)
after maximum brightness. SNe~IIb show prominent hydrogen Balmer lines
in their early-time spectra, but morph into SNe~Ib at late times.  In
addition, the prototypical SN IIb 1993J showed a double-peaked light
curve (Richmond et al. 1994), with a very early first peak, which we
now think is most likely due to black-body emission from the expanding
and cooling shock-heated stellar envelope (e.g., Waxman et al. 2007),
and the regular Ni$^{56}$-powered main maximum. This double-peak light
curve behaviour has most recently also been seen in the Type Ib
SN~2008D (Soderberg et al. 2008; Modjaz et al. 2009). While it is not
clear how common and pronounced the early first peak is among other
SNe~IIb besides the well-studied SN 1993J (e.g., Chevalier \&
Soderberg 2009), we use the smoothed light curve of SN 1993J as the
light-curve template for SN IIb. SNe~IIn show a strong ``narrow''
(actually, generally an intermediate width of $\sim 1000$ km s$^{-1}$)
component to their hydrogen Balmer lines and a wide variety of light
curves.  See Filippenko (1997) for a detailed discussion of the
classifications of these different subtypes.

The distinction between a SN~II-P and a SN~II-L in terms of
photometric evolution is not well documented in the literature,
especially in the $R$ band.  The collection of light curves for the
SNe~II-L in Barbon et al. (1979) and Young \& Branch (1989) are all in
the $B$ band. For our application, we define a SN~II as being a
SN~II-L if it declines by more than 0.5 mag in the $R$ band during the
first 50~d after explosion.

The left panels of Figure 3 show how the light curves of the SNe~II
are constructed. The top panels show the light curves of 15 SNe~II-P
(dots) in our photometry database that have been published by
Poznanski et al. (2009).  As seen here, and also noted by Hamuy
(2003), SNe~II-P vary in the durations of their plateau phase.  We use
the average light curve (solid line) as the template.  The second
panel shows the light curves of 5 SNe~II-L in our unpublished
photometry database; again, an average is derived as the template.
Due to the lack of data, the late-time behaviour of the SN~II-L
template is not well constrained and may have relatively large
uncertainty.  The third panel shows the light curves of 3 SNe~IIb: the
prototypical SN~IIb 1993J (Richmond et al. 1994), and the unfiltered
light curves of SNe 2003gu and 2005em from our photometry database. We
use the smoothed light curve of SN 1993J as the template. The rising
portion of the first peak is not well observed, so our manual
construction is quite arbitrary after considering the earliest
nondetections and detections (e.g., Wheeler et al. 1993).

The bottom-left panel of Figure 3 shows the construction of the
template light curves for SNe~IIn. Eight well-observed SNe~IIn from
our photometry database are plotted, displaying a great degree of
heterogeneity.  This mirrors what has been reported in the literature
about the photometric behaviour of this class of objects: SNe~IIn can
range from very slowly evolving objects such as SN 1988Z (e.g.,
Turatto et al. 1993) and SN 1995G (Pastorello et al. 2002), to more
typical objects like SN 1994W (Sollerman, Cumming, \& Lundqvist 1998),
to very rapidly evolving objects such as SN 1998S (Fassia et
al. 2000).  We use the light curve of SN 1998S (dash-dotted line) as
the template of a fast-declining SN~IIn, that of SN 2003dv (dashed
line) as the template for a slow-evolving SN~IIn, and the average of
the remaining seven objects (solid line) for the average SN~IIn.

\subsection{Photometry of the LF SNe}

It is important to collect photometry for {\it every} SN in the LF
sample to study the light-curve shape and derive the peak absolute
magnitude; otherwise, the sample will not be complete.  Since our
unfiltered survey images are most closely matched to the $R$ band, we
use the follow-up $R$-band photometry for the SNe whenever possible.
This is because the images taken during the follow-up campaigns have a
higher cadence (every 1--2~d near maximum light, every 2--4~d
thereafter) than the unfiltered images taken during the SN
search. Moreover, accurate photometric calibrations for the fields
have been obtained with the 0.76~m Katzman Automatic Imaging Telescope
(KAIT) and the 1.0~m Nickel telescope at Lick Observatory on many
photometric nights. The reduction details are described by
Ganeshalingam et al. (2010), where the filtered photometry for the
SNe~Ia is also provided. An important step in the reduction is the
careful removal of the host-galaxy contamination in the SN flux by
subtracting a template image taken long after the SN has faded.

For SNe~Ia, 62 of the 74 SNe (84\%) in the LF sample have filtered
follow-up photometry.  This large fraction is due to the combined
effect of the luminous nature of SNe~Ia relative to most other SNe,
the early discovery, and our emphasis on studying them.  For several
SNe (details listed in Table 3), the follow-up photometry is adopted
from Jha et al. (2006a; hereafter ``CfA-2") and Hicken et al.  (2009;
hereafter ``CfA-3").  Only 7 out of the 25 SNe~Ibc (28\%) have
follow-up photometry, and for SNe~II the corresponding numbers are 18
out of 76 (24\%).

For the SNe that do not have filtered follow-up photometry, we derive
unfiltered light curves from the SN search images. As discussed in
Paper I, our search has a relatively short observation interval, so we
cover the photometric evolution of the SNe rather well. This is
especially true for the SNe in the LF sample, as their host galaxies
are mostly in the sample that has a designed observational interval of
every 5~d. To reduce the unfiltered images, a high signal-to-noise
ratio template image without the SN is selected. The host-galaxy
contamination is then cleanly removed after image subtraction, similar
to what is done in the follow-up data reduction described by
Ganeshalingam et al. (2010). For photometric calibration, we use the
red magnitudes for the stars in the SN fields in the USNO~B1 catalog
(Monet et al. 2003).  Although the accuracy of this calibration is
only $\sim 0.2$--0.3 mag for an individual star, there are usually
more than 10 stars available in each field, so the uncertainty due to
calibration is $< 0.1$ mag.

We have good unfiltered light curves for a majority of the SNe without
follow-up filtered photometry. However, for a small fraction of the
SNe (13 out of 175, or 7\%), our photometric coverage is relatively
poor. Some of them were discovered near the end of an observing
season, so the search images did not cover the whole period around
maximum light. A few others are faint and the search images do not go
deep enough to yield a constraint on the light-curve shape.  The
majority of them, however, are due to a combination of bad weather and
relatively low cadence. For two objects (SNe 2005W and 2006dy in Table
3), we adopted the photometry measured by amateur astronomers posted
on SNWeb\footnote{http://www.astrosurf.com/}, with good coverage
around maximum light. For the other SNe, we pool all of the
information on the SNe together (discovery magnitude, spectral
identification and age estimate, unfiltered and filtered photometry in
our database and published elsewhere) and constrain the light curves
as much as possible.  Some of them still have large uncertainties, as
reflected in the error bars for their peak magnitudes. We also use the
average light curves according to their types for these poorly
observed SNe.

\subsection{The Light-Curve Fitting Method} 

We use a $\chi^2$-minimizing technique to fit light curves constructed
in \S 2.2 to the photometry collected in \S 2.3, to determine the
light-curve shape and peak magnitude for each SN, as demonstrated in
the right-hand panels of Figures 1 to 3. Because we attempt to use a
small set of light curves to describe the complicated observed variety
of photometric behaviour for the different types of SNe, the fit to
the data is not always perfect, and the reduced $\chi^2$ of the fit
can be several times larger than unity. Whenever possible, the peak
magnitudes are directly measured from a spline fit to the data near
maximum brightness rather than measured from the light-curve fit. As
noted by Cappellaro et al. (1993), the control-time calculation for a
SN search is more sensitive to the adopted peak luminosity of a SN
than to its light-curve shape.  The imperfections in the light-curve
fits also have a chance to cancel each other out when many SNe are
combined in the LF. So, the uncertainty in the light-curve shape
likely has little effect on the final control-time calculation.

We visually check the fits, especially the ones with relatively large
reduced $\chi^2$, to make sure they are a reasonable representation of
the data, and if not, to determine the possible causes. By doing this,
we identified two misclassifications in the LF SNe, SNe 2002au and
2006P, as detailed in Paper I. Both SNe were originally classified as
possible SNe~Ia, but their light-curve fits suggest SN~IIb and SN~Ic,
respectively.  An analysis of their observed spectra confirms the
suggestion from the light-curve fit. This exercise partly validates
our constructed light-curve families and the light-curve fitting
process.

The peak apparent magnitudes measured for the SNe are converted to
absolute magnitudes using distances measured from the recession
velocities corrected for the infall of the Local Group toward the
Virgo Cluster. To account for peculiar velocities in the local flow,
we adopt 300 km s$^{-1}$ as the uncertainty for the recession
velocities.  The uncertainties of the absolute magnitudes include the
photometry measurement error, the light-curve fit uncertainty, and the
distance uncertainty added in quadrature.  Columns 3 and 4 of Tables
3--5 list the results for different types of SNe.

\subsection{The Completeness of Each LF SN} 

It is important to correct for possible incompleteness of the SNe in
the LF sample.  For a particular SN in the LF sample, the peak
absolute magnitude and light-curve shape are given in \S 2.4. With
this information, we can calculate the control time for this SN for
the LOSS galaxies in the ``full-nosmall" sample (the control galaxy
sample for the LF SNe; see Paper I) using their monitoring history log
files [see Paper III (Li et al. 2011a) for details of the control-time
  calculation]. The completeness of of our search to a particular SN
at a given distance is then defined as the sum of the control time of
that particular SN for all of the galaxies within that distance
divided by the sum of the observing season time for these galaxies. To
correct a SN to 100\% completeness within the cutoff distance of the
LF sample, one just needs to use the reciprocal of the completeness as
the corrected number for the SN.

Figure 4 shows the completeness measurements for the SNe in the LF
sample. Each curve represents a SN, and some of the notable SNe are
labeled. The vertical dashed lines indicate the cutoff distance where
the sample is constructed.  The top panel shows the completeness
measurements for the SNe~Ia.  We achieved a completeness higher than
98\% for all of the SNe~Ia because of their extreme luminosity at
peak. The total number of SNe after correction for the incompleteness
is 74.70, only a 1\% change from the input number of 74. The middle
and bottom panels show the completeness measurements for the
CC~SNe. The majority of the SNe have completeness higher than 80\% at
the cutoff distance of 60~Mpc, but a few of them have relatively low
completeness due to their extremely low luminosity. For example, SN
1999br (Pastorello et al. 2004) is an intrinsically faint SN~II-P,
while SN 2002hh (Pozzo et al. 2006) is a highly reddened SN~II-P in
the nearby galaxy NGC 6946.

The corrected number for each SN in the LF function after applying the
completeness correction factor (hereafter CCF) is listed in Column 7
of Tables 3--5. The total corrected number of SNe~Ibc is 28.86, an
18\% increase compared to the input number of 24.5. For SNe~II, the
corrected number of 88.50 is a 16\% increase over the input number of
76.5.  We see that even though our search does not have full control
for all of the SNe within the cutoff distance of the LF sample, the
correction to 100\% completeness is small and thus our LF should not
suffer large uncertainties (see additional discussion in \S 5).

\section{The Volume-Limited Sample: LFs and Fractions of SN Types}

\subsection{The Observed LFs of SNe}

The ``pseudo-observed" LFs of the SNe (corrected for Milky Way
extinction but not host-galaxy extinction) are listed in Tables 3--5
for the different types. The following information is included for
each SN: the subtype, the absolute magnitude and its uncertainty, the
distance of the SN, the Hubble type, inclination, and mass of its host
galaxy, the corrected LF number for a volume-limited sample, the
corrected LF number for a magnitude-limited sample (discussed in the
next section), the light-curve shape of the SN, the source of the
photometry, and additional comments.  Each SN constitutes a discrete
point in the LF, with its own peak absolute magnitude, light-curve
shape, and number contribution to the total LF.

Although it would be ideal to construct a LF for galaxies of every
Hubble type, it is impractical with the relatively small total number
of SNe in the LF sample. Instead, the SNe are grouped into two broad
bins for each SN type: E--Sa and Sb--Irr for SNe~Ia, S0--Sbc and
Sc--Irr for the CC~SNe.  The split of the Hubble types is motivated by
an attempt to include reasonable numbers of SNe in each LF, rather
than by physics.  For example, one may argue that splitting the SNe~Ia
by E--S0 (early-type) and Sa--Irr (late-type) galaxies may be more
physically based, but then the E--S0 bin would suffer more from
small-number statistics. As discussed in Paper III, the exact manner
in which the SN~Ia LF is split has negligible effect on the final 
derived SN~Ia rates.

To study the statistical properties of the LFs, we use histograms to
show their luminosity distribution, but we emphasise that the LFs
should be used as discrete points when calculating the control time
for a survey.  We also use the Kolmogorov-Smirnov (K-S) test
exclusively to study whether two groups of objects come from the same
population (in terms of absolute magnitudes only). We note that the
histograms show the distributions of the {\it corrected numbers} of
the LFs; thus, the number of SNe in each bin does not correctly reflect
Poisson statistics. Since the CCFs are always greater than 1, the
Poisson uncertainty of each bin is always larger than that calculated
directly from the number of SNe in the bin. For example, if one bin
has a single SN with a CCF of 2.0, the number of SNe in the bin with
proper Poisson errors is $2.0 \times 1.0^{+2.29}_{-0.83} =
2.0^{+4.58}_{-1.66}$ (i.e., the error is 2.0 times the Poisson
error of 1.0 SN; Gehrels 1986), rather than $2.0^{+2.63}_{-1.29}$
(i.e., the Poisson error calculated directly from 2.0 SNe). In the
same vein, the K-S tests are also somewhat compromised due to the
deviation from Poisson statistics. Fortunately, the CCFs are close to
1.0 for all of the SNe in the LFs except for the objects fainter than
$-15$ mag.  In our subsequent discussions, all significant K-S test
results will be scrutinised by including/excluding the least luminous
objects in the LFs.

To properly consider the effect of the uncertainties of the LF SN
absolute magnitudes on the K-S test results, we run a Monte Carlo
simulation 1000 times to sample the absolute magnitudes according to
their Gaussian errors, and study the scatter of the resultant
cumulative distribution functions (CDFs; also called cumulative
fractions) and the probabilities of the two samples coming from the
same population.

Figures 5--7 display the histograms for the LFs of SNe~Ia, Ibc, and II,
respectively. The second panel of each figure shows the distribution
for the whole LF sample. We note that while a Gaussian distribution is
an acceptable but not ideal description for the LFs of SNe~Ibc and II,
it is a rather poor description for the LF of SNe~Ia. The average
absolute magnitudes are $-18.49 \pm 0.09$ (with a $1\sigma$ dispersion
of 0.76), $-16.09 \pm 0.23$ ($\sigma = 1.24$), and $-16.05 \pm 0.15$
($\sigma = 1.37$) for the SNe~Ia, Ibc, and II, respectively. These
numbers, together with the average absolute magnitudes for several
other combinations, are listed in Table 6.  

Richardson et al. (2002) did an extensive comparative study of the
peak absolute magnitude distribution for the SN discoveries compiled
in the Asiago SN Catalog (Barbon, Cappellaro, \& Turatto 1989; Barbon
et al. 1999). Their study was done in the $B$ band, although they did
not distinguish among the different photometric bands for some
SNe. They derived an absolute magnitude (without extinction
corrections, and converting to $H_0 = 73$ km s$^{-1}$ Mpc$^{-1}$ used
in our study) of $-18.73 \pm 0.07$ ($\sigma = 0.76$) for {\it normal}
SNe~Ia, $-17.49 \pm 0.30$ ($\sigma = 1.29$) for SNe~Ibc, and $-16.18
\pm 0.23$ ($\sigma = 1.23$) for SNe~II-P. We note the significant
difference compared with our result for the average peak absolute
magnitudes of SNe~Ibc: the Richardson et al. sample suggests a much
brighter magnitude relative to SNe~Ia and II. As Richardson et
al. noted, there are considerable observational biases in their
observed SN sample and the completeness is unknown. In particular, the
SN~Ibc subclass may be more heavily biased in the observed sample due
to its low peak luminosity (relative to SNe~Ia) and fast photometric
evolution (relative to SNe~II-P).

Figure 5 shows the histograms for the LFs of SNe~Ia in different
galaxy bins (the two lower panels). The LF in E--Sa galaxies shows an
apparent difference from the LF in Sb--Irr galaxies, with only a
$8.5^{+10.3}_{-5.0}$\% probability that they come from the same
population (the cumulative fractions and their 1$\sigma$ scatters are
plotted in the top panel). This is likely caused by the observed
preference of different subclasses of SNe~Ia in host galaxies of
different Hubble types: the subluminous SN 1991bg-like objects in
early-type galaxies and the overluminous SN 1991T-like objects in
spiral galaxies (e.g., Della Valle \& Livio 1994; Hamuy et al. 1996;
Howell 2001).

Figure 6 shows the histograms for the LFs of SNe~Ibc in different
galaxy bins (the two lower panels).  The K-S test does not provide
evidence for a significant difference between the two LFs: the SNe
come from the same population at a $46.3^{+23.0}_{-21.2}$\%
probability. SNe~Ibc in the early-type spiral galaxies appear on
average marginally fainter (averaging $-15.98 \pm 0.26$ mag; $\sigma =
0.83$ mag) than their counterparts in the late-type spirals (average
of $-16.15 \pm 0.33$ mag; $\sigma = 1.43$ mag).



Figure 7 shows the histograms for the LFs of SNe~II in different
galaxy bins (the two lower panels); there is a marginal difference,
with a $21.0^{+19.5}_{-10.7}$\% probability that they come from the
same population.  Contrary to the trend shown by the SNe~Ibc, in the
early-type spirals SNe~II are marginally brighter (average of $-16.22
\pm 0.21$ mag; $\sigma = 1.39$ mag) than their counterparts in the
late-type spirals, which average $-15.88 \pm 0.20$ mag ($\sigma =
1.34$ mag).  The significance of the difference between the two LFs is
not dramatically affected by the objects fainter than $-15$ mag: when
they are excluded from the statistics, the two LFs come from the same
population with a $28.0^{+27.7}_{-16.0}$\% probability.


It is generally expected that SNe occurring in late-type galaxies
should on average experience more extinction than those in early-type
galaxies because of a dustier environment. This fact should be taken
into account when translating differences in the observed LFs in
various Hubble types into differences in the intrinsic LFs. For
example, SNe~Ia that occurred in Sc--Irr galaxies should be
intrinsically brighter than SNe~Ia in E--Sa galaxies by a bigger
margin than is shown in Figure 5.

In a recent paper, Bazin et al. (2009) derived an overall
core-collapse SN LF from the Supernova Legacy Survey (SNLS).  A
comparison between our combined SN~Ibc and SN~II LF and that reported
by Bazin et al. shows excellent agreement (J. Rich, 2010, private
communication).

\subsection{The Observed Fractions of SNe} 

In the process of analysing the LF SNe in detail, we are able to put
them into different subclass bins. For SNe~Ia, the light-curve fitting
sequence from 1 to 21 is a loose luminosity indicator, as we
demonstrate in a forthcoming paper (Li et al. 2011b). Moreover,
the SNe are categorised into several subclasses: normal SNe~Ia with
normal expansion velocities (``IaN" in Table 3 and hereafter), normal
SNe~Ia with high expansion velocities (``IaHV," see Wang et al.  2009
for our definition of this subclass), SN 1991bg-like objects
(``Ia-91bg"; Filippenko et al. 1992b; Leibundgut et al. 1993), SN
1991T-like objects (``Ia-91T"; Filippenko et al. 1992a; Phillips et
al. 1992), and SN 2002cx-like objects (``Ia-02cx"; Filippenko 2003; Li
et al. 2003b; Jha et al. 2006b; Phillips et al. 2007). This
classification is based on the information published in the IAU
Circulars and/or analysis of the spectra in our spectral database
(Silverman et al. 2011).  As discussed by Li et al. (2001b),
there is a significant ``age bias" for SN 1991T-like objects, caused
by the fact that such objects can only be easily identified with
spectra taken prior to or near maximum brightness.  Because of this,
the fraction of SN 1991T-like objects should be regarded as a lower
limit in this study. As discussed by Wang et al. (2009), a spectrum
(or expansion-velocity measurement) within a week around maximum
brightness is required to classify a normal SN~Ia into the ``IaN" or
``IaHV" subclasses. Fortunately, we were able to secure such
information for all of the SNe~Ia in our LF sample.

For SNe~Ibc, both the fast- and slow-evolving SNe are relatively rare
(10\% for each subclass), but this conclusion is hampered by the
relatively large fraction of SNe~Ibc that are either peculiar or have
poor light-curve coverage. We put the SNe~Ibc into three subclasses:
SN~Ib, SN~Ic, or peculiar Ibc or Ic (``Ibc-pec" or ``Ic-pec," which we
consider as the same subclass).  We note that in general, there is
considerable uncertainty in classifying SNe~Ibc into these
subclasses. Sometimes the SNe are simply reported as ``SN~Ibc" in the
IAU Circulars without a more specific subclass.  Other times, a SN~Ib
would only develop strong He~I lines after a few weeks, so it might be
misclassified as a SN~Ic from an early-time spectrum.  The differences
in the spectra of the different subclasses also become subtle when the
SNe are in the nebular phase. Although there are spectra for 21 out of
the 25 LF SNe~Ibc in our spectral database, and the other 4 SNe were
classified in IAU Circulars by experienced observers, we do not have a
good series of spectra for every SN in the sample to check for a
possible SN~Ic to SN~Ib transition, so the fraction of SNe~Ic should
be regarded as an upper limit in this study.


We attempt to place the SNe~II into four subclasses: II-P, II-L, IIb,
or IIn. For this purpose, SNe~IIn can often be easily distinguished
from the others because of their unique spectral features (a prominent
narrow or intermediate-width emission component in the hydrogen Balmer
lines), although in rare cases a SN~IIn can spectroscopically evolve
into a regular SN~II (e.g., SN 2005gl, Gal-Yam et al. 2007).  It is
difficult to distinguish between the other three subclasses based on
their spectra alone. First, the defining features or spectral
evolution have not been established to distinguish a SN~II-P from a
SN~II-L. Second, even though a SN~IIb can be identified from its early
resemblance to a SN~II and late metamorphosis into a SN~Ib, it is not
clear whether an early SN~II will turn out to be a SN~IIb unless we
have good spectroscopic coverage for {\it every} SN~II.  Fortunately,
these three subclasses have rather different photometric behaviour:
SNe~II-P have a prominent plateau phase, SNe~II-L have a linear
decline (in magnitudes) after maximum, and SNe~IIb have a
double-peaked light curve (Figure 3). Consequently, for the majority of
SNe our light-curve fitting process reports a strong preference for a
certain subtype.  For a few SNe with poor light-curve coverage, the
data can be fit by more than one template light curve, and we assign
equal fractional weights to the subclasses that provide a satisfactory
fit.

One surprising result from the light-curve fitting process is a
possible high fraction of SNe~IIb in the SN~II sample.  Following
identification of the first known SN~IIb, SN 1987K (Filippenko 1988),
detailed studies of only a few SNe~IIb have been published in the
literature. SN 1993J, the prototypical SN~IIb in the nearby galaxy
M81, has been extensively studied (e.g., Matheson et al. 2000, and
references therein). Another SN~IIb, SN 1996cb, was studied by Qiu et
al. (1999).  With the help of the ``Supernova Identification code''
(SNID; Blondin \& Tonry 2007), some recent SNe have been classified as
SN~IIb. The fraction of SN~IIb within the family of SNe~II is very
uncertain, but generally considered to be relatively small.

Figure 8 shows all possible SNe~IIb in our LF sample. Two of the
objects, SNe 2000N and 2004al, can be fit with both a SN~IIb and a
SN~II-L, so they are assigned 0.5 for each subclass.  Foley et
al. (2004) classified SN 2004bm as a probable SN~Ic based on a
low-quality spectrum. The light curve, though with only four points,
shows a distinct dip that is reminiscent of a SN~IIb. Reanalysis of
the spectrum does not provide a confident classification for the SN,
so we assign 0.5 for both IIb and Ibc-pec. The light curve of SN 2005H
is rather poor. The photometric behaviours of the other seven SNe are
best matched by the template SN~IIb light curve. Considering that our
template light curves are only an average of the observations, it is
conceivable that a few of these SNe can be fit by some variations of
SNe~II-L (these SNe are clearly not SNe II-P, and their spectra do not
show narrow emission components so they are also not SNe~IIn); hence,
the list of SNe in Figure 8 should be considered as an upper limit to
possible SNe~IIb in the LF sample. We also note that for four of our
SN~IIb candidates, there is spectroscopic confirmation of our
classification: SN 2000H (Benetti et al. 2000), SN 2003ed (Leonard,
Chornock, \& Filippenko 2003), SN 2005U (Leonard \& Cenko 2005), and
SN 2006T (Blondin et al. 2006). We consider the SN~IIb classification
for these four objects to be solid, but for the rest of the SNe, we do
not have spectra to corroborate the SN~IIb classification from the
light curves.  Overall, we have four solid (5\% of all the SNe~II), or
up to 9 possible (12\% of all the SNe~II), SNe~IIb in our LF sample.

The observed fractions of different subclasses of SNe can be
illustrated with pie charts, as shown in Figure 9. These fractions are
also listed in the second column in Table 7. To calculate the
uncertainties of the fractions, we ran a Monte Carlo simulation to
generate 1000 different versions of the LF according to Poisson
statistics with the {\it observed } total number of SNe. The 1$\sigma$
scatter of the measurements is then reported as the uncertainty in
each case. Despite having a relatively large number of SNe (175) in
the LF sample, many of the fractions are derived from subsets of SNe
in the LF sample and suffer from small-number statistics; thus, there
are considerable uncertainties in the fractions, especially for those
of SNe~Ibc.  The SNe~Ia within 60~Mpc are considered together with the
CC~SNe in the LF sample to derive their relative fractions in the
leftmost pie chart. Clearly, SNe~II are the most abundant (57\% of
all) type of SNe in a volume-limited sample, while SNe~Ia (24\%) and
SN~Ibc (19\%) have roughly equal fractions.

The SN~Ia pie chart is constructed from the SN~Ia LF sample within
80~Mpc. Normal SNe~Ia are about 70\% of the total, while the other
subclasses are 15\% SN 1991bg-like objects, 9\% SN 1991T-like objects,
and 5\% SN 2002cx-like objects. Li et al. (2001b) studied the rate of
peculiar SNe~Ia with a sample of 45 SNe~Ia discovered by LOSS during
the period between 1997 and 1999, and found a fraction of 64\% normal,
16\% SN 1991bg-like, and 20\% SN 1991T-like. The two studies have a
similar fraction for the normal and SN 1991bg-like objects, but a
different fraction for the SN 1991T-like objects. As discussed above,
the fraction of SN 1991T-like objects suffers from the age bias, which
is probably more serious in this analysis than in the Li et
al. (2001b) study.  Moreover, given the relatively small samples
in both studies, the difference is within the error bars of the
fractions, especially considering that the SN 2002cx-like objects can
be loosely grouped with SN 1991T-like objects because they show
similar strong Fe~III features at early times (but with different
expansion velocities).  The normal SNe~Ia are further divided into the
objects with normal (IaN) and high (IaHV) expansion velocities. Their
fractions, not shown in the pie chart, are 50\% for IaN and 20\% for
IaHV in the SN~Ia sample.

We note that the fraction for the SN 2002cx-like objects, $\sim$
5\% of the total SN~Ia sample, is quite uncertain due to the
heterogeneity of the subclass. For example, our SN~Ia LF sample does
not have the rapidly evolving, very subluminous SN 2002cx-like objects
such as SN 2008ha, which, according to Foley et al. (2010a), could
have a fraction as high as $\sim 10$\% of the SN~Ia population.

The SN~Ibc pie chart shows that SNe~Ic are the largest fraction (54\%
of all), followed by SNe~Ibc-pec (24\%) and SNe~Ib (21\%).  Among the
SNe~Ibc-pec, each of SNe 2002ap, 2003id, and 2004bm is $\sim 4$\%
of the total, while the Ca-rich objects are $\sim 13$\%.


The SN~II pie chart demonstrates that the most abundant component is
SNe~II-P (70\% of all), and the other three subclasses have similar
fractions (10\%, 12\%, and 9\% for SNe~II-L, IIb, and IIn,
respectively).

While a future paper will discuss in detail the rates for the various
types of peculiar SNe and transients, we note here the fractions (or
upper limits) for several kinds of objects. Richardson et al. (2002)
suggested a population of luminous SNe~Ibc (with peak absolute
magnitude brighter than $\sim -20$) and II-L (brighter than $\sim
-19$). Recently, several extremely luminous CC~SNe have been reported,
including SN 2003ma (Rest et al. 2009), SN 2005ap (Quimby et
al. 2007), SN 2006gy (Smith et al. 2007; Ofek et al. 2007), SN 2006tf
(Smith et al. 2008), SN 2008es (Miller et al. 2009; Gezari et
al. 2009), and SN 2008fz (Drake et al. 2010).  As listed in Tables 4
and 5, none of the 88.5 CC~SNe in our LF sample is brighter than $-19$
mag. Thus, unless the very luminous CC~SNe have an extreme preference
to occur in low-luminosity galaxies or near galaxy nuclei, making our
survey strongly biased against them, our LFs suggest that they are
rare ($\simlt 2$\% of the total CC~SNe using Poisson
statistics).\footnote{Note that SN 2006gy was imaged in our survey and
  meets the criterion to be a LF SN, but it was missed in our search
  pipeline due to its extreme proximity to the host-galaxy center. We
  could attempt to derive a fraction for the SN 2006gy-like objects
  based on our detection-efficiency simulations, but we elect to
  discuss the details in a future paper.}


Of the subclass of SNe~Ibc-pec, the broad-lined SNe~Ic deserve special
attention because of their link to gamma-ray bursts (GRBs; e.g.,
Galama et al. 1998; Matheson et al. 2003; Modjaz et al. 2006; Pian et
al. 2006). In our LF SN sample, there is only one broad-lined SN~Ic, SN
2002ap (Foley et al. 2003), which is 3.5\% of the total SNe~Ibc. Thus,
broad-lined SNe~Ic appear to be relatively rare. A more detailed
discussion of their rate and a comparison to the published GRB rates
will be provided in a future paper.

We emphasise that this is the first time the observed fractions of the
subclasses of SNe have been measured from a complete, volume-limited
SN sample, with well-understood completeness measurements, and
light-curve information to help with the classification.  These
fractions provide strong constraints on the possible progenitor
systems and their evolutionary paths for the different subclasses of
SNe, which is the topic of another paper (Smith et al. 2011a).

We note that Smartt et al. (2009) recently used a volume-limited
(within 28 Mpc) sample of 132 SNe to investigate the observed
fractions of SNe.  They based their classifications mostly on the
reports in the IAU Circulars. While that study and ours have similar
fractions for the overall SNe~Ia, Ibc, and II, the fractions for the
subclasses of SNe~II are quite different (our study suggests a lower
fraction for SNe~II-P, but higher fractions for the subclasses of
SNe~II-L, IIb, and IIn).  As noted earlier, photometric behaviour is
key to distinguishing SNe~II-L and IIb from SNe~II-P. Without detailed
light curves for the SNe in the Smartt et al. study, some of the SNe
II-L and IIb might not be recognised as such, a possible explanation
for the differences in the two studies. The two SN samples are also
quite different and may involve different selection biases.

\section{The Magnitude-Limited Sample: LFs and Fractions of SN Types}

\subsection{The Observed LFs of SNe}

In contrast to a volume-limited survey in which all of the SNe within
a certain volume have been discovered, a magnitude-limited survey has
a limiting magnitude for the apparent brightness of the discovered
SNe, $m_{\rm lim}$. Consequently, a SN with an observed absolute magnitude,
$M_{\rm abs}$, will have a survey volume within a distance of $\mu =
m_{\rm lim} - M_{\rm abs}$. The observed LFs in
a volume-limited sample discussed in \S 3 can thus be converted to
those in a magnitude-limited sample, with each SN scaled by its survey
volume.

We emphasise that this exercise is for an ideal situation where the
limiting magnitude of the magnitude-limited survey is deep enough to
sample the faintest end of the observed LFs, and to accumulate enough
statistics for the whole range of the LFs. Moreover, the LFs can only
apply to a scenario in which the survey volume is constantly monitored
--- that is, the observation interval is minimal (e.g., daily), and
{\it all} of the SNe that occurred during the survey are discovered
and measured. The effect of different observation intervals is
discussed in more detail in \S 4.3.  It should also be noted that this
is for a {\it nearby} magnitude-limited survey because it is derived
from the nearby volume-limited sample; the LFs and relative fractions
of SNe may evolve with redshift.

Figure 10 shows the histograms of the LFs of SNe in a
magnitude-limited sample showing the percent of the total number of
SNe for each bin (solid lines, with interval = 1~d), while Column 8 of
Tables 3--5 lists the relative fraction of each SN assuming the total
number of SNe is the same as in the volume-limited sample for each
type. Compared to the volume-limited LFs, the magnitude-limited LFs
clearly have an enhanced fraction of more luminous objects due to
their larger survey volume. The average absolute magnitudes are
$-19.00$ ($\sigma = 0.46$) , $-17.29$ ($\sigma = 0.62$), and $-17.70$
($\sigma = 0.85$) for SNe~Ia, Ibc, and II, respectively, which are
about 0.5, 1.2, and 1.6 mag brighter than those in a volume-limited
sample.  We also note that the scatter of the average absolute
magnitude becomes smaller in a magnitude-limited sample, so the SNe
appear to be more ``standard" because of the redistribution of the SN
fractions. In other words, a magnitude-limited search will be strongly
biased in favour of luminous, unextinguished objects. One needs to be
aware of this selection bias before generalizing a result derived from
a magnitude-limited search.  We note the SN~Ibc absolute magnitude is
now more in line with the average of the observed sample in Richardson
et al. (2002), suggesting that a significant fraction of the observed
SNe~Ibc in their sample were discovered in magnitude-limited surveys.

\subsection{The Observed Fractions of SNe}

Because different subclasses of SNe have different absolute
magnitudes, their observed fractions also change in a
magnitude-limited survey, as shown in Figure 11 and listed in Table 7
(the column marked with ``mag-1d"). The uncertainties of the fractions
are derived from the Monte Carlo simulations discussed in \S 3.2. This
is again for an ideal magnitude-limited survey in which the survey
volume is constantly monitored. SNe~Ia, the most luminous type of the
three, now become the most abundant, accounting for 79\% of the total.
SNe~II, the most abundant in a volume-limited sample, are only 17\% of
the total, while SNe~Ibc are just 4\%.

Among SNe~Ia, normal SNe~Ia are 77\% of the total, SN 1991T-like
objects are 18\%, while SN 1991bg-like and SN 2002cx-like objects are
3\% and 2\%, respectively. The slow-evolving objects (SN 1991T-like
objects and some normal SNe~Ia) have enhanced numbers in a
magnitude-limited survey because they are more luminous than the rest
of the SNe~Ia.  The number of fast-evolving SN 1991bg-like objects, on
the other hand, is depressed due to their subluminous nature. We also
note that there may be hints that SN 1991bg-like objects constitute
less than 3\% of the total in some magnitude-limited surveys conducted
at moderate and high redshifts, such as the Sloan Digital Sky Survey
(SDSS; B. Dilday, 2009, private communication; Foley et al. 2009a) and
the SN Legacy Survey (A. Howell, 2009, private communication),
suggesting further discrimination against them at large look-back
times. This, if confirmed, will constrain the progenitors of SN
1991bg-like objects to a tight range of old populations.

The fractions for the different subclasses of the CC~SNe also change
significantly, especially among SNe~II.  The fractions for SNe~IIb and
II-L are enhanced, while that for SNe~II-P is depressed. It is worth
noting that SNe~II-P, the most abundant SN~II component (70\% of all)
in a volume-limited survey, constitute only 30\% of all in a
magnitude-limited survey due to their subluminous nature.


\subsection{The Effect of Observation Intervals} 

The previous two sections discuss the LFs and subclass fractions of
SNe in an ideal magnitude-limited survey, one with the minimum
observation interval (1~d). In practice, the observation intervals
are significantly longer than 1~d in most magnitude-limited surveys,
and we discuss their effect in this section. 

We perform a Monte Carlo simulation similar to that employed by Li,
Filippenko, \& Riess (2001) to achieve this goal.  The limiting
magnitude of the survey is set to be 19, and the survey period is
10~yr. We use 10$^7$ SNe in the simulation, and they are randomly but
evenly distributed in a volume with the boundary set at a distance
modulus $\mu$ = 40.0 mag. This large volume ensures that the survey is
in the magnitude-limited regime even for the most luminous SNe in the
LFs.  Each SN is randomly selected from a LF that is constructed by
combining the SN~Ibc LF, the SN~II LF, and the SN~Ia LF within 80~Mpc
scaled to $D = 60$~Mpc (by a constant equal to the ratio of the total
number of SNe in the two LFs), with a probability proportional to its
number fraction. The SN is also given a random explosion date during
the 10~yr period. The survey then goes through the series of dates of
observations (according to the observation interval) and checks to see
whether the SN is detected. In these simulations, a step function is
used for the detection efficiency; the SN is marked as being detected
when it is brighter than the survey limiting magnitude at any epoch of
its light curve.

The effect of the observation interval on the LFs is shown in Figure
10. The shape of the LFs has subtle changes for all three SN types.
The most significant change, however, is that more SNe~II (with a
higher percentage of total SNe) are discovered when the observation
interval is longer. This is due to the fact that SNe~II-P have a long
plateau phase and their discovery rate is relatively enhanced with
long observation intervals.


The subclass fractions with different observation intervals are shown
in Figure 12 and listed in Table 7. The upper-left panel shows the
overall SN~Ia, Ibc, and II fractions. The SN~Ibc fraction remains
small, $\sim 4$\% for all of the intervals. The SN~Ia fraction decreases
from 79\% to 69\%, while the SN~II fraction increases from 17\% to
27\%, when the observation interval changes from 1~d to 360~d (or a
single snapshot), respectively.  Also shown in the panel is the curve
of the ``detection fraction," which is the total number of SNe
detected at a given observation interval divided by that with an
observation interval of 1~d.  The detection fraction remains high ($>
94$\%) when the observation interval is smaller than 10~d, and then
declines dramatically with longer intervals. This is likely due to the
fact that most SN light curves do not change much during the 10~d
near maximum brightness. In a snapshot survey (i.e., with an interval
of 360~d), only 8.6\% of the SNe are detected.

The other panels show the subclass fractions with different
observation intervals for SNe~Ia, Ibc, and II, respectively.  We note
that when the observation interval is shorter than 10~d, all
subclass fractions remain nearly unchanged. At longer intervals, the
fractions of the SNe with relatively slow light curves are enhanced,
e.g., SN 1991T-like objects among SNe~Ia and SNe~II-P among SNe~II.
In a snapshot survey, nearly 40\% of the SNe~II are SNe~II-P, much
higher than the fraction of 30\% in an ideal magnitude-limited survey.

\subsection{Comparisons to the Observed Magnitude-Limited Samples}

To check whether our predicted subclass fractions of SNe in a
magnitude-limited sample match observations, we compare our results to
those of several actual magnitude-limited samples.

Although LOSS is a search with a targeted list of nearby galaxies, the
random galaxies projected in the background of the LOSS target fields
have a wide range of redshift, so the SNe discovered in them should
only be limited by the depth of our images; they belong to a
magnitude-limited sample. Gal-Yam et al. (2008) compiled a list of 32
such events discovered during the years 1999--2006. Here we update the
list to include all of the SNe discovered during the years
2007--2008. We also revise the list of Gal-Yam et al. to exclude three
objects (SNe 2002ct, 2003im, and 2004X; all occurred in targeted
galaxies with relatively high redshift), and include six additional
objects (SNe 2001ew, 2002je, 2002ka, 2004as, 2004eb, and 2005bu; all
occurred in the background galaxies).

The full list has 47 SNe and is reported in Table 7. Only 1 object (SN
2001es) does not have a spectroscopic classification. For the rest of
the SNe, 34 (74\%) are SNe~Ia, 4 (9\%) are SNe~Ibc, and 8 (17\%) are
SNe~II.  As the observation interval of our search is on average
smaller than 10~d (Paper I), the observed fractions should be
compared to those predicted by an ideal magnitude-limited search
(79\%, 4\%, and 17\% for SNe~Ia, Ibc, and II, respectively), and they
show excellent agreement. Comparison with the detailed subclasses is
not possible because we do not have good light-curve coverage for
these SNe, and the total number of CC~SNe (12) is small.

The Palomar Transient Factory (PTF, Law et al. 2009) is a wide-field
survey aimed at a systematic exploration of the optical transient sky,
and is a classical magnitude-limited search for SNe. Two batches of
SNe have been reported by Kasliwal et al. (2009) and Quimby et
al. (2009). Among the 29 spectroscopically classified SNe (out of 31
total), 21 (72\%) are SNe~Ia, 1 (3\%) is a SN~Ibc, and 7 (24\%) are
SNe~II. Considering the small total number of SNe involved, and the
unknown observation interval, these fractions are in sufficiently good
agreement with our predictions.

\section{Discussion}

In this section, we discuss the dependence of the volume-limited LFs
on the environments and subclasses of SNe. We also consider possible
applications of our LFs.

\subsection{LFs in Galaxies of Different Sizes}

As described in Paper I, the LOSS galaxy sample has an apparent
deficit of low-luminosity galaxies when compared to a complete
sample. It is thus important to study the correlation between the LFs
and the galaxy sizes\footnote{Hereafter, the ``galaxy size" refers to
  the magnitude of both the luminosity and stellar mass, unless
  otherwise specified, because the mass is directly calculated from
  the luminosity, with a small dependence on $B - K$ colour (Paper I;
  Mannucci et al. 2005).}, and investigate whether the LFs we derived
are biased because of this deficit.

Figure 13 shows the correlation of the LFs of SNe~Ia with galaxy
sizes.  The top panel shows the LFs for the total SNe in the E--Sab
(left) and Sb--Irr (right) galaxies, while the middle and bottom
panels split the LFs into two host-galaxy size bins according to their
$K$-band luminosities, with roughly equal numbers of SNe in each bin.
Galaxy size does not play a significant role in the LFs of SNe~Ia: K-S
tests do not provide strong evidence for a significant difference in
the two LFs for different galaxy sizes.  We note that the bigger
Sb--Irr galaxies host more SNe in the two most luminous bins and the
bins at around $-17.5$ mag than the smaller galaxies, suggesting a
possible more extreme LF in the bigger galaxies.


The total number of SNe in the SN~Ibc LF sample is small (28.9). 
While we do not find any significant difference in the LFs for the
galaxies with different sizes, the constraint is not strong due to
small-number statistics.  

Figure 14 shows the correlation of the LFs of SNe~II with galaxy
sizes.  No significant difference is found for the early-type spirals,
with the SNe in the two LFs coming from the same population at a
$32.7^{+17.6}_{-15.1}$\% probability.  For the late-type spirals, this
probability is $4.2^{+9.0}_{-2.6}$\%, suggesting a rather significant
difference. Even when the SNe fainter than $-15$ mag are not
considered, the probability is still small ($4.5^{+11.2}_{-3.2}$\%).
The SNe~II in the bigger late-type spirals are on average brighter
than those in the smaller galaxies (the average is $16.28 \pm 0.35$
[$\sigma = 1.52$] and $-15.42 \pm 0.25$ [$\sigma = 1.12$],
respectively). Inspection of the SNe in the two LFs suggests that the
difference is likely caused by the different composition of
subclasses. For the 18 SNe~II in the smaller late-type spirals, there
are 3 SNe~IIb, 4 SNe~IIn, and 11 SNe~II-P, while for the 19 SNe~II in
the bigger late-type spirals, there are 2 SNe~IIb, 3 SNe~II-L, and 14
SNe~II-P. Thus, it appears that SNe~IIn might prefer smaller galaxies
while SNe~II-L prefer bigger galaxies (but keep in mind the
small-number statistics). When only SNe~II-P are considered, no
significant difference is found in the two LFs.

In summary, we have not found a significant correlation between the
LFs of SNe and their host-galaxy sizes, although some subclasses of
SNe may have a preference to occur in certain galaxy sizes among some
Hubble types. More discussion of this topic can be found in \S 5.4.

\subsection{LFs in Galaxies of Different Inclinations}

It is of interest to check the LFs of SNe in galaxies having different
inclinations, and to investigate the effect of inclination on the
amount of extinction the SNe experienced in their host galaxies.  For
this purpose, the LF SNe are split into three inclination bins
($0^\circ-40^\circ$ [hereafter, ``face-on"], $40^\circ-75^\circ$
[hereafter, ``inclined"], and $75^\circ-90^\circ$ [hereafter,
  ``edge-on"]), and their LFs are plotted in Figure 15. Only the SNe
occurring in spiral galaxies (Types 3 to 7) are considered because the
inclination is not meaningful for an early elliptical or irregular
galaxy, as discussed in Paper I. Because of the limitation of the
total number of SNe in the LF sample, several LFs suffer from
small-number statistics, especially SNe~Ia and Ibc in the face-on bin,
and SNe~Ibc in the edge-on bin.

The LFs of SNe~Ia do not show a significant difference in the three
inclination bins, as reflected in the average absolute magnitudes in
Table 6 and the K-S test results.  The inclined and the edge-on bins
both have a reasonable number of SNe (33.4 and 18.2,
respectively). Moreover, because of the extraordinary luminosity of
SNe~Ia, our survey should have missed very few objects (even for SNe
with moderate to high, but not extreme, extinction), as indicated by
the small corrections to 100\% completeness. Thus, perhaps
surprisingly (given that many SNe~Ia occur in young to
intermediate-age populations; e.g., Maoz et al. 2011, and references
therein), our data do not provide strong evidence for more extinction
in more highly inclined galaxies for SNe~Ia.

The LFs of SNe~Ibc show a strong trend in the three inclination bins:
the average absolute magnitude is the brightest in the face-on bin and
the faintest in the edge-on bin. This is consistent with more
extinction in more inclined galaxies.  However, both the face-on and
edge-on bins suffer from small-number statistics.

The LFs of SNe~II have reasonable numbers of SNe in all three
inclination bins.  An unexpected result is that the LF for the objects
with intermediate host-galaxy inclination ($40^\circ-75^\circ$) shows
a significant difference from the LFs in the other two inclination
bins, with an average absolute magnitude that is 0.7--0.9 mag
brighter (Table 6). This difference becomes insignificant when only
the objects brighter than $-15$ mag are considered. The LFs in the
face-on and edge-on bins, on the other hand, show no significant
difference.  Thus, the LFs of SNe~II do not provide evidence for more
extinction in more highly inclined galaxies, in contrast with 
expectations.

We note that the LFs of SNe~II in the different inclination bins could
be affected by different subclass distributions. To investigate this,
we plot the LFs of the most common subclass in Figure 15 (SNe~II-P;
shaded histogram).  As can be seen, the SN~II-P LFs exhibit a trend
similar to that of the total SN~II LFs.

Overall, our data do not provide evidence for more extinction in more
highly inclined galaxies, a puzzling result. We emphasise, however,
that because of small-number statistics and the deviation from Poisson
statistics (due to the use of the corrected numbers of SNe), this
result should be considered preliminary and needs to be checked with a
significantly larger sample. For example, the lowest luminosity bin in
the face-on SN~II LF has a corrected number of SNe of 5.4, but it
contains only two observed objects, SNe 1999br and 2002hh. When these
two SNe are not considered, the LFs in the face-on and the
$40^\circ-75^\circ$ bins do not show a significant difference and the
LF in the edge-on bin is on average fainter by $\sim 1$ mag,
consistent with a trend due to extinction.

\subsection{LFs for Different SN Subclasses}

Since this is the first census of the subclasses for a complete sample
of SNe, it is of interest to compare the LFs of different subclasses,
as shown in Figure 16. The LFs of the different subclasses of SNe~Ia
show apparent differences. As expected, SN 1991bg-like objects are
subluminous, while SN 1991T-like objects are overluminous.  The two
groups of normal SNe~Ia with different expansion velocities exhibit a
marginal 2--3$\sigma$ difference, as indicated by the cumulative
fractions shown in the top panel.  The LF of SNe~IaHV is more skewed
toward luminous objects, while it also has more objects at the
faintest end. As discussed by Wang et al. (2009), SNe~IaHV may have a
different reddening law or colour evolution, and on average seem to
suffer more extinction than SNe~IaN.  In fact, the two SNe in the
faintest bin of the SN~IaHV LF are SN 1999cl (Blondin et al. 2009) and
SN 2006X (Wang et al. 2008), both highly reddened objects. Thus,
SNe~IaHV may be among the intrinsically brightest SNe~Ia, though
small-number statistics must be kept in mind.

SNe~Ib appear to have a different LF (brighter with a smaller scatter)
than SNe~Ic (the averages are $-17.01 \pm 0.17$ mag [$\sigma = 0.41$]
and $-16.04 \pm 0.31$ mag [$\sigma = 1.28$], respectively). However,
this result is based on small-number statistics (as reflected by the
error bars of the cumulative fractions shown in the top panel), and as
discussed in \S 3.2, the classification of SNe~Ibc into subclasses is
still quite uncertain.  More objects with definitive spectral
classifications are needed to verify this result. The peculiar SNe~Ibc
are represented by only a small number of objects and exhibit a wide
range of luminosities.

The different subclasses of SNe~II have significant differences in
their LFs.  The least to most luminous subclasses are SNe~II-P (with
an average absolute magnitude of $-15.66 \pm 0.16$ [$\sigma = 1.23$]),
SNe~IIb ($-16.65 \pm 0.40$ [$\sigma = 1.30$]), SNe~IIn ($-16.86 \pm
0.59$ [$\sigma = 1.61$]), and SNe~II-L ($-17.44 \pm 0.22$ [$\sigma =
  0.64$]).  The LF of SNe~II-P is different from that of the other
three subclasses (even when the objects fainter than $-15$ mag are not
considered), while there is no significant difference between SNe~IIb
and II-L.  SNe~IIn have a wide range of luminosities, including
several of the most luminous objects.

To investigate whether the different subclasses of SNe have any
preference in their host-galaxy Hubble types, we show the distribution
in Figure 17. While the CC~SNe display significant differences in
their LFs, their host-galaxy Hubble-type distributions do not exhibit
any significant differences. For SNe~Ia, only SN 1991bg-like objects
show a significant difference in their host-galaxy Hubble-type
distribution: they have a strong preference to occur in elliptical and
early-type spiral galaxies.  SN 1991T-like objects, generally thought
to have a strong preference to occur in spiral galaxies, are
represented by only 5 objects in our LF SN sample, so their
host-galaxy distribution is not well constrained.  We also note that
the host galaxy of SN 1998es, a SN 1991T-like object, may be
misclassified as an early S0 galaxy. Van den Bergh, Li, \& Filippenko
(2002), for example, classified the galaxy as an early-type spiral
galaxy (Sab in the DDO system).


\subsection{The Host-Galaxy Properties of the LF SNe}

Paper I discussed the host-galaxy properties of the full SN sample, in
particular the Hubble-type distribution (its \S 4.2.3 and Figure 5).
Here we examine the host-galaxy properties for the SNe in the LF
sample.

Figure 18 illustrates the histograms for the Hubble-type distribution,
the $B - K$ colour, and the absolute $K$-band luminosity $M(K)$. The
top panels show the statistics for the ``full-nosmall'' galaxy sample,
while the lower panels display the statistics for the hosts of SNe~Ia,
SNe~Ibc, and SNe~II, respectively. The histograms for the individual
SN types are drawn with solid lines for the LF SNe, while the dashed
lines are for the ``season-nosmall'' SN sample, scaled to the same number
of SNe as in the LF sample.

We note that in general, the host galaxies of individual SN types
display significant differences in their properties compared to the
``full-nosmall'' galaxy sample.  This suggests that the different SN types
have some degree of preference to occur in certain types of host
galaxies. The SN~Ia host galaxies are more skewed toward red $B - K$
colours and high $K$-band luminosities.  The CC~SNe, on the other hand,
prefer galaxies with late Hubble types, blue $B - K$ colours, and
low $K$ luminosities.

For a given type of SN, there are notable differences between the SNe
in the season-nosmall (dashed lines) and LF (solid lines)
samples. Overall, the host galaxies of the ``season-nosmall'' SN sample
tend to be skewed toward earlier-type, redder, and more luminous
galaxies. This is likely caused by the evolution of the galaxy
properties over distance in our sample due to selection biases, as
discussed in Paper I. The ``season-nosmall'' SN sample includes many SNe
that occurred in the galaxies that are more distant than the cutoff
distance for the LF SN sample, which, as discussed in \S 4.2.4 and
Figure 4 of Paper I, have a higher fraction of bright, early-type
galaxies than the more nearby galaxies.

The SN~Ia hosts in general have properties that differ from those of
the CC~SN hosts.  The SN~Ibc and SN~II hosts, on the other hand, have
similar distributions for the Hubble types and $B - K$ colours, but
different $M(K)$ distributions (with a 1.8\% probability of coming
from the same population).  The host galaxies of SNe~II are typically
less luminous than those of SNe~Ibc, with the average $M(K) = -22.92
\pm 0.12$ mag ($\sigma = 1.13$ mag) and $-23.42 \pm 0.22$ mag ($\sigma
= 1.20$ mag), respectively. If SNe~Ibc come from a similar population
of massive stars (perhaps in binary systems) as those producing
SNe~II, their preference to occur in more luminous galaxies may
indicate a metallicity effect (e.g., Tremonti et al. 2004) on the
evolution of massive stars, such as by affecting the line-driven winds
for the massive star that eventually explodes as the SN (Vink et
al. 2001; Heger et al. 2003; Vink \& de Koter 2005; Crowther
2007). Our suggestion that SNe~Ibc occur in galaxies of higher
luminosity or metallicity than SNe~II is consistent with the findings
of Prantzos \& Boissier (2003), Prieto, Stanek, \& Beacom (2008), and
Boissier \& Prantzos (2009).

We also investigate whether the different SN subclasses have different
host-galaxy $M(K)$ distributions. For SNe~Ia, the only significant
difference is found between the host galaxies of the SNe~IaN and
SNe~Ia-91bg subclasses, with the hosts of SNe~Ia-91bg being more
luminous on average due to the dominance of earlier Hubble types. The
results for the CC~SN subclasses are shown in Figure 19. The left
panels display the histograms of the $M(K)$ distributions while the
right panel shows the cumulative fractions. Several subclasses still
suffer from small-number statistics; nevertheless, we find the
following trends with varying significance.

\begin{enumerate}

\item{No significant difference is found between the host galaxies of
  SNe~II-P and II-L, though the total number of SNe~II-L is small
  (7.5).}

\item{No significant difference is found between the host galaxies of
  SNe~Ib and Ic, with a 28.0\% probability that they come from the
  same population. The average $M(K)$ of the hosts of SNe~Ib ($-24.20
  \pm 0.46$ mag [$\sigma = 1.15$ mag]) appears to be marginally more
  luminous than the hosts of SNe~Ic ($-23.22 \pm 0.34$ mag [$\sigma =
    1.35$ mag]). In the cumulative fraction plot, the two subclasses
  are combined. }

\item{The host galaxies of SNe~IIb are more luminous than those of
  SNe~II-P, with a 6.9\% probability that they come from the same
  population.  The average $M(K)$ values are $-23.54 \pm 0.42$ mag
  ($\sigma = 1.28$ mag) and $-22.84 \pm 0.14$ mag ($\sigma = 1.11$ mag),
  respectively. Furthermore, there is a relatively high probability
  (68.5\%) that the host galaxies of SNe~IIb and Ibc come from the
  same population, as can be seen by the similar cumulative fraction
  curves in the right-hand panel of Figure 19.}

\item{The host galaxies of SNe~IIn are less luminous than those of
  SNe~II-P, with a 10.3\% probability that they come from the same
  population.  The average $M(K)$ value is $-22.08 \pm 0.54$ mag
  ($\sigma = 1.40$ mag) for the SN~IIn host galaxies.}

\end{enumerate} 

As discussed in Paper I, the LOSS galaxy sample involves several
selection biases, and is not complete at the low-luminosity end.
Nevertheless, since all of the SNe were discovered in the same set of
galaxies and thus suffer from the same selection biases, the above
trends still reveal the general preference for the different
subclasses of SNe in terms of host-galaxy $K$-band luminosities.

Recently, Arcavi et al. (2010) reported on the statistics of 70 CC~SNe
found by PTF and suggested that there might be an excess of SNe~IIb
and Ib in dwarf galaxies, which differs from our finding that the host
galaxies of both subclasses prefer more luminous galaxies. As PTF is
conducting an untargeted, magnitude-limited survey and monitors
numerous dwarf galaxies, the study by Arcavi et al. (2010) is
complementary to ours. The differences in the results might be caused
by the relatively small numbers of objects in both studies, although
Poisson statistics have nominally been taken into account. Perhaps the
discrepancy is related to the dissimilar analysis methods; we study
the differences in the $M(K)$ distributions of the SN host galaxies,
while Arcavi et al. (2010) divided the galaxies into two categories
(giant/dwarf) and analysed the fractions of the different SN
subclasses. Further studies are needed to verify the apparent
discrepancy between the two results.

\subsection{Possible Limitations and Caveats for Our LFs}

One limiting factor of our LFs is the total number of SNe in the
sample. Although much effort has been made to analyse the data for the
175 SNe in our LF sample, many analyses still face small-number
statistics, such as the LFs for the different subclasses of SNe.  As
shown in Figure 4, the cutoff distance for the SN~Ia LF sample can be
increased to 120~Mpc to include more SNe without introducing large
corrections due to incompleteness. For the CC~SNe, the inclusion of
the data in the years 2007--2008 will help increase the sample in the
LF.  We are in the process of reducing more data to continue working
along these directions, and the results will be published in a future
paper.

Another limiting factor of our LFs is that they are only available for
the $R$ band. For SNe~Ia, a significant fraction of the LF SNe have
been followed in $BVRI$, and the multi-colour LFs will be presented in
a future paper. For SNe~Ibc and SNe~II, only a small fraction of the
LF SNe have filtered follow-up photometry, so determining their
multi-colour LFs is not possible at this time.  It will require
considerable effort to obtain filtered photometry for all of the
relatively low-luminosity CC~SNe in either a volume-limited or
magnitude-limited survey to construct the LFs in different passbands.

One concern is that the LFs we derived only apply to our galaxy sample
with its specific Hubble type, colour, and luminosity distributions. 
As discussed in detail in Paper I, however, the galaxy sample within 
the cutoff distance for the LFs is probably {\it representative} of 
galaxies with moderate and large sizes, and only has an apparent 
deficit for small galaxies.  This deficit may cause our LFs to be 
biased against those SNe having a preference to occur in small 
galaxies. In our study, we only find a possible preference for 
SNe~IIn to occur in small, late-type spiral galaxies.

We excluded the SNe that occurred in small (major axis $< 1\arcmin$),
early-type galaxies because of uncertainties in the detection
efficiencies.  One concern is whether this exclusion introduces an
observational bias. No SNe~Ibc or II are excluded because CC~SNe in
early-type galaxies are rare (\S 4.2.3 in Paper I).  For SNe~Ia, only
two objects (SNe 2000dk [IaN] and 2006H [Ia-91bg]) are
excluded. Considering that the total SN~Ia LF has 74 SNe~Ia, the
inclusion of the two additional SNe will have negligible effect on the
overall properties of the SN~Ia LF.

One important question is whether there is a sizable fraction of
highly reddened SNe that are missed in our search.  Some SNe certainly
experience a large amount of extinction; for example, the SN~Ia
2002cv\footnote{ SN 2002cv was not discovered (directly or
  independently) in our search even though its host galaxy, NGC 3190,
  is in our galaxy sample. However, the reason we missed the SN is not
  high host-galaxy extinction; rather, our scheduler considered NGC
  3190 to be too far toward the west at the beginning of night and
  terminated the monitoring of the galaxy for the season. SN 2002cv
  would have been discovered in our search if NGC 3190 were actively
  being monitored at the time of discovery, because the unfiltered
  peak magnitude of SN 2002cv is $\sim 1$ mag brighter than our
  typical limiting magnitude.}  (Di Paola et al. 2002; Elias-Rosa et
al. 2008) has $A_V \approx 9$ mag, while the SN~II-P 2002hh suffered
$A_V \approx 5$ mag (Pozzo et al. 2006). Searches done at
near-infrared and radio wavelengths also suggest that the vast
majority of SNe in massive {\it starburst} galaxies, such as luminous
infrared galaxies (LIRGs) and ultraluminous infrared galaxies
(ULIRGs), are missed by the optical searches due to dust obscuration
(e.g., Mannucci et al. 2003).

We argue, however, that our LFs are not significantly affected by
host-galaxy extinction for the following reasons.  First, LIRGs/ULIRGs
constitute only a very small fraction of the galaxy
population.\footnote{The {\it Infrared Astronomical Satellite} 
({\it IRAS}) Revised Bright Galaxy Sample (Sanders et al. 2003) only
  contains 13 galaxies within 60~Mpc having far-infrared luminosities
  characteristic of LIRGs ($> 10^{11}$\,L$_{\odot}$).}  Second, for a
non-starburst galaxy, although the theoretical studies of Hatano et
al. (1998) and Riello \& Patat (2005) suggest that SNe should
experience more extinction in more highly inclined galaxies, our
investigation of the LFs in different inclination bins (\S 5.2) does
not provide strong supporting evidence.  Third, statistics provided by
the observed SN sample (Jha et al. 2006a; Hicken et al. 2009) indicate
that the majority of the observed SNe~Ia do not suffer significant
amounts of extinction. Finally, our own LFs provide additional
evidence for the scarcity of highly reddened events: only a few of the
175 SNe in the LF sample suffer a large amount of extinction (SNe
2001ci, 2002hh, 2003bk, 2003cg, and 2005bb). For our typical search
limiting magnitude of 19, we can detect SNe brighter than $-12.5$,
$-13.4$, and $-14.9$ mag within 20, 30, and 60~Mpc, respectively.
These are several magnitudes fainter than the average SNe in the LF,
so we should have discovered many more moderately reddened objects
($A_V$ of a few mag) near our detection limit if highly reddened SNe
were quite common.  We conclude that our LFs are not significantly
affected by the missing SNe due to high extinction in the targeted
sample galaxies.

\subsection{Possible Applications of our LFs}

Our LF data (tables and template light curves) are made available
to interested parties electronically. These LFs have the following
potential applications.

\begin{enumerate}

\item{The LFs can be used to calculate the control times for the
  different types of SNe in a SN search, which is a critical step in
  deriving the SN rates. This is the main motivation for our
  study. Paper III will discuss the details of how the LFs are used to
  calculate the control times in our SN search. Any other SN searches
  conducted without filters or using a passband that is similar to $R$
  could use our LFs to help with the control-time and/or
  survey-completeness calculation.  A major advantage of using our LFs
  to calculate the control times is that they are ``pseudo-observed"
  and account for the poorly known host-galaxy extinction.  }

\item{The LFs can be used to simulate the expected SN subclass and
  brightness distribution in a SN search (with known cadence and
  depth), to help the coordination of follow-up efforts. For this
  purpose, the SN rates derived in Paper III are needed as well. }

\item{The LFs, the light-curve distributions, and the observed
  subclass fractions can be used as priors in a photometry-based
  classification scheme, as in Poznanski et al. (2002) and Poznanski,
  Maoz, \& Gal-Yam (2007).}

\item{The LFs can be used to constrain the possible progenitor systems
  and their evolutionary paths for the different types of SNe. Viable
  models should be able to explain both the luminosity distribution
  and the various subclass fractions (e.g., Smith et al. 2011a). }

\end{enumerate}

\section{Conclusions} 

Historically, SN rate calculations have been plagued by two issues:
the intrinsic luminosity distribution of SNe and the host-galaxy
extinction toward SNe. In other words, the calculations were limited
by our knowledge of the {\it observed} luminosity functions of SNe. In
this Paper II of a series aimed to derive a precise nearby SN rate
from the Lick Observatory SN Search, a volume-limited SN sample is
constructed for the first time, and the observed luminosity functions
of SNe are derived.

We first select a volume-limited sample of 175 SNe (with a cutoff
distance of 80~Mpc for SNe~Ia, and 60~Mpc for SNe~Ibc and SNe~II), and
then collect photometry for {\it every} object. Families of light
curves for each SN type are constructed from the literature and/or our
own photometry database, and are used to fit the light curves of the
SNe, to generate peak absolute magnitude and light-curve shape
distributions. We further study the completeness of each SN in the LF,
and correct them to 100\% completeness within the considered volume.

The volume-limited LFs of SNe indicate that a Gaussian scatter around
an average luminosity is generally not a good representation of the
data. There are also significant differences for the LFs in different
host-galaxy Hubble types. For SNe~Ia, the SNe in E--Sab galaxies are
generally fainter than those in Sb--Irr galaxies due to the prominence
of subluminous SN 1991bg-like objects in the former galaxies. For
SNe~Ibc, the objects in early-type spirals are, on average, slightly
fainter than those in late-type spirals.  For SNe~II, the objects in
early-type spirals are, on average, brighter than those in late-type
spirals. These observed trends have significant implications for their
possible progenitor systems and evolutionary paths.


We also have detailed subclass information for all SNe in the LF
sample. While spectral series are adequate to classify SNe~Ia and Ibc
into different subclasses, detailed light-curve information is
necessary to discriminate the different subclasses of SNe~II,
especially SNe~II-P, II-L, and IIb. In a volume-limited sample, SNe~II
are the most abundant type (57\% of all), while SNe~Ia and Ibc
constitute 24\% and 19\%, respectively. For SNe~Ia, normal objects are
70\% of all, SN 1991bg-like objects are 18\%, and the rest are SN
1991T-like and SN 2002cx-like objects (12\%).  The normal SNe~Ia are
further split into objects with normal (50\%) and high (20\%)
expansion velocities. SNe~Ic are the most abundant SNe~Ibc (54\% of
all), while SNe~II-P are the most abundant SNe~II (70\% of all). Among
SNe~II, there are significant fractions of SNe II-L, IIb, and IIn
(10\%, 12\%, and 9\%, respectively).

We further derive the observed LFs and SN subclass fractions for an
ideal magnitude-limited search (i.e., with a short observation
interval) by scaling the SNe with their survey volume. Compared to the
volume-limited LFs, the magnitude-limited LFs have an enhanced
fraction of luminous objects, as well as reduced scatter in the
average luminosity. The observed fractions of SNe have also
dramatically changed. SNe~Ia are the most numerous SNe (79\%) of the
three types, while the fractions of the core-collapse SNe shrink to
17\% and 4\% for SNe~II and Ibc, respectively. Within SNe~Ia, normal
SNe~Ia are 77\% of all, and SN 1991T-like objects are boosted to 18\%.
SNe~Ibc become as abundant as SNe~Ic among the SNe~Ibc. The fractions
of SNe~II-L, IIb, and IIn are enhanced due to their higher
luminosities than those of SNe~II-P.  We compare our predicted
subclass fractions to two observed magnitude-limited samples, one in
the random background galaxies in our own search, and the other from
PTF, and find good agreement.

We also investigate the effect of the observation interval in a
magnitude-limited search on the observed LFs and SN
fractions. Searches done with an observation interval smaller than
10~d have similar LFs and SN fractions, and discover a high fraction
of the SNe in an ideal magnitude-limited search. When the observation
interval is long, the fractions for the SNe with relatively slow light
curves are enhanced. In a search with a very long interval (or a
single snapshot), only $\sim 9$\% of the SNe in an ideal
magnitude-limited search are discovered, and SNe~II-P become the
dominant subclass (40\% of the total) among SNe~II.

We discuss how the LFs we derived change with different environments
and subclasses of SNe. We have not found a persistent correlation
between the LFs of SNe and their host-galaxy sizes, although some
subclasses of SNe seem to have a preference to occur in certain galaxy
sizes in some Hubble types (e.g., SNe~IIn prefer small, late-spiral
galaxies). Surprisingly, the LFs in galaxies of different inclination
do not provide strong evidence in support of greater extinction toward
SNe in more highly inclined galaxies. The different subclasses of SNe
display significant differences in their LFs. For SNe~Ia, the SN
1991bg-like objects are subluminous, while the SN 1991T-like objects
are overluminous. The normal SNe~Ia with high expansion velocities
display a more extreme LF than the normal SNe~Ia having normal
expansion velocities, suggesting that they may belong to two distinct
groups.  SNe~Ib are, on average, more luminous and have a smaller
scatter than SNe~Ic, but this result should be reexamined in a larger
sample with more definitive spectral identifications than our current
sample.  The least to most luminous SNe~II are II-P, IIb, IIn, and
II-L.  Despite the significant difference in the LFs, the different
subclasses of core-collapse SNe show similar host-galaxy Hubble-type
distributions.  For SNe~Ia, SN 1991bg-like objects prefer to occur in
elliptical and early-type spiral galaxies. We note that some of these
results have been found in previous work (e.g., Della Valle \& Livio
1994; Hamuy et al. 1996; Howell 2001; Jha et al. 2006a).

We also compare the host-galaxy properties of the LF SNe, and find a
significant difference in the galaxy luminosity distributions for
SNe~II and Ibc. SNe~Ibc prefer more massive galaxies than SNe~II,
suggesting an influence of metallicity on the mass-loss history in
their evolution. We also find that SNe~IIb prefer more massive
galaxies than SNe~II-P, while SNe~IIn prefer less massive galaxies.

We discuss possible limitations of our LFs; small-number statistics
are the primary one. Caution should be used when applying our LFs to
low-luminosity galaxies, but our LFs do not appear to be seriously
affected by SNe missing due to large extinction.

Our LFs can be used to help with SN rate determinations for any
searches using a passband similar to the $R$ band. Other applications
of the LFs are to coordinate follow-up efforts in large surveys, help
photometry-based classification methods, and constrain viable models
for the SNe.


\section*{Acknowledgments}

We thank the referee, Enrico Cappellaro, for useful comments and
suggestions which improved the paper.  We are grateful to the many
students, postdocs, and other collaborators who have contributed to
the Katzman Automatic Imaging Telescope and the Lick Observatory
Supernova Search over the past two decades, and to discussions
concerning the determination of supernova rates --- especially Frank
Serduke, Jeffrey Silverman, Thea Steele, and Richard R. Treffers.  We
thank the Lick Observatory staff for their assistance with the
operation of KAIT.  LOSS, conducted by A.V.F.'s group, has been
supported by many grants from the US National Science Foundation (NSF;
most recently AST-0607485 and AST-0908886), the TABASGO Foundation, US
Department of Energy SciDAC grant DE-FC02-06ER41453, and US Department
of Energy grant DE-FG02-08ER41563. KAIT and its ongoing operation were
made possible by donations from Sun Microsystems, Inc., the
Hewlett-Packard Company, AutoScope Corporation, Lick Observatory, the
NSF, the University of California, the Sylvia \& Jim Katzman
Foundation, and the TABASGO Foundation.  We give particular thanks to
Russell M. Genet, who made KAIT possible with his initial special
gift; former Lick Director Joseph S. Miller, who allowed KAIT to be
placed at Lick Observatory and provided staff support; and the TABASGO
Foundation, without which this work would not have been completed.
J.L. is grateful for a fellowship from the NASA Postdoctoral Program.
D.P. is supported by an Einstein Fellowship.  X.W.  acknowledges NSFC
grants (10673007, 11073013) and the China-973 Program 2009CB824800.
M.M. acknowledges NSF grants AST-0205808 and AST-0606772, as well as
the Miller Institute for Basic Research in Science (UC Berkeley), for
support during the time over which part of this work was conducted.
We made use of the NASA/IPAC Extragalactic Database (NED), which is
operated by the Jet Propulsion Laboratory, California Institute of
Technology, under contract with NASA. We acknowledge use of the
HyperLeda database (http://leda.univ-lyon1.fr).

\newpage

\begin{figure*}
\includegraphics[scale=0.7,angle=270,trim=0 0 0 0]{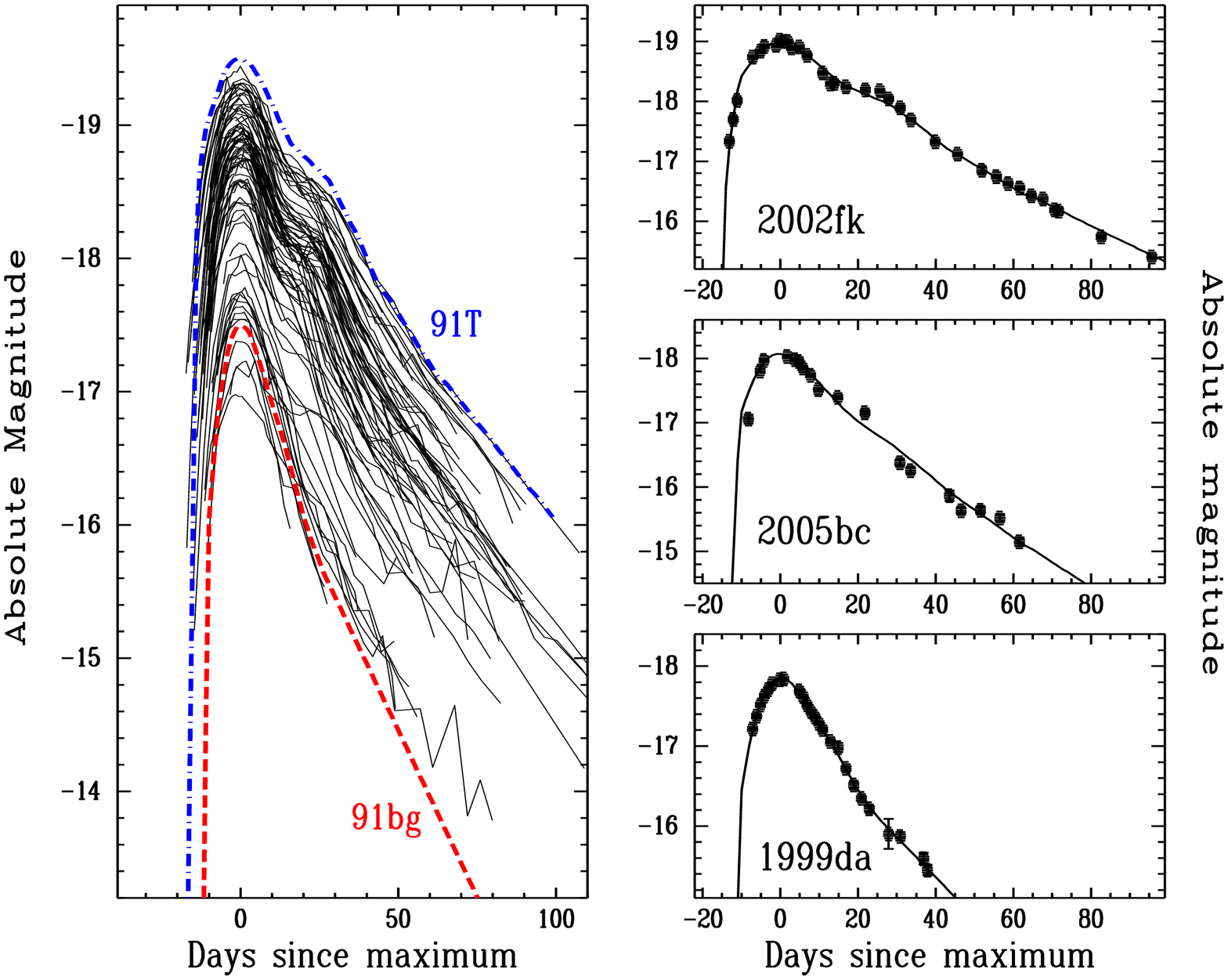}
\caption[] { The light-curve fitting process for the SNe~Ia. In the
  left panel, the solid lines are the observed $R$-band light curves
  in our photometry database (Ganeshalingam et al. 2010), while the
  smoothed light curves of SNe 1991T (dash-dotted line; Lira et
  al. 1998) and 1999by (dashed line, marked as ``91bg"; Garnavich et
  al. 2004) are placed with an arbitrary peak absolute magnitude of
  $-19.5$ and $-17.5$ mag, respectively. A family of 21 light curves
  is interpolated between these two extreme curves, and is used to fit
  the individual objects shown in the right-hand panels.
  }
\label{1}
\end{figure*}

\begin{figure*}
\includegraphics[scale=0.7,angle=270,trim=0 0 0 0]{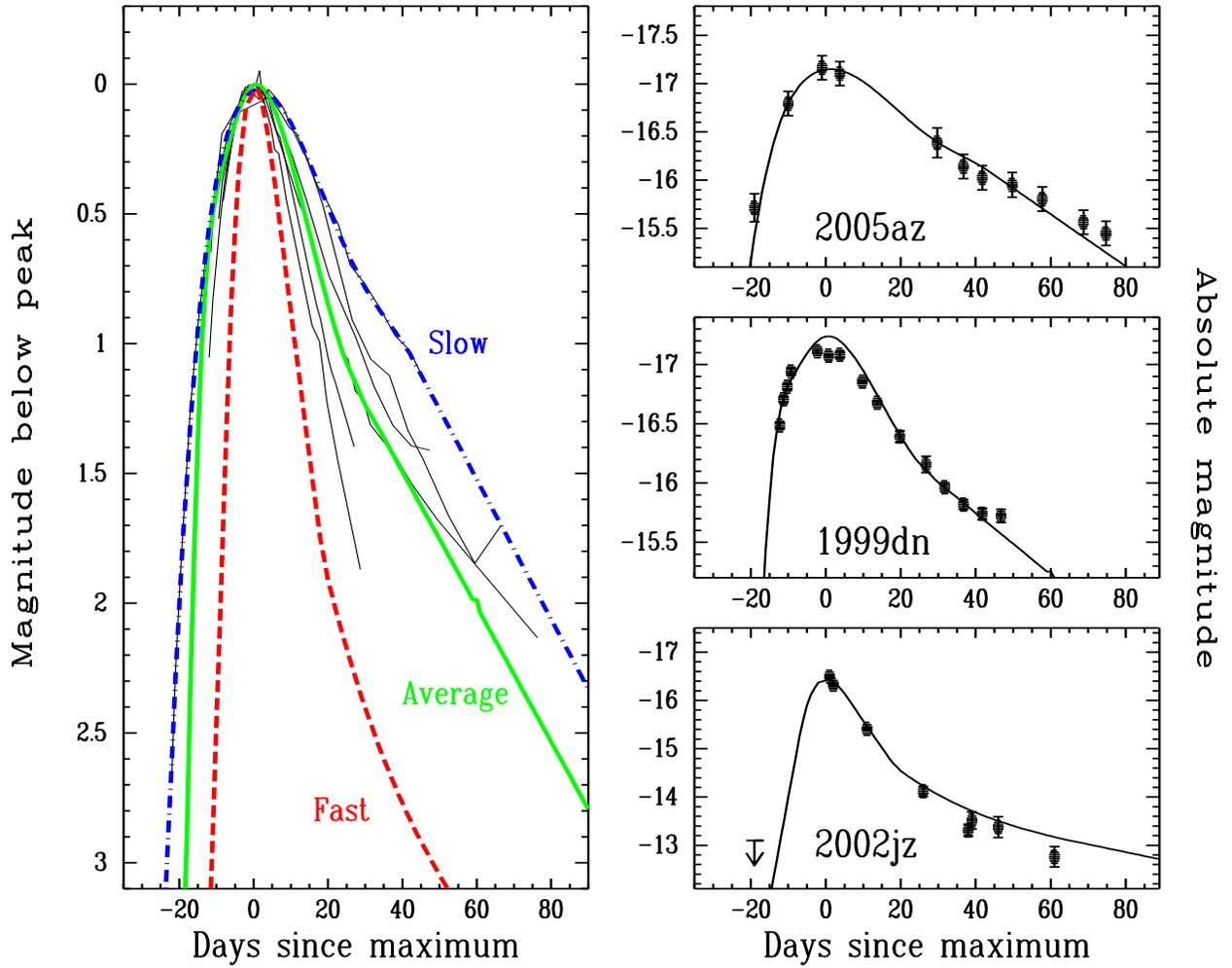}
\caption[] { Same as Figure 1 but for the SNe~Ibc. A family of three
  light curves (fast, average, and slow) is constructed (see text for
  details), and is used to fit the individual objects shown in the
  right panels. }
\label{2}
\end{figure*}

\begin{figure*}
\includegraphics[scale=0.9,angle=270,trim=0 80 0 0]{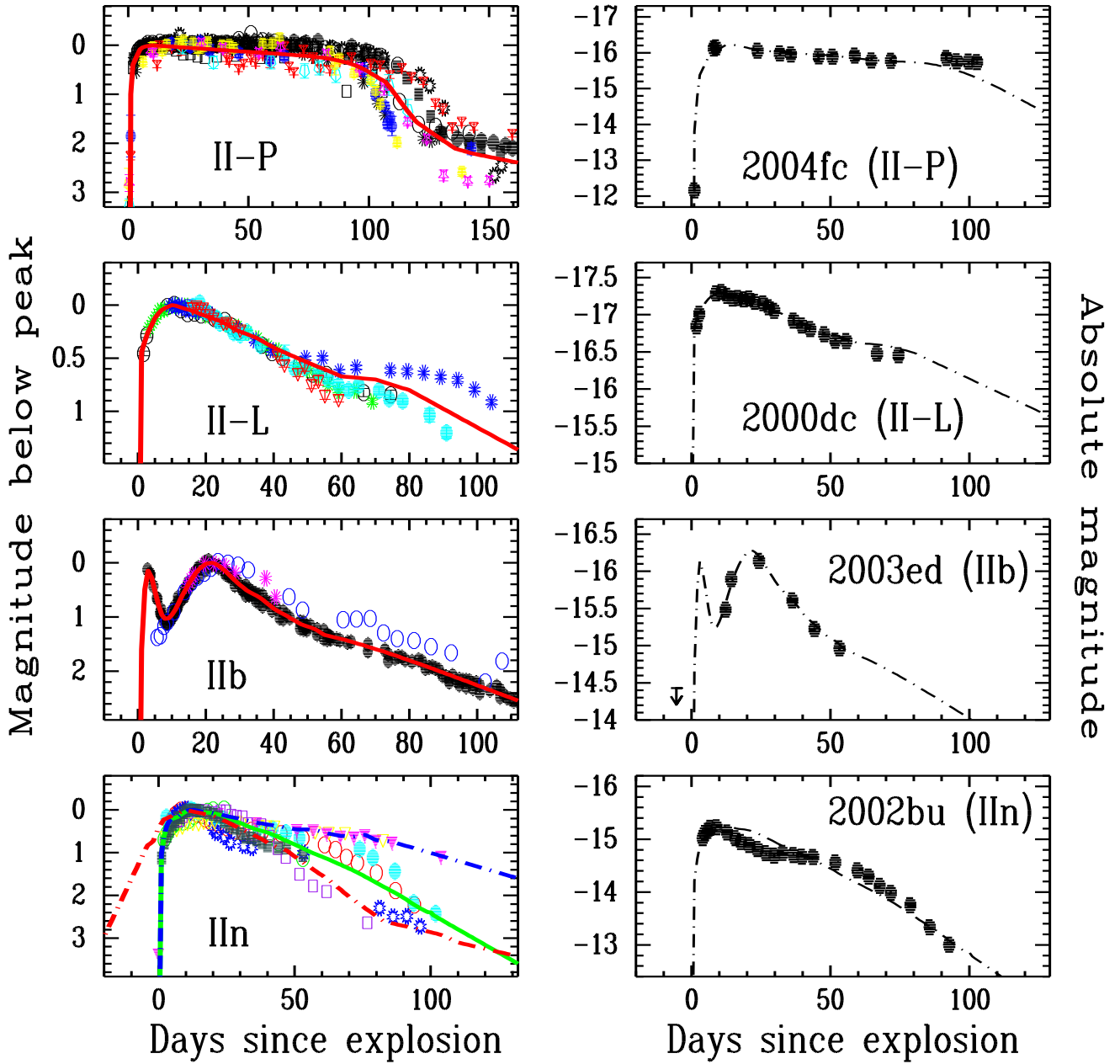}
\caption[] { Same as Figure 1 but for the SNe~II. A single average
  light curve is constructed for the subclasses of SNe~II-P, II-L, and
  IIb, while three light curves (fast, average, and slow) are for 
  SNe~IIn. The fast SN~IIn light curve (dash-dotted line) is plotted
  relative to days since maximum brightness. The right panels show an
  example fit for each subclass.  }
\label{3}
\end{figure*}

\clearpage

\begin{figure*}
\includegraphics[scale=1.0,angle=270,trim=0 80 0 0]{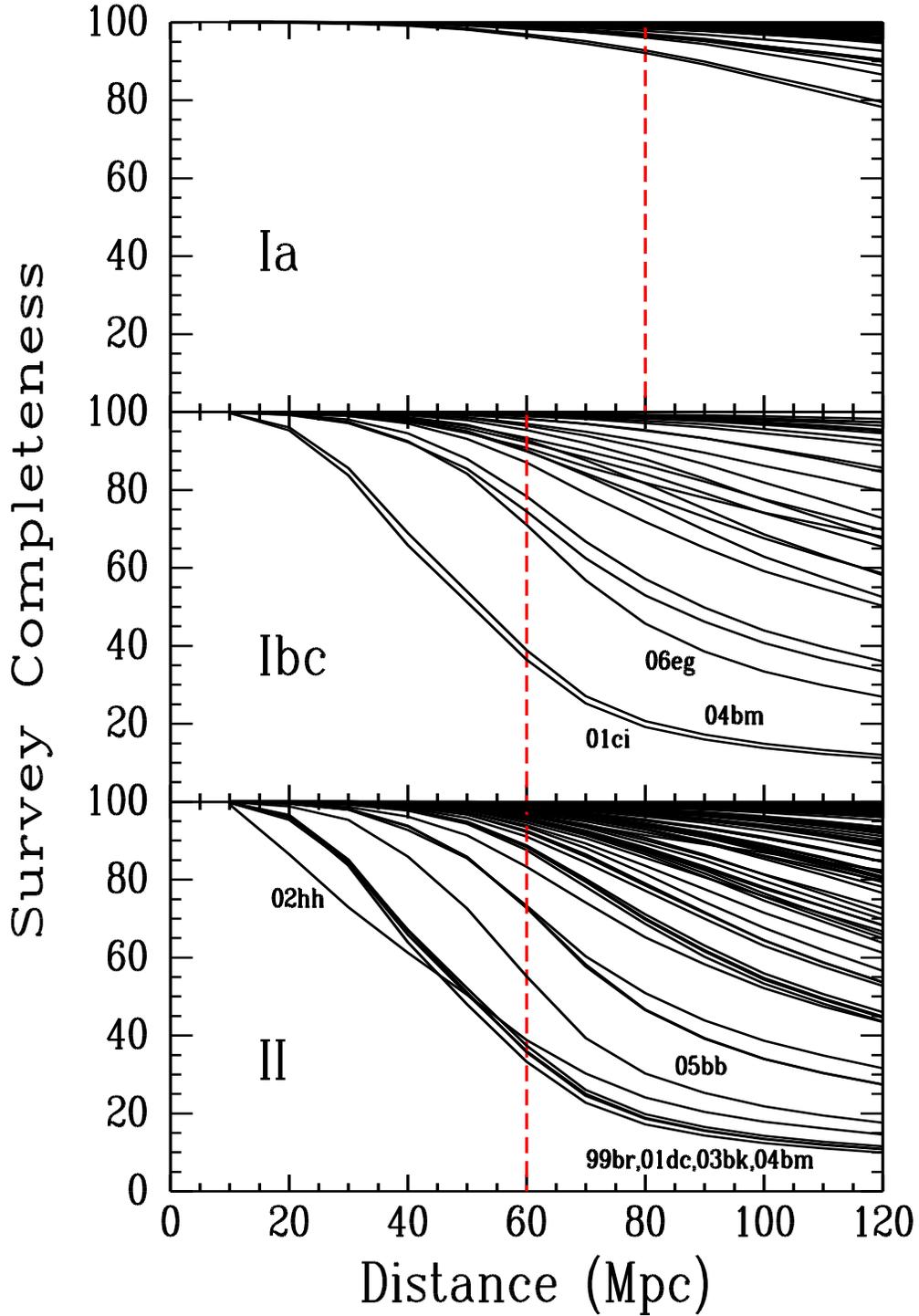}
\caption[] { The completeness of each SN in the LF sample in our SN
  search. The completeness is defined as the ratio of the total
  control time divided by the total season time. Some notable SNe are
  marked. The dashed lines mark the cutoff distances within which the
  LF samples are constructed.  }
\label{4}
\end{figure*}

\clearpage 

\begin{figure*}
\includegraphics[scale=0.9,angle=270,trim=0 30 0 0]{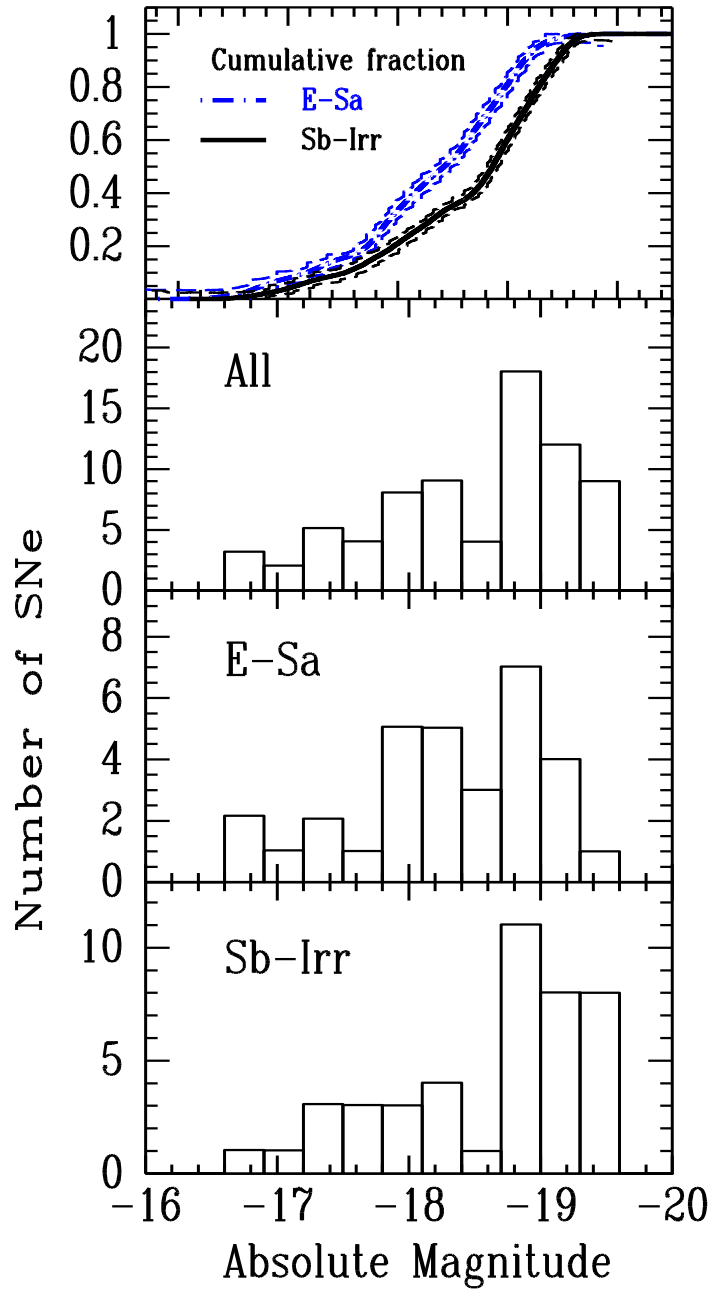}
\caption[] { The pseudo-observed LFs of SNe~Ia. The top panel shows
  the cumulative fractions for the LFs in two different galaxy
  Hubble-type bins (E--Sa and Sb--Irr).  The dashed lines show the
  $1\sigma$ spread in the cumulative fractions considering only the
  uncertainties in the peak absolute magnitudes. The bottom panels
  show the LFs in all, E--Sa, and Sb--Irr galaxies. }
\label{5}
\end{figure*}

\clearpage 

\begin{figure*}
\includegraphics[scale=0.9,angle=270,trim=0 30 0 0]{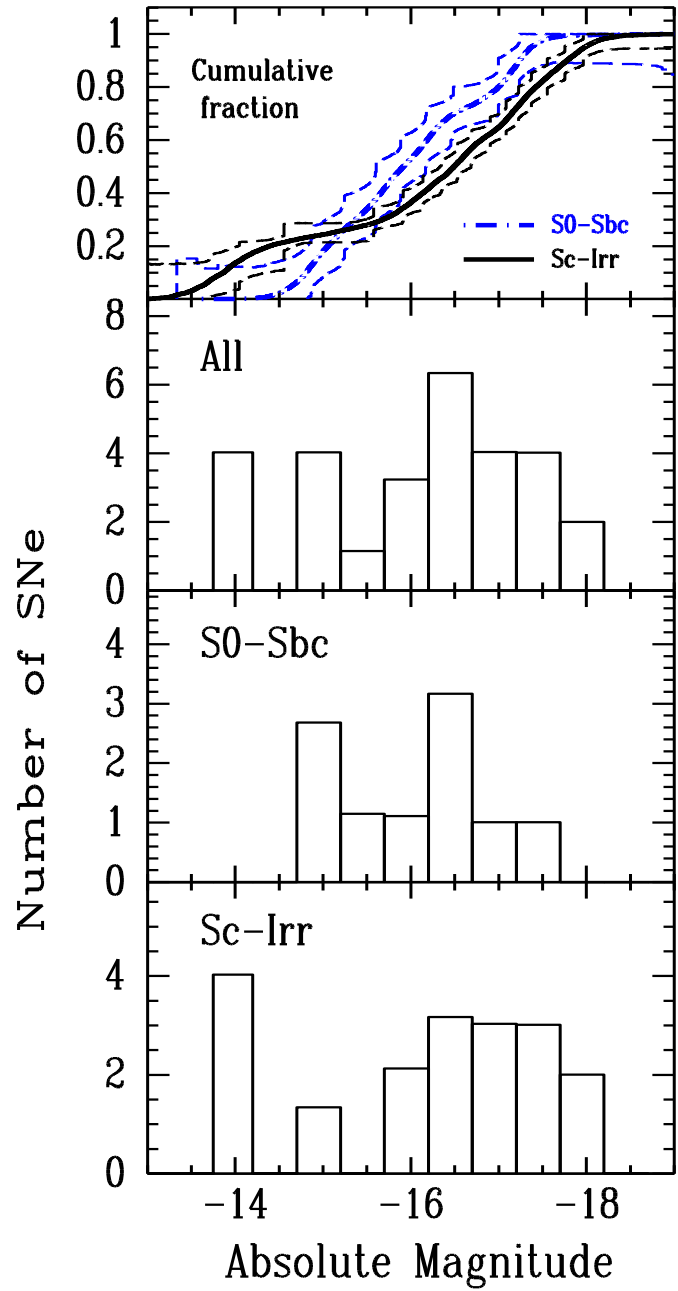}
\caption[] { Same as Figure 5 but for SNe~Ibc. The galaxies are split
  into S0--Sbc and Sc--Irr bins. }
\label{6}
\end{figure*}

\clearpage

\begin{figure*}
\includegraphics[scale=0.9,angle=270,trim=0 30 0 0]{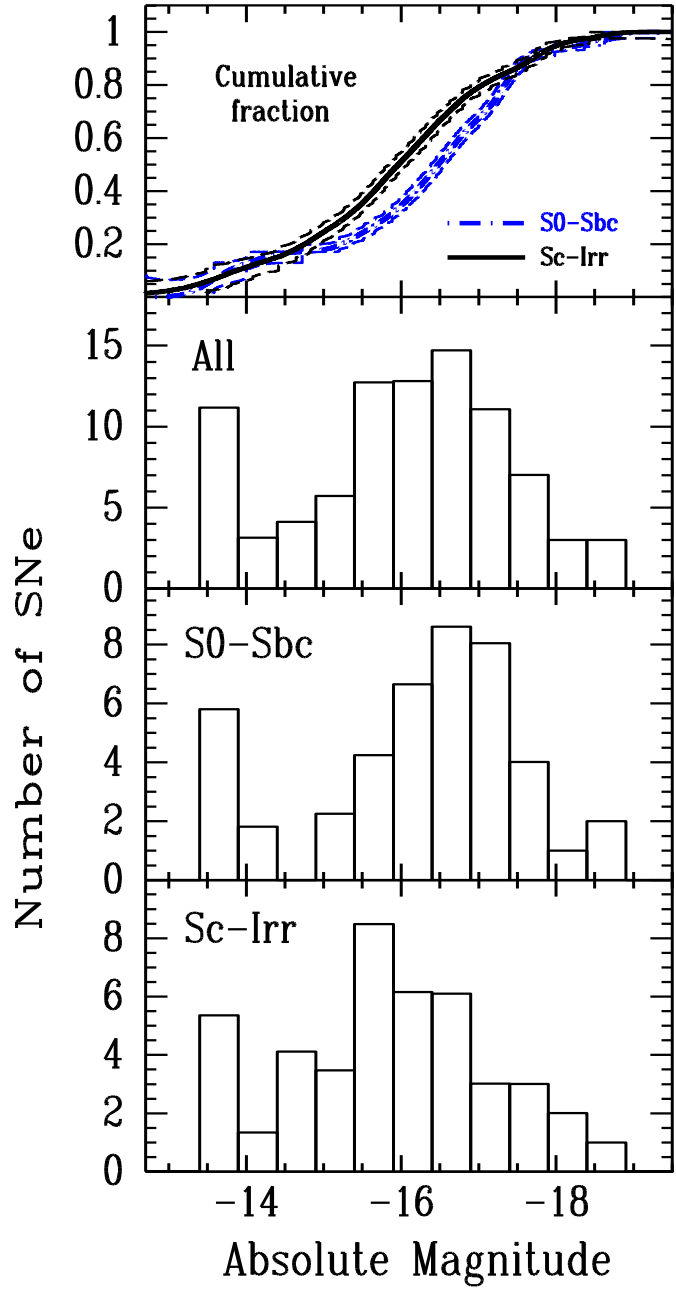}
\caption[] {
Same as Figure 6 but for SNe~II.
}
\label{7}
\end{figure*}

\clearpage

\begin{figure*}
\includegraphics[scale=0.7,angle=270,trim=0 30 0 0]{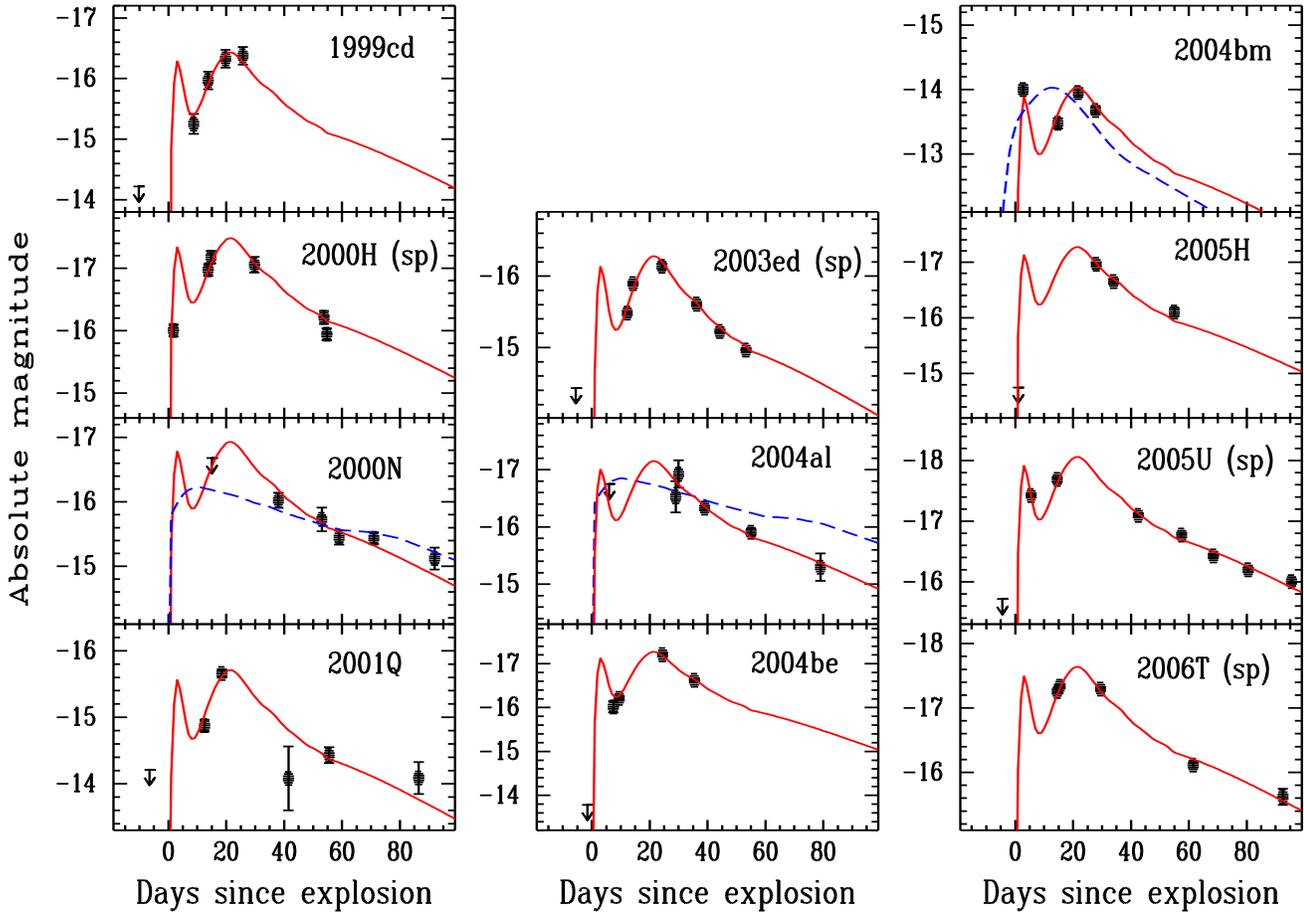}
\caption[] { Possible SNe~IIb in the LF sample. All of these objects
  were spectroscopically classified as SNe~II based on the presence of
  hydrogen lines, but their light curves are best fit with a SN~IIb
  template. The SNe labeled with ``(sp)" were also spectroscopically
  confirmed as SNe~IIb. The dashed lines are the fits with template
  light curves of SNe~II-L (for SNe 2000N and 2004al) and SNe~Ibc (for
  SN 2004bm). See text in \S 3.2 for more details. }
\label{8}
\end{figure*}

\clearpage

\begin{figure*}
\includegraphics[scale=0.14,angle=0,trim=0 0 0 0]{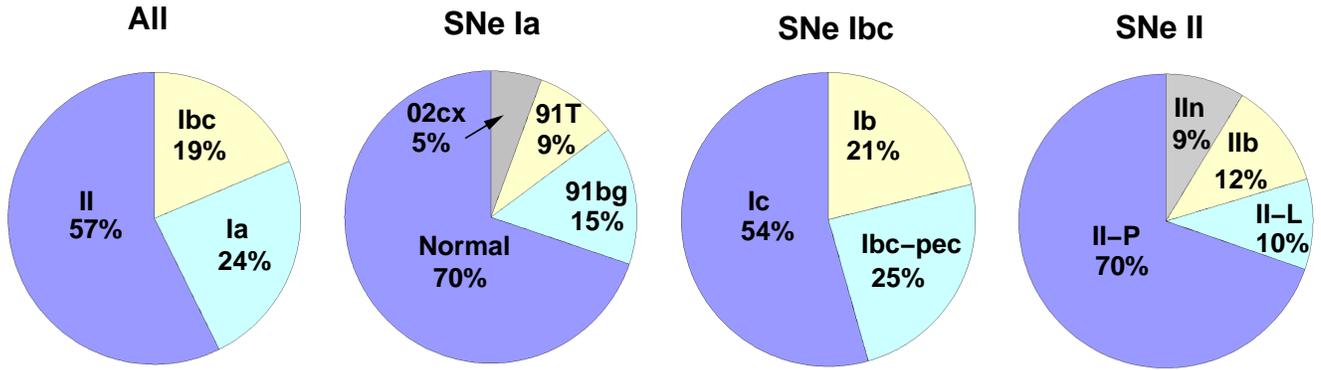}
\caption[] { The observed fractions of the subclasses of SNe in a
  volume-limited sample, illustrated as pie charts. The fractions of
  SNe~Ic and IIb are upper limits, while that of SN 1991T-like objects
  is a lower limit. Also, the subclass of SNe~Ibc-pec consists of
  broad-lined SNe~Ic, peculiar objects, and the ``Ca-rich" objects (see
  text for more details). }
\label{9}
\end{figure*}

\clearpage

\begin{figure*}
\includegraphics[scale=1.0,angle=270,trim=0 80 0 0]{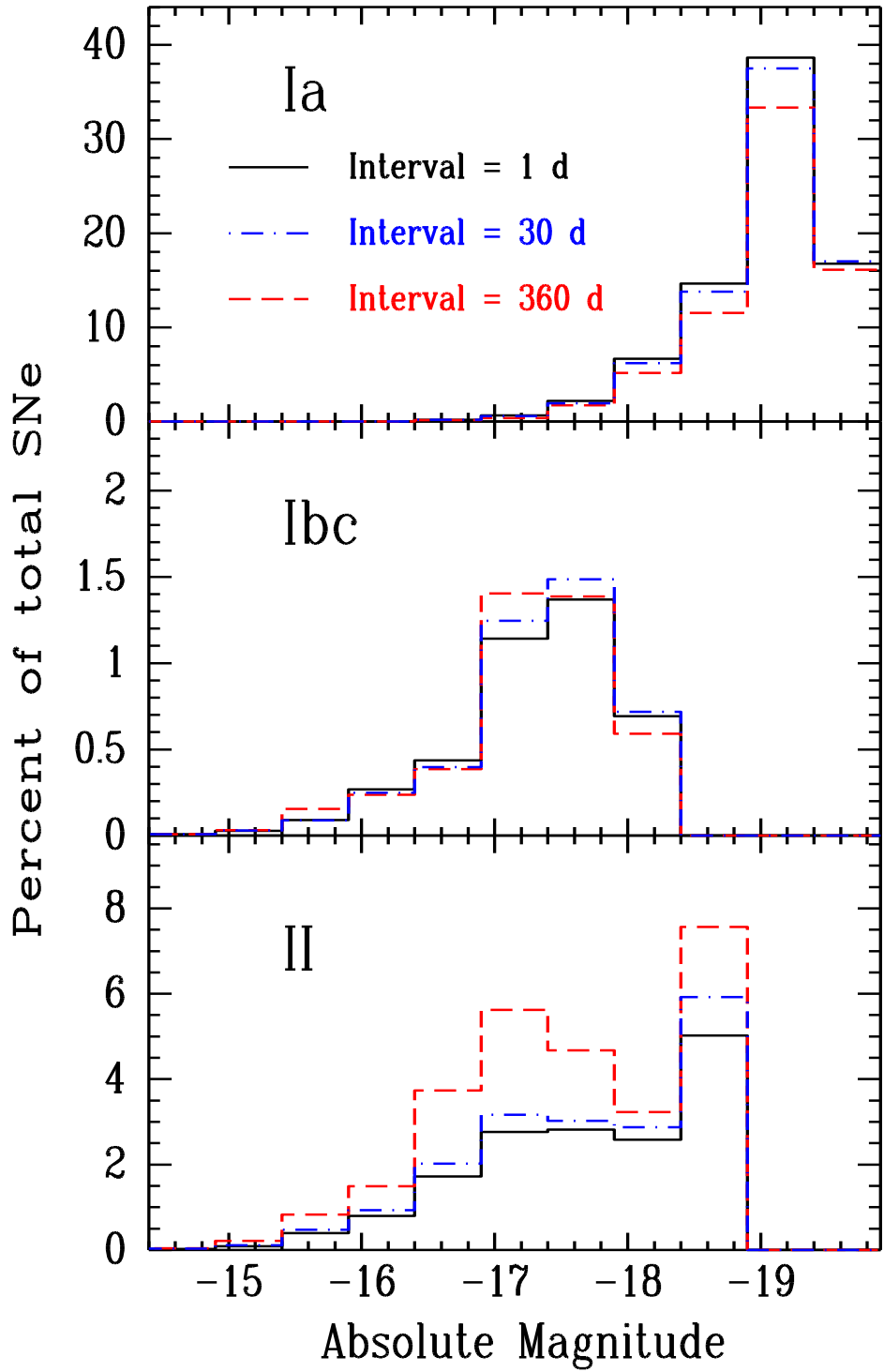}
\caption[] { The observed LFs in a magnitude-limited sample. The solid
  line shows the results of an ideal magnitude-limited survey (with an
  observation interval of 1~d), while the dot-dashed and dashed lines
  show the results with longer observation intervals. }
\label{10}
\end{figure*}

\clearpage

\begin{figure*}
\includegraphics[scale=0.14,angle=0,trim=0 0 0 0]{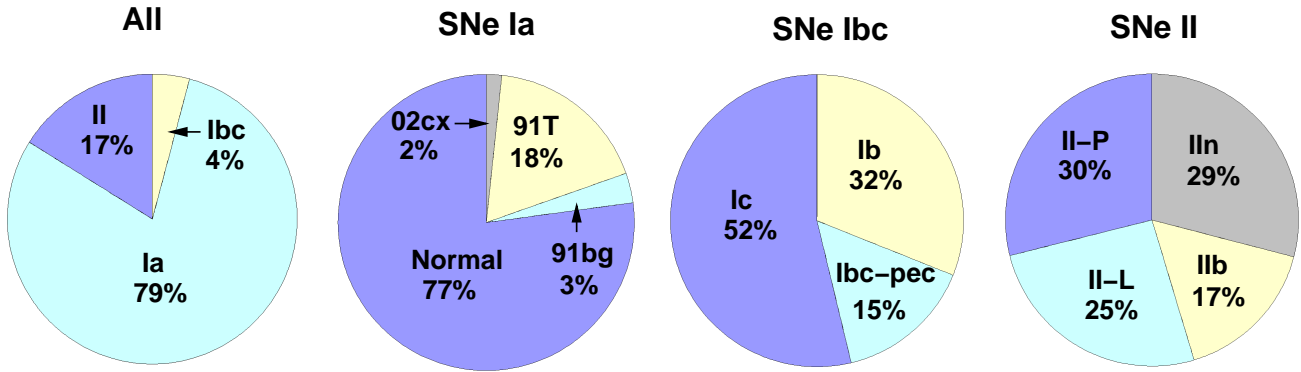}
\caption[] { The observed fractions of the subclasses of SNe in an
  ideal magnitude-limited sample, illustrated as pie charts. }
\label{11}
\end{figure*}

\clearpage

\begin{figure*}
\includegraphics[scale=0.8,angle=270,trim=0 30 0 0]{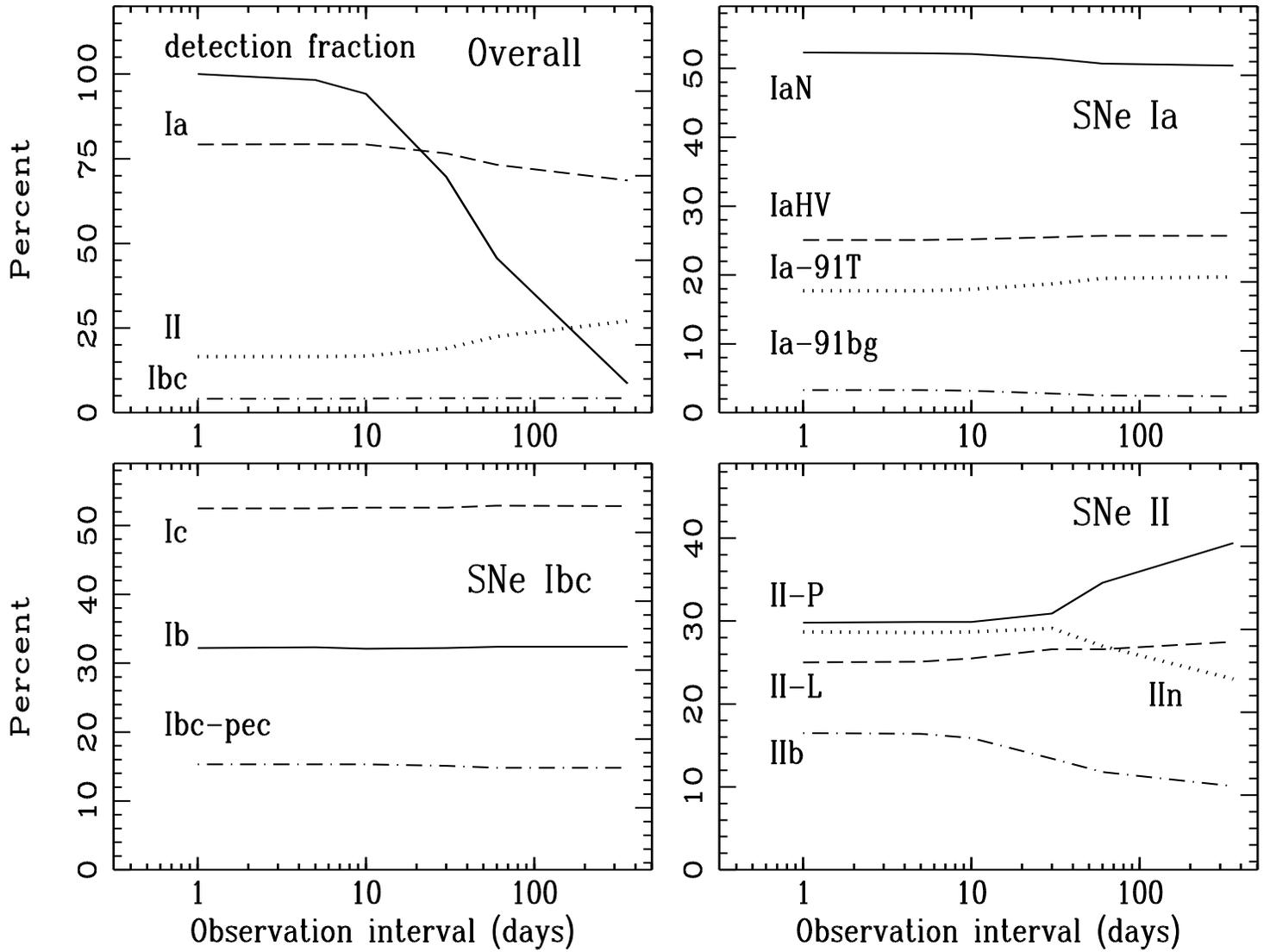}
\caption[] { The observed subclass fractions in a magnitude-limited
  sample as a function of observation interval. Also shown in the
  upper-left panel is the ``detection fraction" curve. See text for
  details.  }
\label{10}
\end{figure*}

\clearpage

\begin{figure*}
\includegraphics[scale=0.8,angle=270,trim=0 40 0 0]{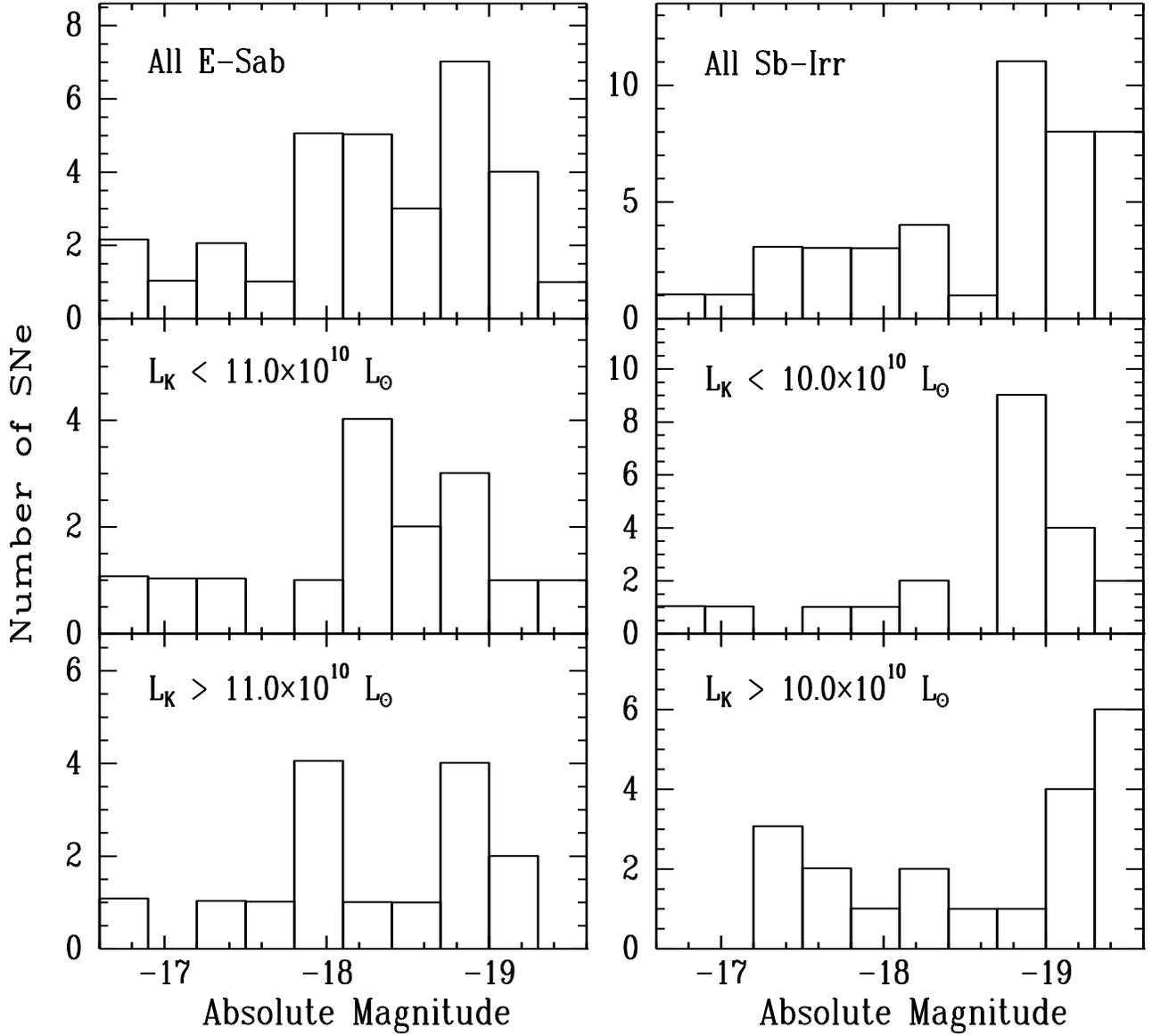}
\caption[] { The luminosity functions of SNe~Ia in galaxies of
  different luminosities. The top panels show the LFs in the total
  sample, while the bottom two panels split the LF into two bins
  according to the $K$-band luminosities of their host galaxies.  }
\label{12}
\end{figure*}

\clearpage

\begin{figure*}
\includegraphics[scale=0.8,angle=270,trim=0 40 0 0]{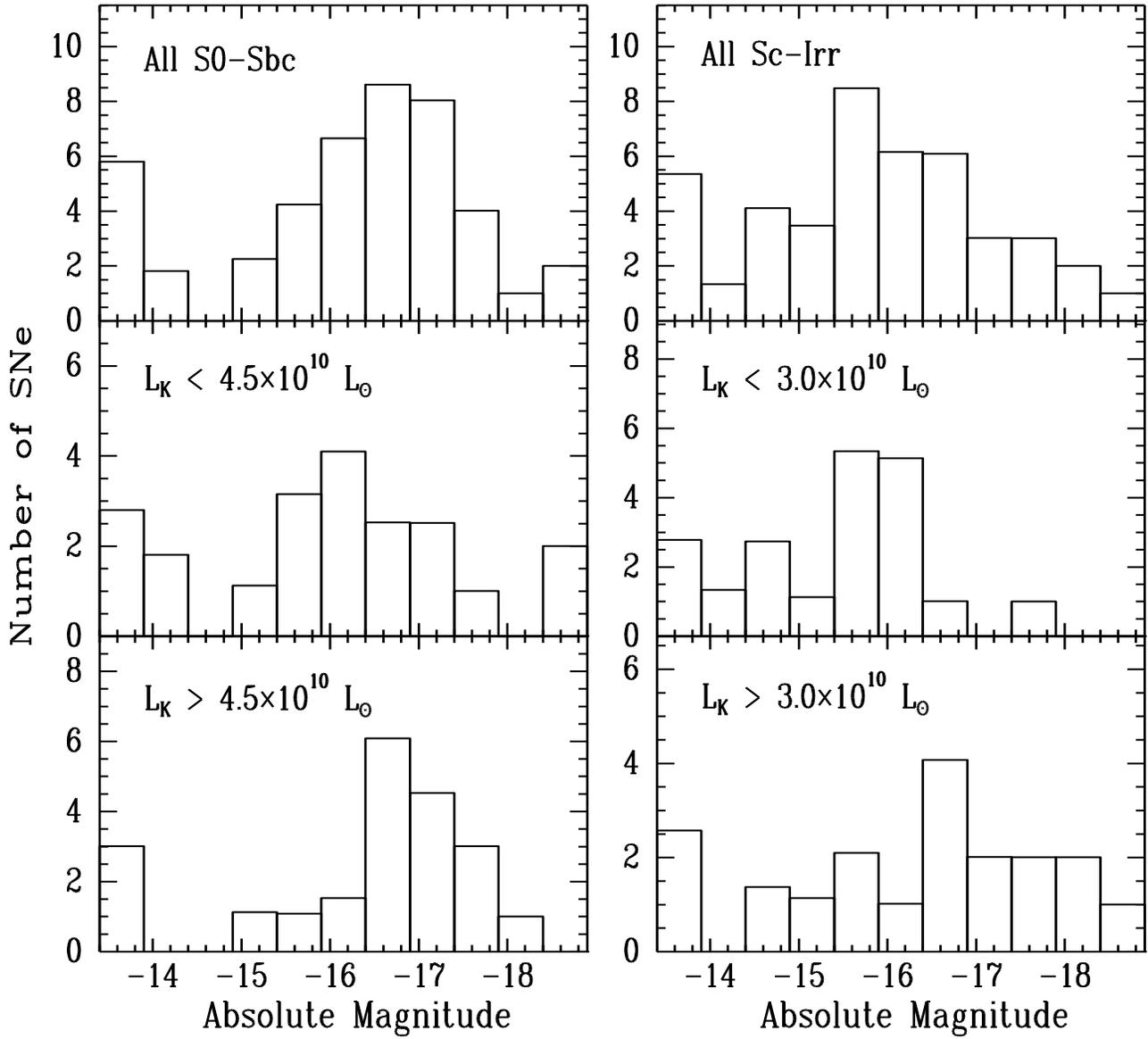}
\caption[] {
Same as Figure 12 but for the LFs of SNe~II.
}
\label{13}
\end{figure*}

\clearpage

\begin{figure*}
\includegraphics[scale=0.7,angle=270,trim=0  0 0 0]{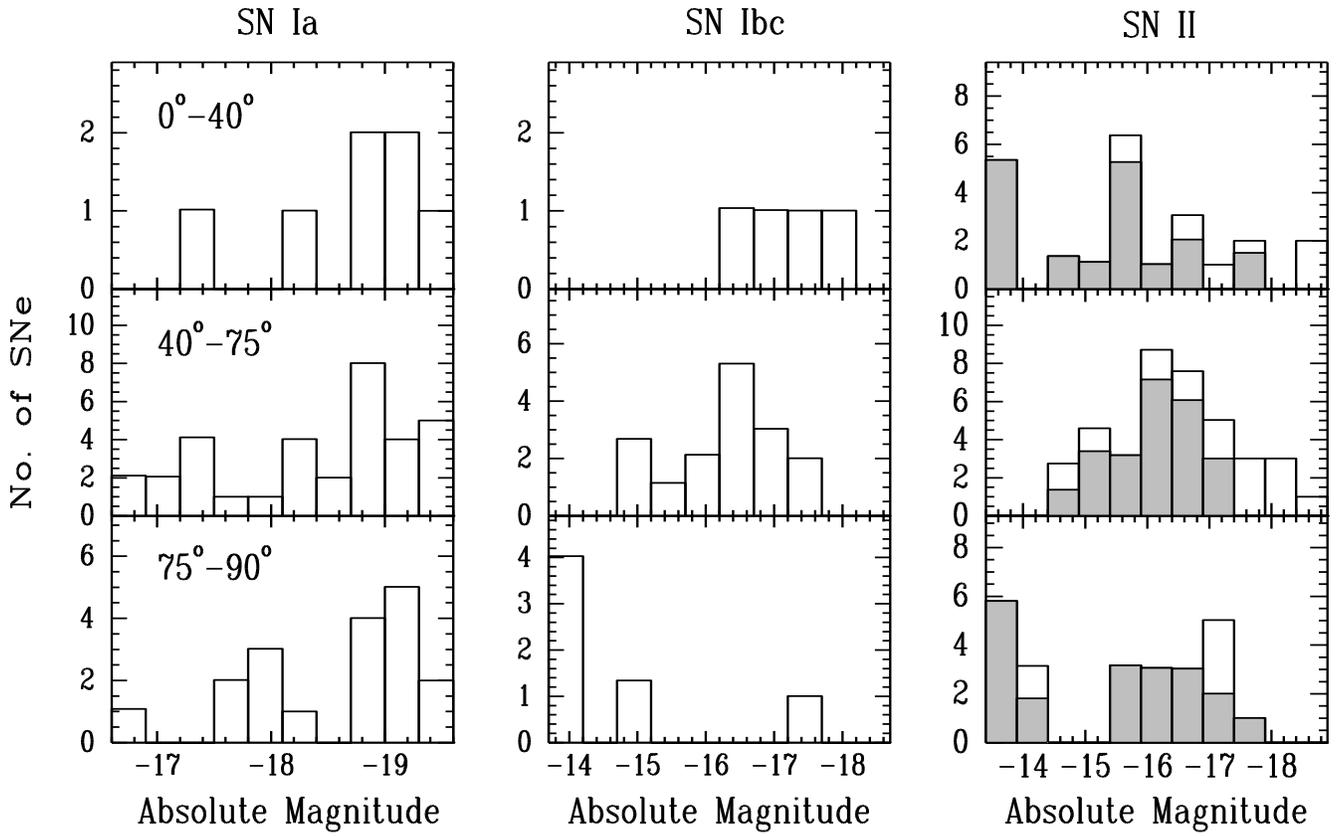}
\caption[] { The LFs of SNe in host galaxies with different
  inclinations.  For the SN~II LFs, the shaded histogram is for the LF
  of the SNe~II-P in the sample.  }
\label{14}
\end{figure*}

\clearpage

\begin{figure*}
\includegraphics[scale=0.7,angle=270,trim=0  0 0 0]{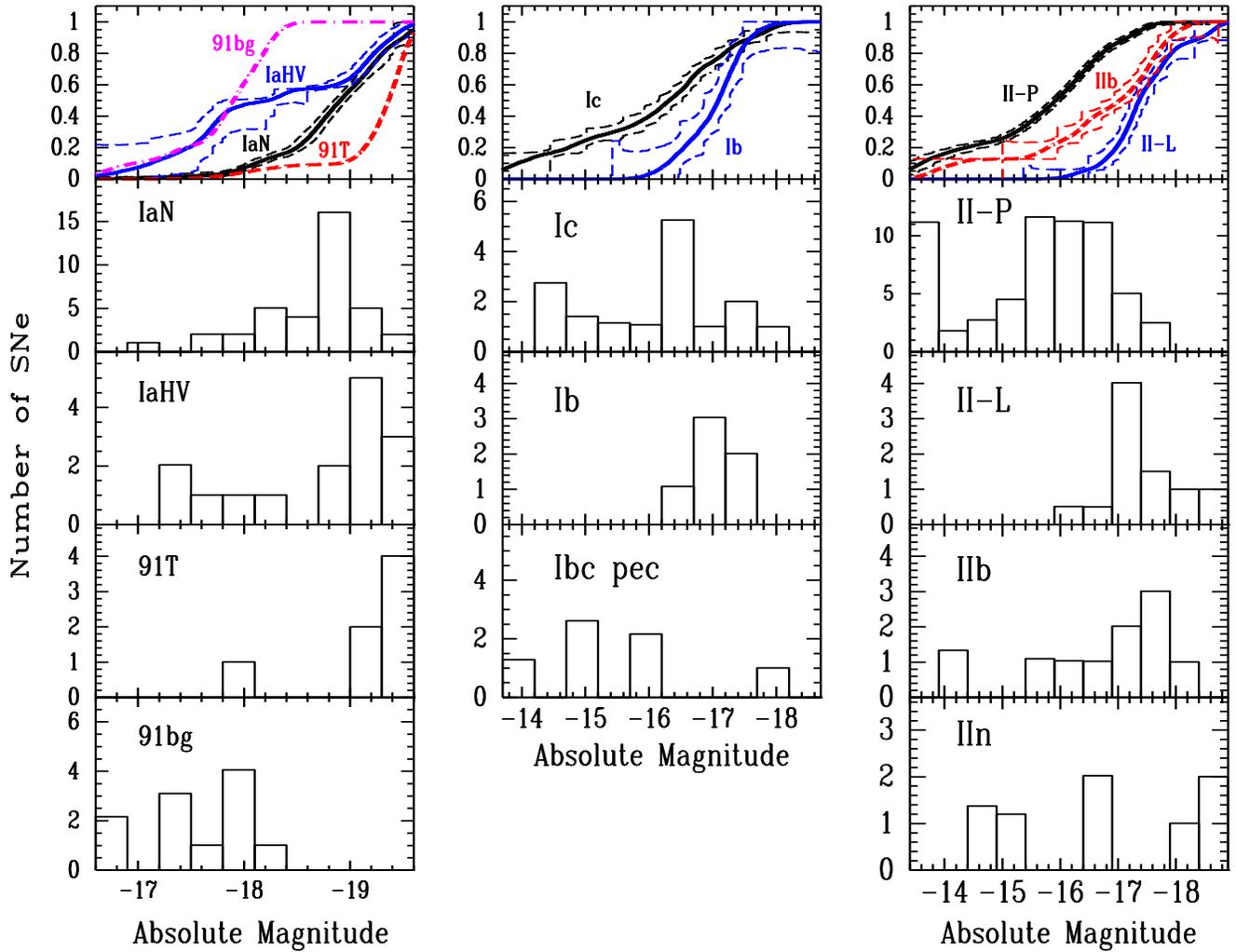}
\caption[] { The LFs of SNe in different subclasses. The top panels
  show the cumulative fractions for selected SN subtypes.  The dashed
  lines show the $1\sigma$ spread in the cumulative fractions
  considering only the uncertainties in the peak absolute magnitudes.
}
\label{15}
\end{figure*}
\newpage

\begin{figure*}
\includegraphics[scale=0.7,angle=270,trim=0  0 0 0]{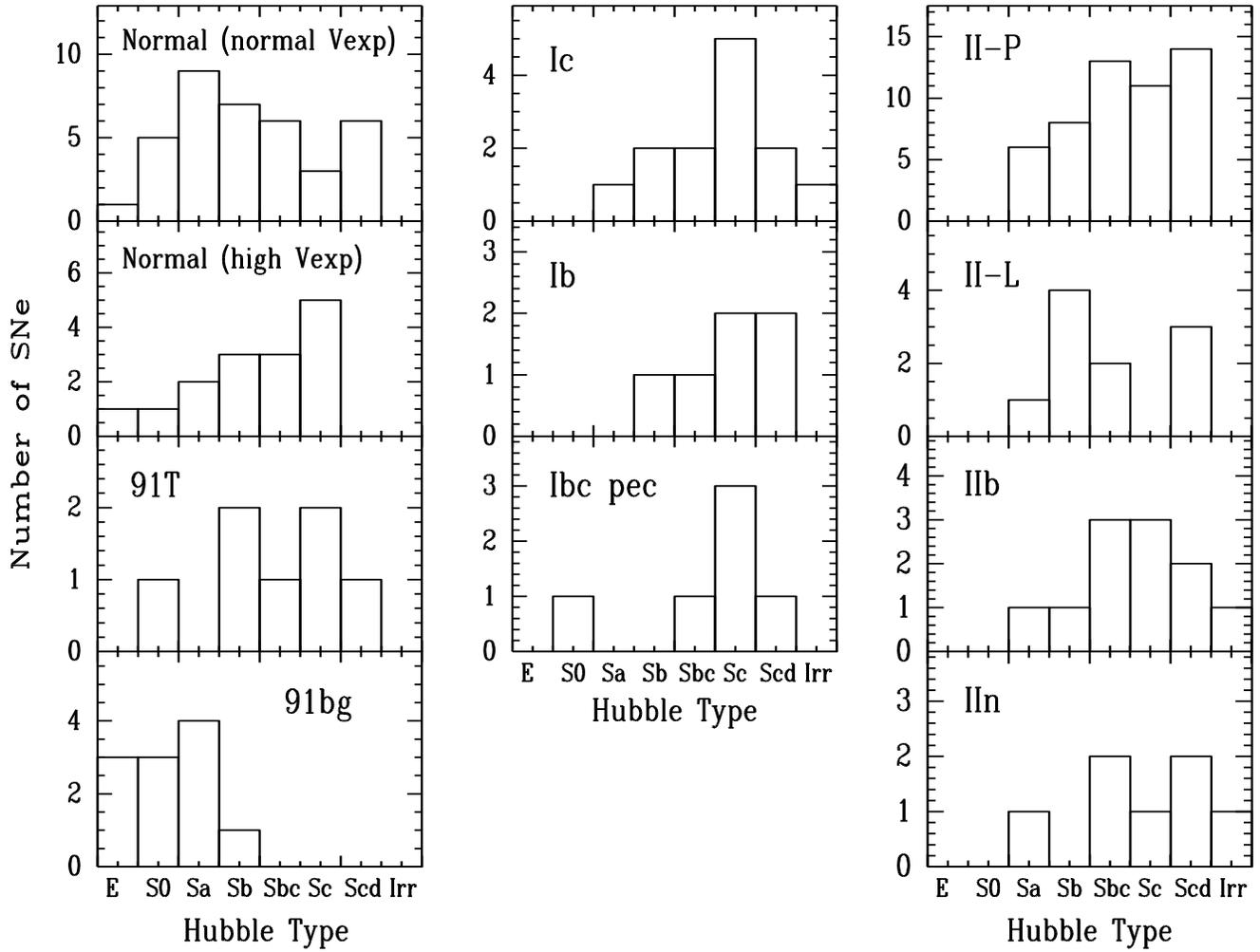}
\caption[] { The Hubble-type distribution of the different SN
  subclasses in the LF sample.  }
\label{16}
\end{figure*}
\newpage

\newpage

\begin{figure*}
\includegraphics[scale=0.7,angle=270,trim=0  0 0 0]{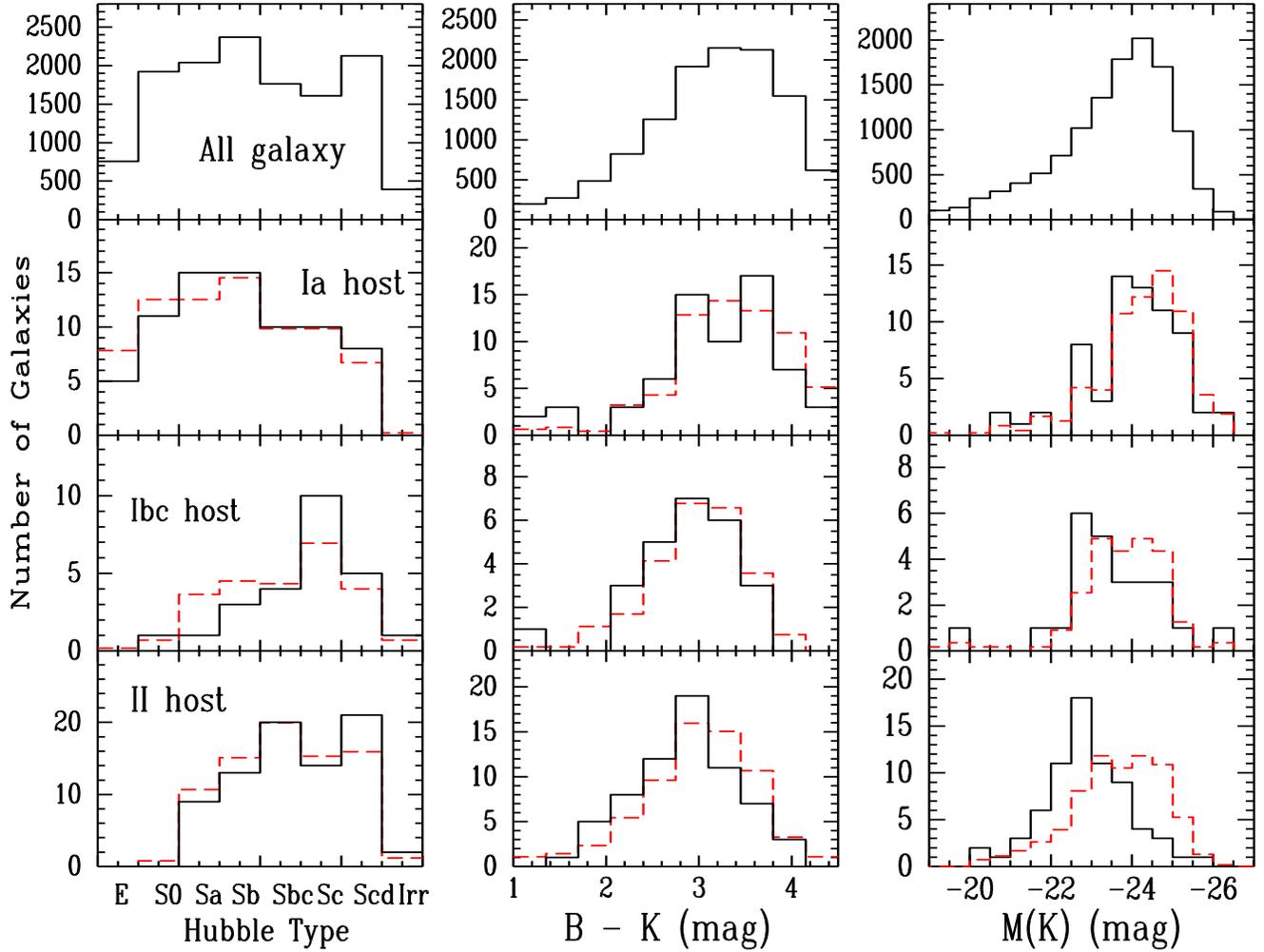}
\caption[] { The SN host-galaxy properties. The left panels show the
  Hubble-type distribution, the middle panels the $B - K$ colour,
  and the right panels the $K$-band absolute magnitude, $M(K)$.
  From top to bottom, the statistics are shown for the ``full-nosmall''
  galaxy sample, the SN~Ia hosts, the SN~Ibc hosts, and the SN~II
  hosts. For the hosts of individual SN types, the solid lines are for
  the LF sample, while the dashed lines are for the ``season-nosmall'' SN
  sample scaled to the same number of SNe as in the LF sample.  }
\label{17}
\end{figure*}

\clearpage

\newpage

\begin{figure*}
\includegraphics[scale=0.7,angle=270,trim=0  0 0 0]{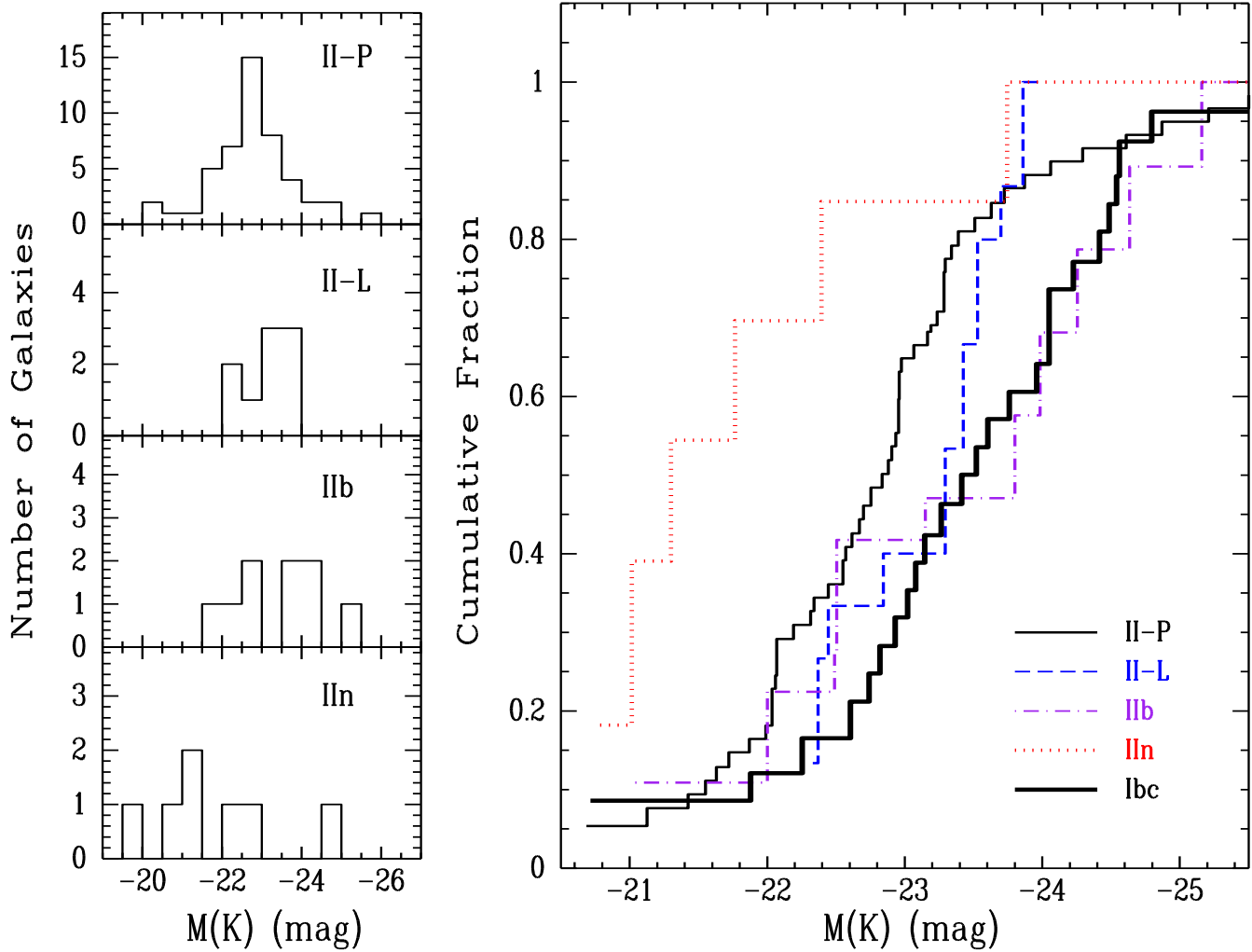}
\caption[] { The host-galaxy $M(K)$ distribution for the different SN
  subclasses (left panels) and the cumulative fractions (right panel)
  for the LF SN sample.  }
\label{17}
\end{figure*}

\renewcommand{\arraystretch}{0.70}

\clearpage
\newpage
\begin{table*}
\begin{minipage}{140mm}
\caption{Host galaxies of supernovae in the luminosity function sample.} 
\begin{tabular}{lllrrrrrr}
\hline
\hline
{SN} & {Type} &{Host Galaxy} &{$h$}
&{Dist (Mpc)} &{$B_0$ (mag)} &{$B_0$(err)}
&{$K$ (mag)} &{$K$(err)} \\ 
\hline
1998dm	&IaN	&UGCA-017	&6	&25.8	&12.128	&0.683	&10.442	&0.055\\
1999cp	&   IaN	& NGC-5468	& 7	& 39.4	& 12.767	& 0.208	& 10.396	& 0.059\\
1999ej	&   IaN	& NGC-0495	& 2	& 57.8	& 13.237	& 0.391	& 9.965	& 0.031\\
1999ek	&   IaN	& UGC-03329	& 5	& 72.0	& 11.960	& 0.500	& 9.780	& 0.032\\
1999gd	&IaN	&NGC-2623	&3	&77.0	&13.163	&0.146	&10.427	&0.027\\
2000dm	&IaN	&UGC-11198	&3	&64.2	&13.500	&0.407	&10.532	&0.034\\
2000dr	&   IaN	& IC-1610	& 2	& 75.6	& 13.685	& 0.359	& 9.844	& 0.038\\
2001L	&IaN	&MCG-01-30-011	&4	&62.5	&13.058	&0.419	&9.916	&0.039\\
2001dn	&   IaN	& NGC-0662	& 5	& 79.3	& 14.081	& 0.148	& 10.797	& 0.045\\
2001ep	&   IaN	& NGC-1699	& 4	& 52.0	& \nodata	& \nodata	& 10.629	& 0.050\\
2001fh	&IaN	&Anon.-Gal.	&7	&57.8	&11.174	&0.153	&8.531	&0.029\\
2002cr	&   IaN	& NGC-5468	& 7	& 39.4	& 12.767	& 0.208	& 10.396	& 0.059\\
2002do	&   IaN	& MCG-+07-41-001	& 1	& 68.4	& 13.262	& 0.116	& 9.076	& 0.024\\
2002fk	&   IaN	& NGC-1309	& 5	& 27.2	& 11.730	& 0.106	& 9.102	& 0.029\\
2002ha	&   IaN	& NGC-6962	& 3	& 58.5	& 12.351	& 0.079	& 8.786	& 0.029\\
2002hw	&   IaN	& UGC-00052	& 6	& 72.3	& 14.139	& 0.321	& 10.405	& 0.060\\
2002jg	&IaN	&NGC-7253B	&6	&63.8	&13.200	&0.893	&99.999	&0.000\\
2003F	&   IaN	& UGC-03261	& 7	& 70.9	& 13.620	& 0.366	& 10.635	& 0.074\\
2003cg	&   IaN	& NGC-3169	& 3	& 16.9	& 10.897	& 0.086	& 7.283	& 0.021\\
2003du	&   IaN	& UGC-09391	& 7	& 30.1	& 14.652	& 0.080	& \nodata	& \nodata\\
2003gt	&IaN	&NGC-6930	&3	&63.8	&12.683	&0.200	&99.999	&0.000\\
2003kf	&IaN	&MCG-02-16-002	&4	&28.6	&12.576	&0.500	&10.935	&0.054\\
2004ab	&   IaN	& NGC-5054	& 5	& 23.4	& 10.875	& 0.130	& \nodata	& \nodata\\
2004bd	&   IaN	& NGC-3786	& 3	& 39.2	& 13.027	& 0.172	& 9.338	& 0.025\\
2004bl	&IaN	&CGCG-013-112	&7	&71.5	&13.700	&0.381	&12.684	&0.153\\
2005W	&   IaN	& NGC-0691	& 5	& 37.2	& 11.687	& 0.202	& 8.822	& 0.038\\
2005am	&IaN	&NGC-2811	&3	&30.7	&11.596	&0.108	&7.976	&0.015\\
2005as	&   IaN	& NGC-3450	& 4	& 53.7	& 12.314	& 0.204	& 8.501	& 0.048\\
2005bc	&   IaN	& NGC-5698	& 4	& 53.5	& 13.208	& 0.307	& 10.287	& 0.049\\
2005bo	&   IaN	& NGC-4708	& 3	& 56.2	& 13.488	& 0.290	& 10.140	& 0.053\\
2005cf	&   IaN	& MCG--01-39-003	& 2	& 27.1	& 14.293	& 0.512	& 11.293	& 0.073\\
2005de	&IaN	&UGC-11097	&4	&65.6	&13.276	&0.410	&10.434	&0.040\\
2005el	&   IaN	& NGC-1819	& 2	& 60.5	& 13.110	& 0.346	& 9.227	& 0.031\\
2005kc	&   IaN	& NGC-7311	& 3	& 62.6	& 12.405	& 0.523	& 8.937	& 0.015\\
2006ax	&   IaN	& NGC-3663	& 5	& 68.3	& 12.650	& 0.312	& 9.894	& 0.068\\
2006dy	&   IaN	& NGC-5587	& 2	& 33.3	& 13.277	& 0.320	& 9.684	& 0.029\\
2006lf	&   IaN	& UGC-03108	& 4	& 56.1	& 10.898	& 0.500	& 9.533	& 0.033\\
1998dh	&IaHV	&NGC-7541	&5	&37.0	&11.386	&0.087	&8.351	&0.007\\
1998dk	&   IaHV	& UGC-00139	& 6	& 53.8	& 13.613	& 0.330	& 11.044	& 0.069\\
1998ef	&   IaHV	& UGC-00646	& 4	& 74.2	& 13.841	& 0.327	& 10.428	& 0.038\\
1999cl	&   IaHV	& MESSIER-088	& 4	& 32.6	& 9.563	& 0.129	& 6.267	& 0.017\\
1999dk	&   IaHV	& UGC-01087	& 6	& 61.7	& 14.483	& 0.365	& 11.096	& 0.082\\
2001E	&   IaHV	& NGC-3905	& 6	& 78.4	& 12.963	& 0.148	& 9.884	& 0.068\\
2001en	&IaHV	&NGC-0523	&5	&66.8	&12.333	&0.066	&9.714	&0.022\\
2002bo	&IaHV	&NGC-3190	&3	&19.1	&11.397	&0.226	&99.999	&0.000\\
2002dj	&   IaHV	& NGC-5018	& 1	& 37.8	& 11.220	& 0.193	& 7.734	& 0.014\\
2002er	&   IaHV	& UGC-10743	& 3	& 36.9	& 13.405	& 0.324	& 10.375	& 0.037\\
2004ca	&   IaHV	& UGC-11799	& 6	& 76.1	& 13.141	& 0.500	& 10.300	& 0.066\\
2005A	&IaHV	&NGC-0958	&6	&77.5	&11.857	&0.049	&8.800	&0.020\\
2006X	&   IaHV	& MESSIER-100	& 5	& 23.0	& 9.840	& 0.154	& \nodata	& \nodata\\
2006ef	&   IaHV	& NGC-0809	& 2	& 72.1	& 14.406	& 0.421	& 10.597	& 0.035\\
2006le	&   IaHV	& UGC-03218	& 4	& 74.4	& 12.321	& 0.560	& 9.208	& 0.024\\
1998de	&   Ia-91bg	& NGC-0252	& 2	& 69.2	& 12.900	& 0.410	& 9.044	& 0.025\\
1999by	&   Ia-91bg	& NGC-2841	& 4	& 11.4	& 9.537	& 0.116	& 6.062	& 0.019\\
1999da	&   Ia-91bg	& NGC-6411	& 1	& 54.3	& 12.512	& 0.115	& 9.126	& 0.023\\
2002cf	&   Ia-91bg	& NGC-4786	& 1	& 63.4	& 12.492	& 0.090	& 8.717	& 0.027\\
2002dk	&   Ia-91bg	& NGC-6616	& 3	& 78.6	& 13.826	& 0.382	& 9.397	& 0.025\\
2002fb	&   Ia-91bg	& NGC-0759	& 1	& 65.3	& 13.387	& 0.079	& 9.139	& 0.019\\
\hline
\hline
\end{tabular}
\end{minipage}
\end{table*}

\clearpage
\newpage

\setcounter{table}{0}
\begin{table*}
\begin{minipage}{140mm}
\caption{continued ...}
\begin{tabular}{lllrrrrrr}
\hline
\hline
{SN} 	& {Type} 	&{Host} 	&{$h$}
	&{Dist} 	&{$B_0$ (mag)} 	&{$B_0$(err)}
	&{$K$ (mag)} 	&{$K$(err)} \\
\hline
2002jm	&   Ia-91bg	& IC-0603	& 3	& 73.5	& 14.334	& 0.409	& 10.431	& 0.052\\
2003Y	&   Ia-91bg	& IC-0522	& 2	& 72.4	& 13.499	& 0.317	& 9.973	& 0.025\\
2005ke	&   Ia-91bg	& NGC-1371	& 3	& 17.4	& 11.296	& 0.099	& 7.630	& 0.039\\
2005mz	&   Ia-91bg	& NGC-1275	& 2	& 73.9	& 11.819	& 0.126	& 8.126	& 0.038\\
2006ke	&Ia-91bg	&UGC-03365	&3	&73.7	&13.458	&0.317	&9.990	&0.056\\
1998es	&   Ia-91T	& NGC-0632	& 2	& 43.0	& 13.481	& 0.446	& 10.096	& 0.028\\
1999aa	&   Ia-91T	& NGC-2595	& 6	& 60.1	& 12.672	& 0.200	& 9.661	& 0.046\\
1999ac	&   Ia-91T	& NGC-6063	& 7	& 40.7	& 13.354	& 0.345	& 10.550	& 0.078\\
1999dq	&   Ia-91T	& NGC-0976	& 6	& 59.4	& 12.679	& 0.059	& 9.114	& 0.021\\
2001V	&Ia-91T	&NGC-3987	&4	&63.5	&12.989	&0.182	&9.046	&0.017\\
2004bv	&   Ia-91T	& NGC-6907	& 5	& 42.6	& 11.470	& 0.056	& 8.370	& 0.020\\
2006cm	&Ia-91T	&UGC-11723	&4	&67.5	&13.702	&0.341	&9.988	&0.028\\
1999bh	&   Ia-02cx	& NGC-3435	& 4	& 74.0	& 13.525	& 0.410	& 10.726	& 0.046\\
2002es	&   Ia-02cx	& UGC-02708	& 2	& 75.6	& 14.450	& 0.489	& \nodata	& \nodata\\
2005cc	&   Ia-02cx	& NGC-5383	& 4	& 33.4	& 12.006	& 0.168	& 8.536	& 0.038\\
2005hk	&   Ia-02cx	& UGC-00272	& 7	& 52.9	& 14.288	& 0.321	& 12.983	& 0.201\\
1998dt	&   Ib	& NGC-0945	& 6	& 59.8	& 12.568	& 0.074	& 9.361	& 0.043\\
1999dn	&   Ib	& NGC-7714	& 4	& 38.3	& 12.530	& 0.141	& 9.762	& 0.027\\
2001is	&   Ib	& NGC-1961	& 6	& 57.1	& 10.971	& 0.089	& 7.730	& 0.035\\
2004dk	&   Ib	& NGC-6118	& 7	& 22.5	& 11.060	& 0.077	& 8.703	& 0.019\\
2004gq	&   Ib	& NGC-1832	& 5	& 24.4	& 10.658	& 0.516	& 8.388	& 0.025\\
2006F	&   Ib	& NGC-0935	& 7	& 57.6	& 12.546	& 0.410	& 9.322	& 0.039\\
1999bu	&   Ic	& NGC-3786	& 3	& 39.2	& 13.027	& 0.172	& 9.338	& 0.025\\
2000C	&   Ic	& NGC-2415	& 8	& 53.5	& 12.329	& 0.235	& 9.776	& 0.020\\
2001M	&   Ic	& NGC-3240	& 4	& 46.6	& 13.481	& 0.186	& 10.588	& 0.052\\
2001ci	&Ic	&NGC-3079	&6	&18.4	&9.970	&0.253	&99.999	&0.000\\
2002J	&   Ic	& NGC-3464	& 6	& 49.7	& 12.490	& 0.076	& 9.464	& 0.046\\
2002jj	&Ic	&IC-0340	&6	&55.1	&14.260	&0.406	&10.866	&0.071\\
2002jz	&   Ic	& UGC-02984	& 7	& 20.9	& 13.150	& 0.500	& 12.142	& 0.110\\
2003aa	&   Ic	& NGC-3367	& 6	& 42.4	& 11.845	& 0.031	& 8.755	& 0.028\\
2004C	&   Ic	& NGC-3683	& 6	& 26.6	& 12.238	& 0.385	& 8.666	& 0.022\\
2004cc	&   Ic	& NGC-4568	& 5	& 32.0	& 10.959	& 0.100	& 7.516	& 0.026\\
2005az	&   Ic	& NGC-4961	& 7	& 37.0	& 13.714	& 0.057	& 10.845	& 0.052\\
2005lr	&   Ic	& ESO-492-G002	& 4	& 32.9	& 11.630	& 0.200	& 9.224	& 0.032\\
2006eg	&   Ic	& CGCG-462-023	& 5	& 55.0	& 14.175	& 0.325	& 11.977	& 0.089\\
2002ap	&   Ic-pec	& MESSIER-074	& 6	& 9.4	& 9.345	& 0.259	& 6.845	& 0.054\\
2003H	&   Ibc-pec	& NGC-2207	& 5	& 35.2	& 11.328	& 0.303	& 8.190	& 0.037\\
2003dr	&Ibc-pec	&NGC-5714	&6	&34.1	&12.691	&0.327	&9.968	&0.033\\
2003id	&   Ic-pec	& NGC-0895	& 7	& 30.1	& 11.875	& 0.104	& 9.405	& 0.051\\
2004bm	&Ibc-pec/IIb	&NGC-3437	&6	&18.9	&11.673	&0.421	&8.878	&0.015\\
2005E	&   Ibc-pec	& NGC-1032	& 2	& 36.5	& 12.134	& 0.097	& 8.379	& 0.018\\
1999D	&   II-P	& IC-0694	& 7	& 44.3	& 12.454	& 1.593	& 8.422	& 0.024\\
1999an	&II-P	&NGC-4019	&4	&22.0	&13.107	&0.634	&11.334	&0.056\\
1999bg	&   II-P	& IC-0758	& 7	& 21.1	& 13.456	& 0.303	& \nodata	& \nodata\\
1999br	&   II-P	& NGC-4900	& 6	& 13.9	& 11.762	& 0.093	& 8.638	& 0.038\\
1999em	&   II-P	& NGC-1637	& 6	& 8.4	& 11.267	& 0.154	& 7.974	& 0.045\\
1999gi	&   II-P	& NGC-3184	& 7	& 10.5	& 10.312	& 0.155	& 7.225	& 0.067\\
2000L	&   II-P	& UGC-05520	& 7	& 48.8	& 13.783	& 0.273	& 11.686	& 0.078\\
2000cb	&   II-P	& IC-1158	& 6	& 27.6	& 12.819	& 0.111	& 10.238	& 0.076\\
2000el	&   II-P	& NGC-7290	& 5	& 41.1	& 13.015	& 0.037	& 10.739	& 0.051\\
2000ex	&   II-P	& ESO-419-G003	& 6	& 53.4	& 13.245	& 0.321	& 10.945	& 0.072\\
2001J	&   II-P	& UGC-04729	& 7	& 54.0	& 14.485	& 0.392	& 12.164	& 0.113\\
2001K	&   II-P	& IC-0677	& 5	& 45.4	& 12.932	& 0.363	& 10.712	& 0.040\\
2001bq	&   II-P/II-L	& NGC-5534	& 3	& 36.4	& 12.858	& 0.143	& 9.629	& 0.032\\
2001cm	&II-P	&NGC-5965	&4	&50.6	&11.520	&0.101	&8.608	&0.031\\
2001dc	&II-P	&NGC-5777	&5	&33.1	&12.741	&0.328	&9.314	&0.016\\
2001fz	&   II-P	& NGC-2280	& 7	& 23.6	& 9.976	& 0.366	& 8.255	& 0.031\\
2002bx	&II-P	&IC-2461	&5	&32.9	&13.502	&0.420	&10.052	&0.020\\
2002ca	&   II-P	& UGC-08521	& 3	& 45.7	& 13.965	& 0.324	& 10.352	& 0.050\\
2002ce	&   II-P	& NGC-2604	& 7	& 29.8	& 13.497	& 0.427	& 11.044	& 0.060\\
2002dq	&   II-P	& NGC-7051	& 3	& 34.5	& 12.963	& 0.375	& 9.467	& 0.025\\
2002ds	&II-P	&UGCA-402	&7	&30.7	&11.736	&0.243	&9.117	&0.021\\
\hline
\hline
\end{tabular}
\end{minipage}
\end{table*}

\setcounter{table}{0}
\begin{table*}
\begin{minipage}{140mm}
\caption{continued ...}
\begin{tabular}{lllrrrrrr}
\hline
\hline
{SN} 	& {Type} 	&{Host} 	&{$h$}
	&{Dist} 	&{$B_0$ (mag)} 	&{$B_0$(err)}
	&{$K$ (mag)} 	&{$K$(err)} \\
\hline
2002gd	&II-P	&NGC-7537	&5	&36.9	&12.678	&0.059	&10.213	&0.027\\
2002gw	&   II-P	& NGC-0922	& 7	& 39.9	& 12.165	& 0.088	& 10.023	& 0.068\\
2002hh	&   II-P	& NGC-6946	& 7	& 4.4	& 8.237	& 0.217	& 5.369	& 0.034\\
2003E	&II-P	&ESO-485-G004	&5	&57.7	&14.329	&0.270	&99.999	&0.000\\
2003Z	&   II-P	& NGC-2742	& 6	& 20.8	& 11.385	& 0.079	& 8.808	& 0.014\\
2003ao	&   II-P	& NGC-2993	& 3	& 30.4	& 12.747	& 0.055	& 10.131	& 0.041\\
2003bk	&II-P	&NGC-4316	&5	&18.2	&12.491	&0.090	&9.246	&0.027\\
2003br	&   II-P	& ESO-447-G023	& 7	& 50.7	& 13.088	& 0.212	& 10.391	& 0.051\\
2003bw	&   II-P	& IC-1077	& 5	& 46.5	& 13.064	& 0.197	& 9.560	& 0.032\\
2003ef	&   II-P	& NGC-4708	& 3	& 56.2	& 13.488	& 0.290	& 10.140	& 0.053\\
2003hg	&   II-P	& NGC-7771	& 3	& 59.9	& 12.296	& 0.168	& 8.348	& 0.017\\
2003hl	&   II-P	& NGC-0772	& 4	& 33.9	& 10.000	& 0.539	& \nodata	& \nodata\\
2003iq	&   II-P	& NGC-0772	& 4	& 33.9	& 10.000	& 0.539	& \nodata	& \nodata\\
2003ld	&II-P	&UGC-00148	&5	&57.6	&15.267	&2.517	&10.531	&0.040\\
2004aq	&II-P	&NGC-4012	&4	&58.2	&13.416	&0.322	&10.430	&0.050\\
2004ci	&II-P	&NGC-5980	&5	&58.2	&12.372	&0.560	&9.441	&0.024\\
2004dd	&   II-P	& NGC-0124	& 6	& 54.8	& 13.264	& 0.357	& 10.755	& 0.068\\
2004er	&   II-P	& UGCA-036	& 6	& 59.3	& 13.585	& 0.384	& 10.734	& 0.081\\
2004et	&   II-P	& NGC-6946	& 7	& 4.4	& 8.237	& 0.217	& 5.369	& 0.034\\
2004fc	&   II-P	& NGC-0701	& 6	& 23.7	& 12.174	& 0.069	& 9.170	& 0.028\\
2004fx	&II-P	&MCG-02-14-003	&6	&34.8	&99.999	&99.999	&10.413	&0.042\\
2005ad	&   II-P	& NGC-0941	& 6	& 20.8	& 12.625	& 0.090	& 10.694	& 0.080\\
2005ay	&   II-P	& NGC-3938	& 6	& 13.9	& 10.802	& 0.033	& 7.809	& 0.048\\
2005bb	&II-P	&UGC-08067	&4	&39.3	&13.621	&0.678	&10.420	&0.061\\
2005ci	&   II-P	& NGC-5682	& 4	& 34.7	& 14.040	& 0.163	& 12.250	& 0.059\\
2005io	&   II-P	& UGC-03361	& 7	& 47.5	& 14.363	& 0.500	& 11.894	& 0.079\\
2005mg	&II-P	&UGC-00155	&4	&54.5	&13.481	&0.410	&9.764	&0.022\\
2006be	&II-P	&IC-4582	&5	&32.1	&13.749	&0.410	&10.574	&0.032\\
2006bp	&   II-P	& NGC-3953	& 5	& 17.6	& 9.881	& 0.292	& 7.047	& 0.026\\
2006ca	&   II-P	& UGC-11214	& 7	& 38.2	& \nodata	& \nodata	& 10.938	& 0.089\\
2006qr	&   II-P	& MCG--02-22-023	& 5	& 58.0	& 13.781	& 0.500	& 10.962	& 0.057\\
1999go	&   II-L	& NGC-1376	& 7	& 55.5	& 12.640	& 0.221	& 9.804	& 0.061\\
2000dc	&   II-L	& ESO-527-G019	& 4	& 41.8	& 13.241	& 0.392	& 10.713	& 0.042\\
2001do	&   II-L	& UGC-11459	& 7	& 46.0	& 12.862	& 0.332	& 9.775	& 0.046\\
2001hf	&II-L	&ESO-564-G015	&5	&59.6	&12.934	&0.200	&99.999	&0.000\\
2002an	&   II-L	& NGC-2575	& 7	& 53.9	& 13.350	& 0.250	& 10.226	& 0.068\\
2005J	&II-L	&NGC-4012	&4	&58.2	&13.416	&0.322	&10.430	&0.050\\
2005an	&II-L	&ESO-506-G011	&4	&43.7	&14.797	&0.179	&10.885	&0.049\\
1999cd	&   IIb	& NGC-3646	& 5	& 59.6	& 11.112	& 0.153	& 8.484	& 0.025\\
2000H	&   IIb	& IC-0454	& 3	& 53.8	& \nodata	& \nodata	& 9.387	& 0.023\\
2001Q	&   IIb	& UGC-06429	& 6	& 54.6	& 13.655	& 0.090	& 11.212	& 0.084\\
2003ed	&   IIb	& NGC-5303	& 6	& 22.3	& 12.279	& 0.410	& 10.227	& 0.026\\
2004be	&   IIb	& ESO-499-G034	& 7	& 29.3	& 14.650	& 0.200	& \nodata	& \nodata\\
2005H	&   IIb	& NGC-0838	& 8	& 51.2	& 13.370	& 0.117	& 9.743	& 0.023\\
2005U	&   IIb	& NGC-3690	& 7	& 46.3	& 13.142	& 3.595	& \nodata	& \nodata\\
2006T	&   IIb	& NGC-3054	& 5	& 31.2	& 11.668	& 0.145	& 8.343	& 0.029\\
2000N	&   IIb/II-L	& MCG--02-34-054	& 5	& 54.4	& 12.866	& 0.390	& 9.909	& 0.058\\
2004al	&   IIb/II-L	& ESO-565-G025	& 4	& 58.1	& 14.280	& 0.297	& 11.319	& 0.061\\
1999el	&   IIn	& NGC-6951	& 5	& 23.3	& 10.022	& 0.214	& 7.220	& 0.025\\
2000eo	&   IIn	& MCG--02-09-003	& 3	& 41.2	& 13.318	& 0.366	& 11.046	& 0.075\\
2002bu	&   IIn	& NGC-4242	& 7	& 10.1	& 11.260	& 0.185	& \nodata	& \nodata\\
2003G	&   IIn	& IC-0208	& 5	& 47.7	& 14.291	& 0.318	& 10.679	& 0.081\\
2003dv	&   IIn	& UGC-09638	& 8	& 34.9	& 15.779	& 1.055	& \nodata	& \nodata\\
2005aq	&   IIn	& NGC-1599	& 6	& 53.4	& 13.870	& 0.030	& 12.182	& 0.115\\
2006bv	&   IIn	& UGC-07848	& 7	& 37.9	& 13.700	& 0.310	& 11.768	& 0.064\\
1999bw	&IIni	&NGC-3198	&6	&11.7	&9.947	&0.202	&7.779	&0.042\\
2000ch	&IIni	&NGC-3432	&7	&10.7	&10.583	&0.120	&9.061	&0.050\\
2001ac	&   IIni	& NGC-3504	& 3	& 22.9	& 11.434	& 0.133	& 8.273	& 0.014\\
2002kg	&   IIni	& NGC-2403	& 7	& 5.1	& 8.114	& 0.098	& 6.191	& 0.039\\
2003gm	&   IIni	& NGC-5334	& 6	& 19.6	& 12.430	& 0.729	& 9.935	& 0.047\\
\hline
\hline
\end{tabular}
\end{minipage}
\end{table*}

\renewcommand{\arraystretch}{1.00}

\begin{table*}
\begin{minipage}{180mm}
\caption{Average light curves of the
supernovae (the numbers are magnitudes below peak brightness) $^a$ }
\begin{tabular}{rrrrrrrrrrrrr}
\hline
\hline
{t(day)} &{Ia.01} &{Ia.11} &{Ia.21}
&{Ibc.fast} &{Ibc.ave} &{Ibc.slow}
&{II-P} &{II-L} &{IIb} 
&{IIn.fast} &{IIn.ave} &{IIn.slow} \\ 
\hline
-30.0&57.530&44.540&31.550&10.488&9.580&4.445& \nodata & \nodata & \nodata &2.961& \nodata & \nodata \\
-29.0&54.660&42.160&29.660&10.065&8.974&4.134&\nodata&\nodata&\nodata&2.827&\nodata&\nodata\\
-28.0&51.800&39.790&27.780&9.642&8.368&3.823&\nodata&\nodata&\nodata&2.693&\nodata&\nodata\\
-27.0&48.930&37.410&25.890&9.219&7.762&3.513&\nodata&\nodata&\nodata&2.558&\nodata&\nodata\\
-26.0&46.060&35.030&24.010&8.796&7.156&3.202&\nodata&\nodata&\nodata&2.424&\nodata&\nodata\\
-25.0&43.190&32.660&22.120&8.373&6.550&2.891&\nodata&\nodata&\nodata&2.290&\nodata&\nodata\\
-24.0&40.320&30.280&20.240&7.950&5.944&2.580&\nodata&\nodata&\nodata&2.155&\nodata&\nodata\\
-23.0&37.450&27.900&18.350&7.527&5.338&2.269&\nodata&\nodata&\nodata&2.021&\nodata&\nodata\\
-22.0&34.580&25.530&16.470&7.104&4.732&1.985&\nodata&\nodata&\nodata&1.887&\nodata&\nodata\\
-21.0&31.710&23.150&14.580&6.681&4.126&1.727&\nodata&\nodata&30.334&1.753&\nodata&\nodata\\
-20.0&28.840&20.770&12.700&6.258&3.520&1.505&\nodata&\nodata&1.599&1.618&\nodata&\nodata\\
-19.0&25.980&18.390&10.810&5.835&2.914&1.283&\nodata&\nodata&0.495&1.484&\nodata&\nodata\\
-18.0&23.110&16.020&8.930&5.412&2.308&1.095&\nodata&\nodata&0.146&1.350&\nodata&\nodata\\
-17.0&20.240&13.640&7.040&4.989&1.793&0.927&\nodata&\nodata&0.277&1.215&\nodata&\nodata\\
-16.0&17.370&11.260&5.160&4.566&1.369&0.779&\nodata&\nodata&0.521&1.081&\nodata&\nodata\\
-15.0&14.500&8.890&3.270&4.143&1.019&0.649&37.436&\nodata&0.776&0.947&\nodata&66.472\\
-14.0&11.630&6.510&1.390&3.720&0.782&0.535&19.931&\nodata&0.960&0.824&\nodata&50.113\\
-13.0&8.760&4.820&0.870&3.297&0.614&0.437&2.427&56.200&1.033&0.778&\nodata&33.753\\
-12.0&5.890&3.290&0.680&2.874&0.497&0.353&1.447&38.800&1.025&0.733&17.394&17.394\\
-11.0&3.020&1.780&0.530&2.451&0.416&0.281&0.847&21.400&0.959&0.576&1.035&1.035\\
-10.0&1.450&0.950&0.450&2.028&0.354&0.225&0.680&4.000&0.859&0.412&0.622&0.622\\
-9.0&1.170&0.770&0.380&1.605&0.298&0.170&0.514&0.440&0.740&0.315&0.378&0.378\\
-8.0&0.880&0.600&0.310&1.183&0.245&0.128&0.347&0.340&0.615&0.220&0.257&0.257\\
-7.0&0.650&0.440&0.240&0.827&0.193&0.093&0.293&0.253&0.491&0.192&0.210&0.210\\
-6.0&0.470&0.320&0.160&0.540&0.145&0.064&0.238&0.187&0.380&0.165&0.190&0.190\\
-5.0&0.330&0.210&0.090&0.322&0.102&0.041&0.184&0.122&0.280&0.137&0.160&0.160\\
-4.0&0.200&0.130&0.060&0.187&0.065&0.023&0.131&0.075&0.193&0.110&0.120&0.120\\
-3.0&0.110&0.070&0.030&0.068&0.036&0.013&0.072&0.046&0.122&0.082&0.077&0.077\\
-2.0&0.050&0.030&0.010&0.044&0.014&0.002&0.026&0.028&0.067&0.055&0.038&0.038\\
-1.0&0.010&0.010&0.000&0.021&0.002&0.001&0.005&0.013&0.020&0.027&0.010&0.011\\
0.0&0.000&0.000&0.000&0.000&0.000&0.000&0.000&0.000&0.000&0.000&0.000&0.000\\
1.0&0.010&0.010&0.010&0.044&0.007&0.007&0.009&0.004&0.001&0.014&0.003&0.003\\
2.0&0.030&0.030&0.030&0.112&0.023&0.014&0.021&0.015&0.025&0.028&0.014&0.014\\
3.0&0.070&0.060&0.040&0.196&0.046&0.027&0.039&0.025&0.065&0.041&0.029&0.029\\
4.0&0.130&0.100&0.070&0.294&0.075&0.041&0.058&0.034&0.116&0.055&0.043&0.043\\
5.0&0.200&0.160&0.110&0.402&0.112&0.061&0.075&0.044&0.171&0.069&0.053&0.053\\
6.0&0.280&0.210&0.150&0.518&0.152&0.080&0.091&0.056&0.230&0.083&0.061&0.061\\
7.0&0.350&0.270&0.190&0.633&0.194&0.105&0.108&0.067&0.291&0.096&0.068&0.068\\
8.0&0.430&0.330&0.240&0.749&0.242&0.129&0.122&0.079&0.350&0.110&0.076&0.076\\
9.0&0.510&0.400&0.290&0.864&0.292&0.158&0.135&0.090&0.406&0.124&0.088&0.088\\
10.0&0.590&0.470&0.340&0.975&0.346&0.186&0.147&0.102&0.459&0.138&0.105&0.102\\
\hline
\hline
\end{tabular}

\medskip

$^a$Only three representative SN~Ia light curves are listed,
and only parts of the light curves are shown. The entire set of light
curves is available electronically.

\end{minipage}
\end{table*}

\clearpage
\newpage

\renewcommand{\arraystretch}{0.50}

\begin{table*}
\begin{minipage}{180mm}
\caption{The luminosity function of SNe~Ia in two Hubble-type bins.}
\begin{tabular}{@{}lrrrrrrrrrrll}
\hline 
\hline
{SN} &{Type} &{Abs. mag} &{err}
&{$D$ (Mpc)} &
{$h^a$}
&Incl.$^a$  & Mass$^a$ 
& $N$(SN)$_{\rm Vol}^b$
&$N$(SN)$_{\rm Mag}^c$
&{LC$^d$} & Src$^e$
& {Comment}  \\
\hline
\hline
\multicolumn{12}{c}{LF in E--Sa} \\
\hline
1999ej &     IaN & -18.58 & 0.11 & 57.8 & 2 & 49.1  & 4.438	 & 1.0044 & 0.7564 & Ia.11 & followup & \\
1999gd &     IaN & -18.15 & 0.13 & 77.0 & 3 & 80.5  & 3.915	 & 1.0058 & 0.4182 & Ia.19 & CfA-2/unfilt & \\
2000dm &     IaN & -19.02 & 0.10 & 64.2 & 3 & 84.9  & 2.907	 & 1.0025 & 1.3865 & Ia.13 & followup & \\
2000dr &     IaN & -18.64 & 0.09 & 75.6 & 2 & 34.3  & 11.004	 & 1.0039 & 0.8213 & Ia.13 & followup & \\
2002do &     IaN & -18.75 & 0.13 & 68.4 & 1 & 3.0   & 23.905	& 1.0044 & 0.9566 & Ia.04 & followup & \\
2002ha &     IaN & -19.17 & 0.11 & 58.5 & 3 & 46.7  & 15.700	 & 1.0020 & 1.7050 & Ia.14 & followup & \\
2003cg &     IaN & -17.02 & 0.33 & 16.9 & 3 & 56.9  & 5.253	 & 1.0350 & 0.0903 & Ia.17 & followup & \\
2003gt &     IaN & -19.20 & 0.10 & 63.8 & 3 & 78.3  & $-$	 & 1.0018 & 1.7767 & Ia.18 & followup & \\
2004bd &     IaN & -18.22 & 0.25 & 39.2 & 3 & 64.2  & 4.380	 & 1.0074 & 0.4614 & Ia.09 & followup & \\
2005am &     IaN & -18.90 & 0.19 & 30.7 & 3 & 76.8  & 9.226	 & 1.0032 & 1.1755 & Ia.10 & followup & \\
2005bo &     IaN & -18.50 & 0.11 & 56.2 & 3 & 47.7  & 3.671	 & 1.0041 & 0.6771 & Ia.17 & followup & \\
2005cf &     IaN & -18.96 & 0.22 & 27.1 & 2 & 82.9  & 0.253	 & 1.0024 & 1.2761 & Ia.19 & followup & \\
2005el &     IaN & -18.97 & 0.11 & 60.5 & 2 & 51.1  & 13.090	 & 1.0026 & 1.2941 & Ia.14 & followup & \\
2005kc &     IaN & -18.79 & 0.14 & 62.6 & 3 & 67.9  & 15.045	 & 1.0035 & 1.0100 & Ia.11 & CfA-3 & \\
2006dy &     IaN & -18.37 & 0.35 & 33.3 & 2 & 81.7  & 2.201	 & 1.0059 & 0.5667 & Ia.13 & SNWeb & \\
2002bo &    IaHV & -17.97 & 0.29 & 19.1 & 3 & 77.1  & 6.673	 & 1.0077 & 0.3267 & Ia.19 & followup & \\
2002dj &    IaHV & -19.11 & 0.16 & 37.8 & 1 & 56.0  & 16.576	 & 1.0019 & 1.5691 & Ia.20 & followup & \\
2002er &    IaHV & -18.86 & 0.16 & 36.9 & 3 & 68.0  & 1.136	 & 1.0028 & 1.1119 & Ia.16 & followup & \\
2006ef &    IaHV & -18.84 & 0.17 & 72.1 & 2 & 43.4  & 4.926	 & 1.0027 & 1.0815 & Ia.19 & CfA-3 & \\
1998de & Ia-91bg & -17.74 & 0.10 & 69.2 & 2 & 48.4  & 19.666	 & 1.0167 & 0.2399 & Ia.03 & followup & \\
1999da & Ia-91bg & -17.85 & 0.15 & 54.3 & 1 & 53.7  & 8.921	 & 1.0151 & 0.2788 & Ia.02 & followup & \\
2002cf & Ia-91bg & -17.99 & 0.10 & 63.4 & 1 & 44.5  & 21.308	 & 1.0118 & 0.3372 & Ia.03 & followup & \\
2002dk & Ia-91bg & -17.31 & 0.12 & 78.6 & 3 & 70.9  & 24.977	 & 1.0324 & 0.1345 & Ia.02 & unfilter & \\
2002fb & Ia-91bg & -17.88 & 0.13 & 65.3 & 1 & 23.1  & 19.627	 & 1.0134 & 0.2901 & Ia.04 & followup & \\
2002jm & Ia-91bg & -17.35 & 0.26 & 73.5 & 3 & 59.2  & 6.316	 & 1.0335 & 0.1423 & Ia.01 & unfilter &  poor coverage\\
 2003Y & Ia-91bg & -17.81 & 0.09 & 72.4 & 2 & 41.0  & 7.762	 & 1.0157 & 0.2640 & Ia.02 & followup & \\
2005ke & Ia-91bg & -16.67 & 0.37 & 17.4 & 3 & 52.4  & 4.135	 & 1.0774 & 0.0580 & Ia.12 & CfA-3 & \\
2005mz & Ia-91bg & -18.22 & 0.13 & 73.9 & 2 & 52.0  & 49.948	 & 1.0082 & 0.4617 & Ia.05 & CfA-3 & \\
2006ke & Ia-91bg & -16.71 & 0.17 & 73.7 & 3 & 90.0  & 8.059	 & 1.0851 & 0.0617 & Ia.01 & CfA-3/unfilt & \\
1998es &  Ia-91T & -19.44 & 0.14 & 43.0 & 2 & 34.3  & 2.273	 & 1.0012 & 2.4737 & Ia.21 & followup & \\
2002es & Ia-02cx & -18.31 & 0.13 & 75.6 & 2 & 24.2  & 9.314	 & 1.0059 & 0.5216 & Ia.05hk & followup & \\
\hline
\hline
\multicolumn{12}{c}{LF in Sb--Irr} \\
\hline
1998dm &     IaN & -17.75 & 0.22 & 25.8 & 6 & 90.0  & 0.260	 & 1.0095 & 0.2415 & Ia.20 & followup & \\
1999cp &     IaN & -19.03 & 0.15 & 39.4 & 7 & 24.2  & 0.879	 & 1.0023 & 1.4055 & Ia.17 & followup & \\
1999ek &     IaN & -18.64 & 0.13 & 72.0 & 5 & 72.8  & 5.667	 & 1.0034 & 0.8209 & Ia.19 & unfilter & \\
2001dn &     IaN & -18.92 & 0.12 & 79.3 & 5 & 50.9  & 3.881	 & 1.0025 & 1.2076 & Ia.18 & unfilter & \\
2001ep &     IaN & -18.89 & 0.12 & 52.0 & 4 & 48.7  & $-$	 & 1.0027 & 1.1588 & Ia.17 & followup & \\
2001fh &     IaN & -19.25 & 0.11 & 57.8 & 7 & 77.4  & 15.365	 & 1.0018 & 1.9039 & Ia.14 & followup & \\
 2001L &     IaN & -18.96 & 0.18 & 62.5 & 4 & 82.2  & 5.034	 & 1.0022 & 1.2759 & Ia.21 & unfilter & \\
2002cr &     IaN & -18.83 & 0.15 & 39.4 & 7 & 24.2  & 0.879	 & 1.0027 & 1.0667 & Ia.19 & followup & \\
2002fk &     IaN & -18.97 & 0.21 & 27.2 & 5 & 23.4  & 1.573	 & 1.0023 & 1.2937 & Ia.19 & followup & \\
2002hw &     IaN & -18.14 & 0.13 & 72.3 & 6 & 45.3  & 5.865	 & 1.0087 & 0.4136 & Ia.06 & CfA-3 & \\
2002jg &     IaN & -17.93 & 0.10 & 63.8 & 6 & 76.4  & $-$	 & 1.0097 & 0.3098 & Ia.11 & followup & \\
2003du &     IaN & -18.82 & 0.20 & 30.1 & 7 & 40.8  & $-$	 & 1.0027 & 1.0520 & Ia.20 & followup & \\
 2003F &     IaN & -18.91 & 0.17 & 70.9 & 7 & 43.5  & 3.416	 & 1.0024 & 1.1909 & Ia.21 & unfilter & \\
2003kf &     IaN & -18.90 & 0.22 & 28.6 & 4 & 90.0  & 0.217	 & 1.0025 & 1.1747 & Ia.20 & followup/unfilt & \\
2004ab &     IaN & -17.78 & 0.31 & 23.4 & 5 & 59.6  & 6.542	 & 1.0099 & 0.2518 & Ia.19 & unfilter & \\
2004bl &     IaN & -19.34 & 0.13 & 71.5 & 7 & 86.1  & 0.182	 & 1.0015 & 2.1553 & Ia.17 & unfilter & \\
2005as &     IaN & -18.38 & 0.18 & 53.7 & 4 & 32.1  & 19.096	 & 1.0050 & 0.5741 & Ia.15 & unfilter & \\
2005bc &     IaN & -18.07 & 0.12 & 53.5 & 4 & 70.0  & 2.333	 & 1.0078 & 0.3752 & Ia.12 & followup & \\
2005de &     IaN & -18.82 & 0.10 & 65.6 & 4 & 78.8  & 2.894	 & 1.0029 & 1.0522 & Ia.17 & followup & \\
 2005W &     IaN & -18.82 & 0.25 & 37.2 & 5 & 50.6  & 4.319	 & 1.0034 & 1.0528 & Ia.11 & SNWeb    & \\
2006ax &     IaN & -18.89 & 0.31 & 68.3 & 5 & 48.0  & 5.103	 & 1.0027 & 1.1588 & Ia.18 & CfA-3    & \\
2006lf &     IaN & -19.55 & 0.12 & 56.1 & 4 & 48.1  & 3.306	 & 1.0016 & 2.8809 & Ia.20 & CfA-3    & \\
1998dh &    IaHV & -19.03 & 0.16 & 37.0 & 5 & 77.4  & 7.149	 & 1.0021 & 1.4053 & Ia.19 & followup & \\
1998dk &    IaHV & -19.16 & 0.23 & 53.8 & 6 & 70.5  & 1.001	 & 1.0023 & 1.6820 & Ia.11 & CfA-2    & \\
1998ef &    IaHV & -19.52 & 0.09 & 74.2 & 4 & 73.4  & 5.120	 & 1.0011 & 2.7626 & Ia.19 & followup & \\
1999cl &    IaHV & -17.35 & 0.51 & 13.1 & 4 & 64.7  & 42.622	 & 1.0229 & 0.1408 & Ia.19 & followup & \\
1999dk &    IaHV & -19.27 & 0.11 & 61.7 & 6 & 25.5  & 1.875	 & 1.0015 & 1.9566 & Ia.20 & followup & \\
 2001E &    IaHV & -18.36 & 0.09 & 78.4 & 6 & 51.4  & 7.924	 & 1.0045 & 0.5582 & Ia.20 & followup & \\
2001en &    IaHV & -19.15 & 0.10 & 66.8 & 5 & 82.1  & 5.392	 & 1.0019 & 1.6584 & Ia.17 & followup & \\
2004ca &    IaHV & -19.34 & 0.21 & 76.1 & 6 & 50.9  & 5.204	 & 1.0015 & 2.1552 & Ia.18 & unfilter & poor coverage \\
 2005A &    IaHV & -17.70 & 0.11 & 77.5 & 6 & 80.9  & 20.698	 & 1.0102 & 0.2256 & Ia.19 & CSP      & \\
2006le &    IaHV & -19.40 & 0.13 & 74.4 & 4 & 57.9  & 15.328	 & 1.0013 & 2.3409 & Ia.21 & followup & \\
 2006X &    IaHV & -17.47 & 0.43 & 15.2 & 5 & 26.2  & 15.343	 & 1.0165 & 0.1652 & Ia.20 & followup & \\
1999by & Ia-91bg & -17.42 & 0.47 & 11.4 & 4 & 66.6  & 6.785	 & 1.0366 & 0.1572 & Ia.01 & followup & \\
1999aa &  Ia-91T & -19.19 & 0.11 & 60.1 & 6 & 51.8  & 5.521	 & 1.0017 & 1.7522 & Ia.21 & followup & \\
1999ac &  Ia-91T & -19.05 & 0.15 & 40.7 & 7 & 63.4  & 1.011	 & 1.0021 & 1.4445 & Ia.20 & followup & \\
1999dq &  Ia-91T & -19.52 & 0.11 & 59.4 & 6 & 24.1  & 11.991	 & 1.0011 & 2.7624 & Ia.21 & followup & \\
 2001V &  Ia-91T & -19.43 & 0.10 & 63.5 & 4 & 84.1  & 17.022	 & 1.0012 & 2.4399 & Ia.21 & CfA-3    & \\
2004bv &  Ia-91T & -19.39 & 0.14 & 42.6 & 5 & 40.2  & 9.599	 & 1.0013 & 2.3088 & Ia.21 & followup & \\
2006cm &  Ia-91T & -18.04 & 0.31 & 67.5 & 4 & 90.7  & 7.293	 & 1.0066 & 0.3595 & Ia.21 & CfA-3/unfilt & \\
1999bh & Ia-02cx & -17.08 & 0.13 & 74.0 & 4 & 52.3  & 2.810	 & 1.0308 & 0.0977 & Ia.05hk & followup & \\
2005cc & Ia-02cx & -16.73 & 0.20 & 33.4 & 4 & 44.2  & 5.943	 & 1.0406 & 0.0608 & Ia.05hk & followup & 02cx-like\\
2005hk & Ia-02cx & -18.32 & 0.12 & 52.9 & 7 & 74.9  & 0.087	 & 1.0044 & 0.5281 & Ia.05hk & followup & 02cx-like \\
\hline
\hline
\end{tabular}

\medskip

$^a$The Hubble type (in the coding scheme of this series
  of papers; see Table 1 of Paper I for details), inclination (in
    degrees), and mass (in $10^{10}\, {\rm M}_\odot$) of the host galaxies. 

$^b$The number fractions of the SNe in a volume-limited survey.
See text in \S 5.5 for possible limitations of our LFs.

$^c$The number fractions of the SNe in a magnitude-limited survey
with continuous coverage (i.e., very small observation intervals).
See text in \S 5.5 for possible limitations of our LFs. 

$^d$The label for the light-curve shape. The data corresponding to
the different labels are presented in Table 2. 

$^e$The source of the photometry: ``followup" = our own
filtered photometry database; ``unfilter" = our unfiltered 
photometry from the SN survey images; ``CfA-2" = Jha et al. (2006a);
``CfA-3" = Hicken et al. (2009); ``CSP" = Contreras et al. (2009);
``SNWeb" = http://www.astrosurf.com/snweb2/.

\end{minipage}
\end{table*}

\clearpage

\begin{table*}
\caption{The luminosity function of SNe~Ibc in two Hubble-type bins$^a$. }
\begin{minipage}{190mm}
\begin{tabular}{lrrrrrrrrrrll}
\hline
\hline
{SN} &{Type} &{Abs. mag} &{err}
&{$D$ (Mpc)} &{$h$} &{Incl.} &{Mass} 
& {$N$(SN)$_{\rm Vol}$} &{$N$(SN)$_{\rm Mag}$}
&{LC} & {Src} &{Comment}  \\
\hline
\hline
\multicolumn{12}{c}{LF in S0--Sbc}  \\
\hline
      1999dn &      Ib & -17.24 & 0.16 & 38.3 & 4 & 47.4   & 1.827	& 1.0066 & 1.8248 & Ibc.ave  & followup  & \\
      2004gq &      Ib & -17.10 & 0.24 & 24.4 & 5 & 73.7   & 2.069	& 1.0085 & 1.5067 & Ibc.ave  & followup  & \\
      1999bu &      Ic & -16.52 & 0.52 & 39.2 & 3 & 64.2   & 4.380	& 1.0311 & 0.6913 & Ibc.ave  & unfilter  & poor coverage \\
       2001M &      Ic & -16.20 & 0.16 & 46.6 & 4 & 34.3   & 1.347	& 1.0367 & 0.4467 & Ibc.ave  & unfilter  & \\
      2004cc &      Ic & -16.20 & 1.02 & 32.0 & 5 & 69.7   & 13.957	& 1.1012 & 0.4745 & Ibc.ave  & unfilter  & poor coverage \\
      2005lr &      Ic & -15.56 & 0.53 & 32.9 & 4 & 50.9   & 1.976	& 1.1495 & 0.2046 & Ibc.ave  & unfilter  & poor coverage\\
      2006eg &      Ic & -14.86 & 0.23 & 55.0 & 5 & 44.7   & 0.378	& 1.4081 & 0.0953 & Ibc.ave  & unfilter  & \\
       2003H & Ibc-pec & -15.13 & 0.43 & 35.2 & 5 & 60.9   & 7.936	& 1.2759 & 0.1254 & Ibc.ave  & unfilter  & ``Ca-rich"\\
       2005E & Ibc-pec & -15.70 & 0.19 & 36.5 & 2 & 85.1   & 9.528	& 1.1095 & 0.2396 & Ibc.05E  & unfilter  & ``Ca-rich"\\
\hline
\hline
\multicolumn{12}{c}{LF in Sc--Irr}  \\ 
\hline
      1998dt &      Ib & -17.14 & 0.51 & 59.8 & 6 & 31.4   & 7.904	& 1.0107 & 1.5957 & Ibc.ave  & unfilter  & poor coverage\\
      2001is &      Ib & -16.37 & 0.32 & 57.1 & 6 & 49.7   & 34.012	& 1.0840 & 0.5907 & Ibc.fast & unfilter  & \\
      2004dk &      Ib & -17.53 & 0.25 & 22.5 & 7 & 72.1   & 1.421	& 1.0039 & 2.7166 & Ibc.slow & followup  & \\
       2006F &      Ib & -16.74 & 0.41 & 57.6 & 7 & 56.1   & 8.105	& 1.0179 & 0.9248 & Ibc.ave  & unfilter  & poor coverage \\
       2000C &      Ic & -17.94 & 0.19 & 53.5 & 8 & 13.5   & 3.152	& 1.0027 & 4.7809 & Ibc.ave  & unfilter  & \\
      2001ci &      Ic & -13.85 & 0.36 & 18.4 & 6 & 90.0   & 4.024	& 2.7447 & 0.0460 & Ibc.ave  & unfilter  & \\
       2002J &      Ic & -16.61 & 0.15 & 49.7 & 6 & 53.8   & 4.592	& 1.0182 & 0.7730 & Ibc.ave  & unfilter  & \\
      2002jj &      Ic & -17.68 & 0.23 & 55.1 & 6 & 75.4   & 1.853	& 1.0040 & 3.3425 & Ibc.ave  & unfilter  & poor coverage\\
      2002jz &      Ic & -16.50 & 0.33 & 20.9 & 7 & 56.5   & 0.031	& 1.0712 & 0.6986 & Ibc.fast & unfilter  & \\
      2003aa &      Ic & -17.21 & 0.17 & 42.4 & 6 & 19.7   & 6.574	& 1.0057 & 1.7490 & Ibc.slow & unfilter  & \\
       2004C &      Ic & -15.81 & 0.26 & 26.6 & 6 & 71.8   & 3.540	& 1.0774 & 0.2708 & Ibc.ave  & unfilter  & \\
      2005az &      Ic & -17.17 & 0.18 & 37.0 & 7 & 41.8   & 0.652	& 1.0060 & 1.6556 & Ibc.slow & unfilter  & \\
      2003dr & Ibc-pec & -15.10 & 0.43 & 34.1 & 6 & 90.0   & 1.158	& 1.3430 & 0.1266 & Ibc.05E  & unfilter  & ``Ca-rich"\\
      2004bm & Ibc-pec & -13.93 & 0.36 & 18.9 & 6 & 75.9   & 1.005	& 1.2862 & 0.0241 & Ibc.ave  & unfilter  & \\
      2002ap &  Ic-pec & -17.73 & 0.56 &  9.4 & 6 & 24.1   & 1.422	& 1.0049 & 3.5850 & Ibc.ave  & unfilter  & broad-lined\\
      2003id &  Ic-pec & -16.05 & 0.20 & 30.1 & 7 & 52.4   & 1.340	& 1.0489 & 0.3674 & Ibc.ave  & unfilter  & IAUC 8228\\
\hline
\hline
\end{tabular}

\medskip

$^a$The meanings of the different columns are the same as in Table 3.

\end{minipage}
\end{table*}

\begin{table*}
\caption{The luminosity function of SNe~II in two Hubble-type bins$^a$. }
\begin{minipage}{190mm}
\begin{tabular}{lrrrrrrrrrrll}
\hline
\hline
{SN} &{Type} &{Abs. mag} &{err}
&{$D$ (Mpc)} &{$h$} & {Incl.} &{Mass} 
& {$N$(SN)$_{\rm Vol}$} &{$N$(SN)$_{\rm Mag}$}
&{LC} &{Src} &{Comment} \\
\hline
\hline
\multicolumn{12}{c} {LF in S0--Sbc} \\ 
\hline
    1999an &     II-P & -16.39 & 0.32 & 22.0 & 4 & 90.0  & 0.087	& 1.0189 & 0.4373 & II-P & unfilter  & \\
    2000el &     II-P & -16.22 & 0.21 & 41.1 & 5 & 66.5  & 0.674	& 1.0208 & 0.3464 & II-P & unfilter  & \\
2001bq.II-P &    II-P & -17.41 & 0.22 & 36.4 & 3 & 28.6  & 2.330	& 0.5020 & 0.8817 & II-P & followup  & \\
    2001cm &     II-P & -17.40 & 0.19 & 50.6 & 4 & 90.0  & 9.765	& 1.0040 & 1.7392 & II-P & unfilter  & \\
    2001dc &     II-P & -13.53 & 0.26 & 33.1 & 5 & 90.0  & 2.808	& 3.0084 & 0.0248 & II-P & unfilter  & \\
     2001K &     II-P & -16.73 & 0.20 & 45.4 & 5 & 71.1  & 0.808	& 1.0099 & 0.6933 & II-P & unfilter  & \\
    2002bx &     II-P & -16.22 & 0.26 & 32.9 & 5 & 89.2  & 1.421	& 1.0232 & 0.3472 & II-P & unfilter  & \\
    2002ca &     II-P & -15.76 & 0.24 & 45.7 & 3 & 37.7  & 2.257	& 1.0486 & 0.1885 & II-P & followup  & \\
    2002dq &     II-P & -16.25 & 0.19 & 34.5 & 3 & 46.4  & 2.866	& 1.0200 & 0.3608 & II-P & unfilter  & \\
    2002gd &     II-P & -15.89 & 0.25 & 36.9 & 5 & 83.3  & 0.971	& 1.0382 & 0.2233 & II-P & followup  & \\
    2003ao &     II-P & -15.55 & 0.24 & 30.4 & 3 & 35.8  & 0.761	& 1.0695 & 0.1438 & II-P & unfilter  & \\
    2003bk &     II-P & -13.61 & 0.32 & 18.2 & 5 & 85.8  & 0.824	& 2.7988 & 0.0258 & II-P & unfilter  & \\
    2003bw &     II-P & -15.24 & 0.24 & 46.5 & 5 & 37.9  & 4.697	& 1.1299 & 0.0990 & II-P & unfilter  & poor coverage\\
    2003ef &     II-P & -16.85 & 0.32 & 56.2 & 3 & 47.7  & 3.671	& 1.0088 & 0.8174 & II-P & unfilter  & poor coverage \\
     2003E &     II-P & -16.21 & 0.27 & 57.7 & 5 & 90.0  &  $-$ 	& 1.0229 & 0.3423 & II-P & unfilter  & \\
    2003hg &     II-P & -17.36 & 0.14 & 59.9 & 3 & 65.4  & 29.444	& 1.0042 & 1.6460 & II-P & unfilter  & \\
    2003hl &     II-P & -16.72 & 0.18 & 33.9 & 4 & 61.3  & 15.581	& 1.0100 & 0.6838 & II-P & followup  & \\
    2003iq &     II-P & -17.32 & 0.18 & 33.9 & 4 & 61.3  & 15.581	& 1.0044 & 1.5578 & II-P & followup  & \\
    2003ld &     II-P & -16.72 & 0.41 & 57.6 & 5 & 79.1  & 5.338	& 1.0117 & 0.6850 & II-P & unfilter  & poor coverage\\
    2004aq &     II-P & -15.47 & 0.22 & 58.2 & 4 & 79.9  & 2.511	& 1.0841 & 0.1305 & II-P & unfilter  & \\
    2004ci &     II-P & -16.53 & 0.18 & 58.2 & 5 & 78.4  & 6.086	& 1.0130 & 0.5275 & II-P & unfilter  & \\
    2005bb &     II-P & -14.21 & 0.25 & 39.3 & 4 & 75.9  & 1.278	& 1.8123 & 0.0383 & II-P & unfilter  & \\
    2005ci &     II-P & -15.27 & 0.26 & 34.7 & 4 & 74.4  & 0.093	& 1.1266 & 0.1029 & II-P & unfilter  & 1987A-like?\\
    2005mg &     II-P & -17.37 & 0.32 & 54.5 & 4 & 83.6  & 5.832	& 1.0045 & 1.6694 & II-P & unfilter  & \\
    2006be &     II-P & -16.70 & 0.27 & 32.1 & 5 & 83.2  & 0.734	& 1.0114 & 0.6661 & II-P & unfilter  & \\
    2006bp &     II-P & -16.40 & 0.33 & 17.6 & 5 & 64.4  & 4.817	& 1.0173 & 0.4427 & II-P & unfilter  & \\
    2006qr &     II-P & -15.92 & 0.14 & 58.0 & 5 & 69.0  & 1.415	& 1.0341 & 0.2318 & II-P & unfilter  & \\
    2000dc &     II-L & -17.29 & 0.15 & 41.8 & 4 & 51.6  & 0.814	& 1.0047 & 1.4950 & II-L & followup  & \\
 2000N.II-L &    II-L & -16.23 & 0.23 & 54.4 & 5 & 47.6  & 3.540	& 0.5113 & 0.1759 & II-L & unfilter  & \\
2001bq.II-L &    II-L & -17.41 & 0.22 & 36.4 & 3 & 28.6  & 2.330	& 0.5021 & 0.8818 & II-L & followup  & \\
    2001hf &     II-L & -17.26 & 0.32 & 59.6 & 5 & 79.6  &  $-$ 	& 1.0057 & 1.4358 & II-L & unfilter  & \\
2004al.II-L &    II-L & -16.85 & 0.22 & 58.1 & 4 & 62.9  & 1.100	& 0.5043 & 0.4086 & II-L & unfilter  & \\
    2005an &     II-L & -16.92 & 0.24 & 43.7 & 4 & 81.5  & 1.494	& 1.0078 & 0.8995 & II-L & unfilter  & \\
     2005J &     II-L & -17.24 & 0.18 & 58.2 & 4 & 79.9  & 2.511	& 1.0050 & 1.3957 & II-L & unfilter  & \\
    1999el &      IIn & -18.30 & 0.26 & 23.3 & 5 & 54.9  & 7.983	& 1.0015 & 6.0153 & IIn.ave & followup  & \\
    2000eo &      IIn & -18.46 & 0.24 & 41.2 & 3 & 72.4  & 0.513	& 1.0013 & 7.5020 & IIn.fast& unfilter  & \\
     2003G &      IIn & -18.72 & 0.23 & 47.7 & 5 & 35.2  & 1.838	& 1.0009 & 10.7403& IIn.fast& unfilter  & \\
    1999cd &      IIb & -16.43 & 0.18 & 59.6 & 5 & 66.7  & 13.271	& 1.0255 & 0.4651 & IIb & unfilter  & \\
     2000H &      IIb & -17.48 & 0.23 & 53.8 & 3 & 63.8  &  $-$ 	& 1.0050 & 1.9444 & IIb & unfilter  & IAUC 7375\\
 2000N.IIb &      IIb & -16.93 & 0.23 & 54.4 & 5 & 47.6  & 3.540	& 0.5059 & 0.4578 & IIb & unfilter  & \\
2004al.IIb &      IIb & -17.15 & 0.22 & 58.1 & 4 & 62.9  & 1.100	& 0.5042 & 0.6183 & IIb & unfilter  & \\
     2006T &      IIb & -17.64 & 0.24 & 31.2 & 5 & 57.0  & 5.914	& 1.0041 & 2.4232 & IIb & unfilter  & IAUC 8680\\
\hline
\hline
\multicolumn{12}{c}{LF in Sc--Irr} \\
\hline
    1999bg &     II-P & -15.86 & 0.33 & 21.1 & 7 & 68.2  &  $-$ 	& 1.0437 & 0.2154 & II-P & unfilter  & \\
    1999br &     II-P & -13.57 & 0.44 & 13.9 & 6 & 21.9  & 0.794	& 2.7830 & 0.0243 & II-P & unfilter  & \\
     1999D &     II-P & -16.77 & 0.16 & 44.3 & 7 & 54.0  & 15.349	& 1.0092 & 0.7321 & II-P & followup  & \\
    1999em &     II-P & -16.32 & 0.62 &  8.4 & 6 & 34.0  & 0.590	& 1.0312 & 0.4018 & II-P & followup  & \\
    1999gi &     II-P & -15.84 & 0.51 & 10.5 & 7 & 17.2  & 1.635	& 1.0568 & 0.2121 & II-P & followup  & \\
    2000cb &     II-P & -16.37 & 0.23 & 27.6 & 6 & 65.3  & 0.571	& 1.0172 & 0.4246 & II-P & followup  & 1987A-like?\\
    2000ex &     II-P & -15.47 & 0.19 & 53.4 & 6 & 55.7  & 0.935	& 1.0802 & 0.1301 & II-P & unfilter  & \\
     2000L &     II-P & -15.23 & 0.23 & 48.8 & 7 & 55.2  & 0.361	& 1.1316 & 0.0978 & II-P & unfilter  & \\
    2001fz &     II-P & -15.20 & 0.28 & 23.6 & 7 & 69.7  & 1.689	& 1.1412 & 0.0946 & II-P & unfilter  & \\
     2001J &     II-P & -15.73 & 0.23 & 54.0 & 7 & 37.9  & 0.317	& 1.0509 & 0.1812 & II-P & unfilter  & poor coverage\\
    2002ce &     II-P & -14.76 & 0.54 & 29.8 & 7 & 25.2  & 0.289	& 1.3651 & 0.0616 & II-P & unfilter  & poor coverage\\
    2002ds &     II-P & -17.03 & 0.24 & 30.7 & 7 & 90.0  & 2.046	& 1.0065 & 1.0458 & II-P & unfilter  & \\
    2002gw &     II-P & -16.55 & 0.21 & 39.9 & 7 & 36.5  & 1.134	& 1.0128 & 0.5422 & II-P & unfilter  & \\
    2002hh &     II-P & -13.49 & 1.06 &  4.4 & 7 & 19.7  & 1.567	& 2.5745 & 0.0201 & II-P & followup  & \\
    2003br &     II-P & -16.24 & 0.19 & 50.7 & 7 & 62.4  & 1.758	& 1.0207 & 0.3561 & II-P & unfilter  & \\
     2003Z &     II-P & -14.68 & 0.29 & 20.8 & 6 & 64.8  & 1.174	& 1.3741 & 0.0555 & II-P & followup  & \\
    2004dd &     II-P & -16.46 & 0.18 & 54.8 & 6 & 60.2  & 1.309	& 1.0146 & 0.4796 & II-P & unfilter  & \\
    2004er &     II-P & -17.12 & 0.18 & 59.3 & 6 & 45.0  & 1.842	& 1.0059 & 1.1835 & II-P & unfilter  & \\
    2004et &     II-P & -16.69 & 1.06 &  4.4 & 7 & 19.7  & 1.567	& 1.0391 & 0.6749 & II-P & followup  & \\
    2004fc &     II-P & -16.23 & 0.26 & 23.7 & 6 & 65.4  & 1.335	& 1.0214 & 0.3514 & II-P & unfilter  & \\
    2004fx &     II-P & -15.78 & 0.19 & 34.8 & 6 & 90.0  &  $-$ 	& 1.0448 & 0.1930 & II-P & unfilter  & \\
    2005ad &     II-P & -15.67 & 0.37 & 20.8 & 6 & 48.3  & 0.151	& 1.0645 & 0.1690 & II-P & unfilter  & \\
    2005ay &     II-P & -15.87 & 0.40 & 13.9 & 6 & 17.2  & 1.610	& 1.0457 & 0.2188 & II-P & followup  & \\
    2005io &     II-P & -16.09 & 0.23 & 47.5 & 7 & 42.9  & 0.366	& 1.0275 & 0.2913 & II-P & unfilter  & \\
    2006ca &     II-P & -17.55 & 0.18 & 38.2 & 7 & 25.5  &  $-$ 	& 1.0033 & 2.1383 & II-P & unfilter  & \\
    1999go &     II-L & -18.62 & 0.18 & 55.5 & 7 & 19.7  & 3.798	& 1.0009 & 9.3543 & II-L & unfilter  & \\
    2001do &     II-L & -17.76 & 0.14 & 46.0 & 7 & 53.1  & 3.189	& 1.0027 & 2.8561 & II-L & followup  & \\
    2002an &     II-L & -17.95 & 0.23 & 53.9 & 7 & 42.5  & 2.790	& 1.0022 & 3.7117 & II-L & followup  & \\
    2002bu &      IIn & -15.25 & 0.53 & 10.1 & 7 & 49.5  &  $-$ 	& 1.2006 & 0.1067 & IIn.ave & followup  & \\
    2003dv &      IIn & -16.75 & 0.19 & 34.9 & 8 & 41.0  &  $-$ 	& 1.0095 & 0.7124 & IIn.slow& unfilter  & \\
    2005aq &      IIn & -16.83 & 0.51 & 53.4 & 6 & 21.9  & 0.225	& 1.0131 & 0.7985 & IIn.ave & unfilter  & \\
    2006bv &      IIn & -14.78 & 0.21 & 37.9 & 7 & 72.5  & 0.185	& 1.3745 & 0.0638 & IIn.ave & unfilter  & \\
     2001Q &      IIb & -15.71 & 0.23 & 54.6 & 6 & 23.4  & 0.820	& 1.0983 & 0.1842 & IIb & unfilter  & \\
    2003ed &      IIb & -16.28 & 0.32 & 22.3 & 6 & 64.3  & 0.281	& 1.0380 & 0.3827 & IIb & unfilter  & IAUC 8144\\
    2004be &      IIb & -17.27 & 0.36 & 29.3 & 7 & 28.6  &  $-$ 	& 1.0080 & 1.4591 & IIb & unfilter  & \\
2004bm.IIb &      IIb & -13.93 & 0.36 & 18.9 & 6 & 76.0  & 1.005	& 1.3363 & 0.0192 & IIb & unfilter  & \\
     2005H &      IIb & -17.61 & 0.23 & 51.2 & 8 & 48.4  & 5.006	& 1.0042 & 2.3251 & IIb & unfilter  & \\
     2005U &      IIb & -18.06 & 0.24 & 46.3 & 7 & 42.7  & 23.420	& 1.0025 & 4.3223 & IIb & unfilter  & Atel 431\\
\hline
\hline
\end{tabular}

\medskip

$^a$The meaning of the different columns are the same
as Table 3.

\end{minipage}
\end{table*}

\renewcommand{\arraystretch}{1.00}

\begin{table*}
\begin{minipage}{180mm}
\caption{The average absolute magnitudes of supernovae.}
\begin{tabular}{l|ccc|ccc|ccc}
\hline 
\hline
{} &{}  &{Ia}  &{}
&{}  &{Ibc}  &{}
&{}  &{II}  &{} \\
\hline
{Bin} &{Mean} &
$\sigma$\footnote{$\sigma$ is the standard deviation, i.e., root-mean
  square (RMS) of the average.} 
&SDOM\footnote{SDOM is the standard deviation of the mean, i.e.,
  RMS/$\sqrt N$, where $N$ is the number of measurements.} 
&{Mean} &{$\sigma$} &{SDOM}
&{Mean} &{$\sigma$} &{SDOM} \\
\hline
\hline
all   &$-$18.49&0.76&0.09&      &    &    &      &    &    \\
E-Sab &$-$18.29&0.75&0.13&&&&&&\\
Sb-Irr&$-$18.63&0.74&0.11&&&&&&\\
all   &      &    &    &$-$16.09&1.24&0.23&$-$16.05&1.37&0.15\\
S0-Sbc&      &    &    &$-$15.98&0.83&0.26&$-$16.22&1.39&0.21\\
Sc-Irr&      &    &    &$-$16.15&1.43&0.33&$-$15.88&1.34&0.20\\
\hline
E--Sab ($L_K < 11.0 \times 10^{10} {\rm L}_\odot$)  &$-$18.27&0.78&0.20&&&&&&\\
E--Sab ($L_K > 11.0 \times 10^{10} {\rm L}_\odot$)  &$-$18.24&0.73&0.19&&&&&&\\
Sb--Irr ($L_K < 9.0 \times 10^{10} {\rm L}_\odot$)&$-$18.64&0.73&0.16&&&&&&\\
Sb--Irr ($L_K > 9.0 \times 10^{10} {\rm L}_\odot$)&$-$18.66&0.79&0.17&&&&&&\\
S0--Sbc ($L_K <  4.5 \times 10^{10} {\rm L}_\odot$)&      &    &    &      &    &    &$-$16.02&1.46&0.32    \\
S0---Sbc ($L_K >  4.5 \times 10^{10} {\rm L}_\odot$)&      &    &    &      &    &    &$-$16.37&1.38&0.30    \\
Sc--Irr ($L_K <  3.0 \times 10^{10} {\rm L}_\odot$)&      &    &    &      &    &    &$-$15.42&1.12&0.25    \\
Sc--Irr ($L_K >  3.0 \times 10^{10} {\rm L}_\odot$)&      &    &    &      &    &    &$-$16.28&1.52&0.35\\
\hline
Sa--Scd ($i = 0^\circ -  40^\circ$)  &$-$18.78&0.68&0.26&$-$17.06&0.64&0.32 &$-$15.77&1.59&0.33 \\
Sa--Scd ($i = 40^\circ -  75^\circ$) &$-$18.40&0.87&0.15&$-$16.29&0.80&0.20 &$-$16.47&1.00&0.16 \\
Sa--Scd ($i = 75^\circ -  90^\circ$) &$-$18.56&0.76&0.18&$-$14.73&1.49&0.59 &$-$15.55&1.49&0.30 \\ 
\hline
normal    Ia&$-$18.67&0.51&0.08&      &&&&&\\
HV        Ia&$-$18.70&0.74&0.19&      &&&&&\\
91T-like  Ia&$-$19.15&0.52&0.20&      &&&&&\\
91bg-like Ia&$-$17.55&0.53&0.14&      &&&&&\\
Ic    &      &    &    &$-$16.04&1.28&0.31&&& \\
Ib    &      &    &    &$-$17.01&0.41&0.17&&& \\
Ibc-pec&      &    &    &$-$15.50&1.21&0.46&&& \\
II-P  &      &    &    &      &    &    &$-$15.66&1.23&0.16   \\
II-L  &      &    &    &      &    &    &$-$17.44&0.64&0.22    \\
IIb   &      &    &    &      &    &    &$-$16.65&1.30&0.40    \\
IIn   &      &    &    &      &    &    &$-$16.86&1.61&0.59\\
\hline
\hline
\end{tabular}
\end{minipage}
\end{table*}

\renewcommand{\baselinestretch}{1.8}

\begin{table*}
\caption{Relative supernova fractions in two kinds of surveys.}
\begin{minipage}{140mm}
\begin{tabular}{@{}llllllll}
\hline
\hline
{SN} &{Vol-limited$^a$}
&{Mag-1d$^b$}
&{Mag-5d$^b$} 
&{Mag-10d$^b$} 
&{Mag-30d$^b$} 
&{Mag-60d$^b$}
&{Mag-360d/snapshot$^b$} \\
\hline
\hline
\multicolumn{8}{c}{Overall} \\  
Ia     &  $ 24.1_{-3.5}^{+3.7}$ &  $ 79.2_{-5.5}^{+4.2}$ &  $ 79.3_{-5.6}^{+4.2}$ &  $ 79.2_{-5.6}^{+4.2}$ &  $ 76.6_{-6.1}^{+4.7}$ &  $ 73.2_{-6.6}^{+5.2}$ &  $ 68.6_{-7.1}^{+5.8}$ \\
Ibc    &  $ 18.7_{-3.3}^{+3.5}$ &  $  4.1_{-1.3}^{+1.6}$ &  $  4.1_{-1.2}^{+1.6}$ &  $  4.2_{-1.3}^{+1.6}$ &  $  4.3_{-1.3}^{+1.6}$ &  $  4.3_{-1.3}^{+1.6}$ &  $  4.3_{-1.4}^{+1.6}$ \\
II     &  $ 57.2_{-4.1}^{+4.3}$ &  $ 16.6_{-3.9}^{+5.0}$ &  $ 16.6_{-3.9}^{+5.0}$ &  $ 16.7_{-3.9}^{+5.0}$ &  $ 19.0_{-4.4}^{+5.5}$ &  $ 22.5_{-4.9}^{+6.0}$ &  $ 27.1_{-5.5}^{+6.7}$ \\
\hline
\multicolumn{8}{c}{SNe Ia} \\
IaN    &  $ 49.8_{-8.4}^{+9.3}$ &  $ 52.3_{-10.9}^{+12.0}$ &  $ 52.2_{-10.8}^{+12.0}$ &  $ 52.1_{-10.9}^{+11.9}$ &  $ 51.4_{-11.0}^{+12.2}$ &  $ 50.7_{-11.1}^{+12.3}$ &  $ 50.4_{-11.1}^{+12.5}$ \\
IaHV   &  $ 20.2_{-6.6}^{+7.4}$ &  $ 25.1_{-9.6}^{+10.4}$ &  $ 25.1_{-9.6}^{+10.5}$ &  $ 25.2_{-9.7}^{+10.4}$ &  $ 25.5_{-9.7}^{+10.7}$ &  $ 25.7_{-9.9}^{+10.7}$ &  $ 25.7_{-9.8}^{+10.8}$ \\
Ia-91bg&  $ 15.2_{-5.9}^{+6.8}$ &  $  3.3_{-1.5}^{+2.0}$ &  $  3.3_{-1.5}^{+2.0}$ &  $  3.2_{-1.5}^{+2.0}$ &  $  2.8_{-1.3}^{+1.7}$ &  $  2.5_{-1.1}^{+1.6}$ &  $  2.4_{-1.1}^{+1.5}$ \\
Ia-91T &  $  9.4_{-4.7}^{+5.9}$ &  $ 17.7_{-9.3}^{+10.8}$ &  $ 17.7_{-9.2}^{+10.9}$ &  $ 17.9_{-9.4}^{+10.9}$ &  $ 18.7_{-9.8}^{+11.3}$ &  $ 19.5_{-10.1}^{+11.7}$ &  $ 19.7_{-10.2}^{+11.8}$ \\
Ia-02cx&  $  5.4_{-3.3}^{+4.7}$ &  $  1.6_{-1.2}^{+1.9}$ &  $  1.6_{-1.2}^{+1.9}$ &  $  1.6_{-1.2}^{+1.9}$ &  $  1.7_{-1.2}^{+1.9}$ &  $  1.6_{-1.2}^{+1.9}$ &  $  1.8_{-1.3}^{+2.2}$ \\
\hline
\multicolumn{8}{c}{SNe Ibc} \\
Ib     &  $ 21.2_{-7.7}^{+8.4}$ &  $ 32.2_{-12.6}^{+15.0}$ &  $ 32.3_{-12.6}^{+15.0}$ &  $ 32.1_{-12.6}^{+15.0}$ &  $ 32.2_{-13.0}^{+15.3}$ &  $ 32.4_{-13.3}^{+15.7}$ &  $ 32.4_{-14.0}^{+16.4}$ \\
Ic     &  $ 54.2_{-9.8}^{+9.8}$ &  $ 52.5_{-16.7}^{+14.7}$ &  $ 52.5_{-16.4}^{+14.8}$ &  $ 52.6_{-16.4}^{+15.0}$ &  $ 52.6_{-17.1}^{+14.8}$ &  $ 52.9_{-17.3}^{+15.0}$ &  $ 52.8_{-17.6}^{+16.4}$ \\
Ibc-pec&  $ 24.5_{-8.4}^{+9.0}$ &  $ 15.3_{-10.3}^{+13.2}$ &  $ 15.3_{-10.2}^{+13.0}$ &  $ 15.3_{-10.3}^{+12.9}$ &  $ 15.1_{-10.1}^{+13.4}$ &  $ 14.8_{-10.1}^{+13.1}$ &  $ 14.8_{-10.2}^{+13.9}$ \\
\hline
\multicolumn{8}{c}{SNe II} \\
II-P   &  $ 69.9_{-5.8}^{+5.1}$ &  $ 29.8_{-7.2}^{+9.3}$ &  $ 29.9_{-7.1}^{+9.5}$ &  $ 29.9_{-7.1}^{+9.5}$ &  $ 30.9_{-7.4}^{+9.7}$ &  $ 34.6_{-7.9}^{+10.0}$ &  $ 39.4_{-8.5}^{+10.4}$ \\
II-L   &  $  9.7_{-3.2}^{+4.0}$ &  $ 25.0_{-10.0}^{+11.6}$ &  $ 25.1_{-10.1}^{+11.5}$ &  $ 25.5_{-10.3}^{+11.7}$ &  $ 26.6_{-10.6}^{+12.0}$ &  $ 26.6_{-10.4}^{+12.0}$ &  $ 27.5_{-10.7}^{+11.7}$ \\
IIb    &  $ 11.9_{-3.6}^{+3.9}$ &  $ 16.5_{-6.2}^{+7.9}$ &  $ 16.4_{-6.1}^{+7.9}$ &  $ 15.9_{-6.0}^{+7.7}$ &  $ 13.4_{-5.2}^{+6.7}$ &  $ 11.8_{-4.5}^{+6.1}$ &  $ 10.1_{-4.1}^{+5.3}$ \\
IIn    &  $  8.6_{-3.2}^{+3.3}$ &  $ 28.7_{-13.4}^{+13.0}$ &  $ 28.6_{-13.2}^{+13.0}$ &  $ 28.7_{-13.3}^{+13.1}$ &  $ 29.1_{-13.3}^{+13.3}$ &  $ 27.0_{-12.3}^{+12.7}$ &  $ 23.0_{-10.7}^{+11.4}$ \\
\hline
\hline
\end{tabular}

\medskip
$^a$The SN fractions in a volume-limited survey, expressed as a percentage of the corresponding category.

$^b$The SN fractions in a magnitude-limited survey. The different columns correspond to different observation intervals from 1~d to 360~d.

\end{minipage}
\end{table*}

\renewcommand{\baselinestretch}{1.0}

\begin{table*}
\caption{Background supernovae in the KAIT fields. }
\begin{tabular}{llcllcll}
\hline
\hline
{SN} &{Type} &{\,\,\,\,\,\,\,\,} 
&{SN} &{Type}&{\,\,\,\,\,\,\,\,}
&{SN} &{Type} \\
\hline
1999ce&Ia&&2003go&IIn&&2006is&Ia\\
1999co&II&&2003hw&Ia&&2006iu&II\\
2000dd&Ia&&2004U&II&&2006lu&Ia\\
2000Q&Ia-91bg&&2004V&II&&2007aj&Ia\\
2001bp&Ia&&2004Y&Ia&&2007al&Ia-91bg\\
2001ei&Ia-91bg&&2004as&Ia&&2007az&Ib\\
2001es&unknown&&2004dz&Ia&&2007H&Ia\\
2001ew&Ia&&2004eb&II&&2007I&Ic-pec\\
2002cc&Ia&&2005X&Ia&&2007V&Ia\\
2002eu&Ia&&2005ac&Ia&&2007ry&Ia\\
2002ey&Ia-91bg&&2005ag&Ia&&2007ux&Ia\\
2002hi&IIn&&2005bu&Ia&&2008Z&Ia\\
2002je&II&&2005eu&Ia&&2008cf&Ia\\
2002ka&Ia&&2005kf&Ic&&2008fk&Ia\\
2003ah&Ia&&2006bw&Ia&&2008iq&Ia\\
2003ev&Ic&&2006dw&Ia&&\\
\hline
\hline
\end{tabular}
\end{table*}


\begin{thebibliography}{}

\bibitem[Arcavi et al.(2010)]{2010arXiv1004.0615A} Arcavi I., et
  al.\ 2010, ApJ, 721, 777
\bibitem[Barbon, Ciatti, \& Rosino(1979)]{1979A&A....72..287B} 
  Barbon R., Ciatti F., Rosino L.\ 1979, A\&A, 72, 287
\bibitem[Barbon, Cappellaro, \& Turatto(1989)]{1989A&AS...81..421B} 
  Barbon R., Cappellaro E., Turatto M.\ 1989, A\&AS, 81, 421
\bibitem[Barbon et al.(1999)]{1999A&AS..139..531B} Barbon R.,
  Buond{\'{\i}} V., Cappellaro E., Turatto M.\ 1999, A\&AS, 139,
  531
\bibitem[Barris \& Tonry(2006)]{2006ApJ...637..427B} Barris B.~J.,
  Tonry J.~L.\ 2006, ApJ, 637, 427
\bibitem[Bazin et al.(2009)]{2009A&A...499..653B} Bazin G., et
  al.\ 2009, A\&A, 499, 653
\bibitem[Benetti et al.(2000)]{2000IAUC.7375....2B} Benetti S.,
  Cappellaro E., Turatto M., Pastorello A.\ 2000, IAU Circ.,
  7375, 2
\bibitem[Blanc et al.(2004)]{2004A&A...423..881B} Blanc G., et
  al.\ 2004, A\&A, 423, 881
\bibitem[Blondin et al.(2006)]{2006IAUC.8680....2B} Blondin S.,
  Modjaz M., Kirshner R., Challis P., Berlind P.\ 2006,
  IAU Circ., 8680, 2
\bibitem[Blondin et al.(2009)]{2009ApJ...693..207B} Blondin S.,
  Prieto J.~L., Patat F., Challis P., Hicken M., Kirshner R.~P.,
  Matheson T., Modjaz M.\ 2009, ApJ, 693, 207
\bibitem[Blondin \& Tonry(2007)]{2007ApJ...666.1024B} Blondin S.,
  Tonry J.~L.\ 2007, ApJ, 666, 1024
\bibitem[Boissier \& Prantzos(2009)]{2009A&A...503..137B} Boissier
  S., Prantzos N.\ 2009, A\&A, 503, 137
\bibitem[Botticella et al.(2008)]{2008A&A...479...49B} Botticella
  M.~T., et al.\ 2008, A\&A, 479, 49
\bibitem[Cappellaro et al.(1993)]{1993A&A...268..472C} Cappellaro E.,
  Turatto M., Benetti S., Tsvetkov D.~Yu., Bartunov O.~S.,
  Makarova I.~N.\ 1993, A\&A, 268, 472
\bibitem[Cappellaro, Evans, \& Turatto(1999)]{1999A&A...351..459C} 
  Cappellaro E., Evans R., Turatto, M.\ 1999, A\&A, 351, 459 (C99)
\bibitem[Chevalier \& Soderberg(2010)]{2010ApJ...711L..40C} Chevalier
  R.~A., Soderberg A.~M.\ 2010, ApJ, 711, L40
\bibitem[Clocchiatti \& Wheeler(1997)]{1997ApJ...491..375C}
  Clocchiatti A., Wheeler J.~C.\ 1997, ApJ, 491, 375
\bibitem[Crowther(2007)]{2007ARA&A..45..177C} Crowther P.~A.\ 2007,
  ARA\&A, 45, 177
\bibitem[Della Valle \& Livio(1994)]{1994ApJ...423L..31D} Della Valle
  M., Livio M.\ 1994, ApJ, 423, L31
\bibitem[Dilday et al.(2008)]{2008ApJ...682..262D} Dilday B., et
  al.\ 2008, ApJ, 682, 262
\bibitem[Di Paola et al.(2002)]{2002A&A...393L..21D} Di Paola A.,
  Larionov V., Arkharov A., Bernardi F., Caratti o Garatti A.,
  Dolci M., Di Carlo E., Valentini G.\ 2002, A\&A, 393, L21
\bibitem[Drake et al.(2010)]{2009arXiv0908.1990D} Drake A.~J., et
  al.\ 2010, ApJ, 718, L127
\bibitem[Elias-Rosa et al.(2008)]{2008MNRAS.384..107E} Elias-Rosa N.,
  et al.\ 2008, MNRAS, 384, 107
\bibitem[Fassia et al.(2000)]{2000MNRAS.318.1093F} Fassia A., et
  al.\ 2000, MNRAS, 318, 1093
\bibitem[Filippenko 1988]{}Filippenko A. V. 1988, AJ, 96, 1941
\bibitem[Filippenko 1997]{}Filippenko A. V. 1997, ARA\&A, 35, 309
\bibitem[Filippenko(2003)]{2003fthp.conf..171F} Filippenko
  A.~V.\ 2003, in From Twilight to Highlight: The Physics of Supernovae,
   ed. W. Hillebrandt \& B. Leibundgut (Berlin: Springer-Verlag), 171
\bibitem[Filippenko et al.(1992a)]{1992ApJ...384L..15F} Filippenko
  A.~V., et al.\ 1992a, ApJ, 384, L15
\bibitem[Filippenko et al.(1992b)]{1992AJ....104.1543F} Filippenko
  A.~V., et al.\ 1992b, AJ, 104, 1543
\bibitem[Foley et al.(2010a)]{2010ApJ...708L..61F} Foley R.~J., Brown
  P.~J., Rest A., Challis P.~J., Kirshner R.~P., Wood-Vasey
  W.~M.\ 2010a, ApJ, 708, L61
\bibitem[Foley et al.(2010b)]{2010ApJ...708.1748F} Foley R.~J., Narayan 
  G., Challis P.~J., Filippenko A.~V., Kirshner R.~P., Silverman J.~M., 
  \& Steele T.~N.\ 2010b, ApJ, 708, 1748
\bibitem[Foley et al.(2004)]{2004IAUC.8339....2F} Foley R.~J., Wong
  D.~S., Ganeshalingam, M., Filippenko A.~V., Chornock R.\ 2004,
  IAU Circ., 8339, 2
\bibitem[Foley et al.(2003)]{2003PASP..115.1220F} Foley R.~J., et
  al.\ 2003, PASP, 115, 1220
\bibitem[Foley et al.(2009a)]{2009AJ....138..376F} Foley R.~J., et
  al.\ 2009a, AJ, 137, 3731
\bibitem[Foley et al.(2009b)]{2009AJ....138..376F} Foley R.~J., et
  al.\ 2009b, AJ, 138, 376
\bibitem[Foley et al.(2010c)]{2010arXiv1008.0635F} Foley, R.~J., et al.\ 
  2010c, AJ, 140, 1321
\bibitem[Galama et al.(1998)]{1998Natur.395..670G} Galama T.~J., et
  al.\ 1998, Nature, 395, 670
\bibitem[Gal-Yam et al.(2007)]{2007ApJ...656..372G} Gal-Yam A., et
  al.\ 2007, ApJ, 656, 372
\bibitem[Gal-Yam et al.(2008)]{2008ApJ...680..550G} Gal-Yam A., Maoz
  D., Guhathakurta P., Filippenko A.~V.\ 2008, ApJ, 680, 550
\bibitem[Ganeshalingam et al.(2010)]{ganesh2010}Ganeshalingam M., et al.
  2010, ApJS, 190, 418
\bibitem[Ganeshalingam et al.(2011)]{ganesh2010}Ganeshalingam M., et
  al. 2011, in prep.
\bibitem[Garnavich et al.(2004)]{2004ApJ...613.1120G} Garnavich
  P.~M., et al.\ 2004, ApJ, 613, 1120
\bibitem[Gehrels(1986)]{1986ApJ...303..336G} Gehrels N.\ 1986, ApJ,
  303, 336
\bibitem[Gezari et al.(2009)]{2009ApJ...690.1313G} Gezari S., et
  al.\ 2009, ApJ, 690, 1313
\bibitem[Hamuy et al.(1996)]{1996AJ....112.2391H} Hamuy M., Phillips
  M.~M., Suntzeff N.~B., Schommer R.~A., Maza J., Aviles
  R.\ 1996, AJ, 112, 2391
\bibitem[Hamuy(2003)]{2003ApJ...582..905H} Hamuy M.\ 2003, ApJ, 582,
  905
\bibitem[Hamuy \& Roth(2003)]{2003IAUC.8228....2H} Hamuy M., Roth
  M.\ 2003, IAU Circ., 8228, 2
\bibitem[Hardin et al.(2000)]{2000A&A...362..419H} Hardin D., et
  al.\ 2000, A\&A, 362, 419
\bibitem[Hatano, Branch, \& Deaton(1998)]{1998ApJ...502..177H} 
  Hatano K., Branch D., Deaton J.\ 1998, ApJ, 502, 177
\bibitem[Heger et al.(2003)]{2003ApJ...591..288H} Heger A., Fryer
  C.~L., Woosley S.~E., Langer N., Hartmann D.~H.\ 2003, ApJ,
  591, 288
\bibitem[Hicken et al.(2009)]{2009ApJ...700..331H} Hicken M., et
  al.\ 2009, ApJ, 700, 331
\bibitem[Horesh et al.(2008)]{2008MNRAS.389.1871H} Horesh A.,
  Poznanski D., Ofek E.~O., Maoz D.\ 2008, MNRAS, 389, 1871
\bibitem[Howell(2001)]{2001ApJ...554L.193H} Howell D.~A.\ 2001,
  ApJ, 554, L193
\bibitem[Howell et al.(2006)]{2006Natur.443..308H} Howell D.~A., et
  al.\ 2006, Nature, 443, 308
\bibitem[Jha et al.(2006a)]{2006AJ....131..527J} Jha S., et
  al.\ 2006a, AJ, 131, 527
\bibitem[Jha et al.(2006b)]{2006AJ....132..189J} Jha S., Branch D.,
  Chornock R., Foley R.~J., Li W., Swift B.~J., Casebeer D.,
  Filippenko A.~V.\ 2006b, AJ, 132, 189
\bibitem[Kaiser et al.(2002)]{2002SPIE.4836..154K} Kaiser N., et
  al.\ 2002, Proc. SPIE, 4836, 154
\bibitem[Kasliwal et al.(2009)]{2009CBET.1820....2K} Kasliwal M.~M.,
  et al.\ 2009, CBET, 1820, 2
\bibitem[Law et al.(2009)]{2009PASP..121.1395L} Law N.~M., et
  al.\ 2009, PASP, 121, 1395
\bibitem[Leaman et al.(2011)]{paperI} Leaman J., Li W., Chornock
  R., \& Filippenko A.~V. 2011, MNRAS, submitted (arXiv:1006.4611)
 (Paper I)
\bibitem[Leibundgut et al.(1991)]{1991A&AS...89..537L} Leibundgut B.,
  Tammann G.~A., Cadonau R., Cerrito D.\ 1991, A\&AS, 89, 537
\bibitem[Leibundgut et al.(1993)]{1993AJ....105..301L} Leibundgut B.,
  et al.\ 1993, AJ, 105, 301
\bibitem[Leonard \& Cenko(2005)]{2005ATel..431....1L} Leonard D.~C.,
  Cenko S.~B.\ 2005, The Astronomer's Telegram, 431, 1
\bibitem[Leonard, Chornock, \& Filippenko(2003)]{2003IAUC.8144....2L} 
  Leonard D.~C., Chornock R., Filippenko A.~V.\ 2003, IAU Circ., 8144, 2
\bibitem[Li et al.(2003a)]{2003PASP..115..844L} Li W., Filippenko
  A.~V., Chornock R., Jha S.\ 2003a, PASP, 115, 844
\bibitem[Li, Filippenko, \& Riess(2001)]{2001ApJ...546..719L} Li W., 
  Filippenko A.~V., Riess A.~G.\ 2001, ApJ, 546, 719
\bibitem[Li et al.(2001b)]{2001ApJ...546..734L} Li W., Filippenko
  A.~V., Treffers R.~R., Riess A.~G., Hu J., Qiu Y.\ 2001b,
  ApJ, 546, 734
\bibitem[Li et al.(2001a)]{2001PASP..113.1178L} Li W., et al.\ 2001a,
  PASP, 113, 1178
\bibitem[Li et al.(2003b)]{2003PASP..115..453L} Li W., et al.\ 2003b,
  PASP, 115, 453
\bibitem[Li et al.(2011)]{paperIII} Li W., Chornock R., Leaman J.,
  Filippenko A.~V., Poznanski D., Wang X., Ganeshalingam M.,
  Mannucci F. 2011, MNRAS, submitted (arXiv:1006.4612) (Paper III)
\bibitem[Li et al.(2011b)]{Li2011b} Li W., et al. 2011b, in prep.
\bibitem[Lira et al.(1998)]{1998AJ....115..234L} Lira P., et
  al.\ 1998, AJ, 115, 234
\bibitem[Mannucci et al.(2003)]{2003A&A...401..519M} Mannucci F., et
  al.\ 2003, A\&A, 401, 519
\bibitem[Mannucci et al.(2005)]{2005A&A...433..807M} Mannucci F.,
   Della Valle M., Panagia N., Cappellaro E., Cresci G., Maiolino
   R., Petrosian A., Turatto M.\ 2005, A\&A, 433, 807
\bibitem[Maoz et al.(2011)]{2010arXiv1002.3056M} Maoz D., Mannucci F.,
  Li W., Filippenko, A.~V., Della Valle M.,
  Panagia N.\ 2011, MNRAS, in press (arXiv:1002.3056)
\bibitem[Matheson et al.(2000)]{} Matheson T., et
  al.\ 2000, AJ, 120, 1487 
\bibitem[Matheson et al.(2003)]{2003ApJ...599..394M} Matheson T., et
  al.\ 2003, ApJ, 599, 394
\bibitem[McClelland et al.(2010)]{2010ApJ...720..704M} McClelland, C.~M., 
  et al.\ 2010, ApJ, 720, 704
\bibitem[Miller et al.(2009)]{2009ApJ...690.1303M} Miller A.~A., et
  al.\ 2009, ApJ, 690, 1303
\bibitem[Modjaz et al.(2006)]{2006ApJ...645L..21M} Modjaz M., et
  al.\ 2006, ApJ, 645, L21
\bibitem[Modjaz(2007)]{2007PhDT.........7M} Modjaz M.\ 2007,
  Ph.D.~Thesis, Harvard University
\bibitem[Modjaz et al.(2009)]{2009ApJ...702..226M} Modjaz M., et
  al.\ 2009, ApJ, 702, 226
\bibitem[Monet et al.(2003)]{2003AJ....125..984M} Monet D.~G., et
  al.\ 2003, AJ, 125, 984
\bibitem[Narayan et al.(2011)]{2010arXiv1008.4353N} Narayan, G., et al.\ 
   2011, submitted (arXiv:1008.4353)
\bibitem[Neill et al.(2006)]{2006AJ....132.1126N} Neill J.~D., et
  al.\ 2006, AJ, 132, 1126
\bibitem[Ofek et al.(2007)]{2007ApJ...659L..13O} Ofek E.~O., et
  al.\ 2007, ApJ, 659, L13
\bibitem[Pain et al.(2002)]{2002ApJ...577..120P} Pain R., et
  al.\ 2002, ApJ, 577, 120
\bibitem[Pastorello et al.(2002)]{2002MNRAS.333...27P} Pastorello A.,
  et al.\ 2002, MNRAS, 333, 27
\bibitem[Pastorello et al.(2004)]{2004MNRAS.347...74P} Pastorello A.,
  et al.\ 2004, MNRAS, 347, 74
\bibitem[Perets et al.(2010)]{2010Natur.465..322P} Perets H.~B., et
  al. 2010, Nature, 465, 322
\bibitem[Phillips(1993)]{1993ApJ...413L.105P} Phillips M.~M.\ 1993,
  ApJ, 413, L105
\bibitem[Phillips et al.(1992)]{1992AJ....103.1632P} Phillips M.~M.,
  Wells L.~A., Suntzeff N.~B., Hamuy M., Leibundgut B., Kirshner
  R.~P., Foltz C.~B.\ 1992, AJ, 103, 1632
\bibitem[Phillips et al.(2007)]{2007PASP..119..360P} Phillips M.~M.,
  et al.\ 2007, PASP, 119, 360
\bibitem[Pian et al.(2006)]{2006Natur.442.1011P} Pian E., et
  al.\ 2006, Nature, 442, 1011
\bibitem[Poznanski et al.(2002)]{2002PASP..114..833P} Poznanski D.,
  Gal-Yam A., Maoz D., Filippenko A.~V., Leonard D.~C.,
  Matheson T.\ 2002, PASP, 114, 833
\bibitem[Poznanski et al.(2007)]{2007MNRAS.382.1169P} Poznanski D.,
  et al.\ 2007, MNRAS, 382, 1169
\bibitem[Poznanski, Maoz, \& Gal-Yam(2007)]{2007AJ....134.1285P}
  Poznanski D., Maoz D., Gal-Yam A.\ 2007, AJ, 134, 1285
\bibitem[Poznanski et al.(2009)]{2009ApJ...694.1067P} Poznanski D.,
  et al.\ 2009, ApJ, 694, 1067
\bibitem[Pozzo et al.(2006)]{2006MNRAS.368.1169P} Pozzo M., et
  al.\ 2006, MNRAS, 368, 1169
\bibitem[Prantzos \& Boissier(2003)]{2003A&A...406..259P} Prantzos
  N., Boissier S.\ 2003, A\&A, 406, 259
\bibitem[Prieto, Stanek, \& Beacom(2008)]{2008ApJ...673..999P} 
  Prieto J.~L., Stanek K.~Z., Beacom J.~F.\ 2008, ApJ, 673, 999
\bibitem[Qiu et al.(1999)]{1999AJ....117..736Q} Qiu Y., Li W., Qiao
  Q., Hu J.\ 1999, AJ, 117, 736
\bibitem[Quimby et al.(2007)]{2007ApJ...668L..99Q} Quimby R.~M.,
  Aldering G., Wheeler J.~C., H{\"o}flich P., Akerlof C.~W.,
  Rykoff E.~S.\ 2007, ApJ, 668, L99
\bibitem[Quimby et al.(2009)]{2009ATel.2005....1Q} Quimby R., et
  al.\ 2009, The Astronomer's Telegram, 2005, 1
\bibitem[Rest et al.(2009)]{2009arXiv} Rest A., et al.\ 2009, submitted
  (arXiv:0911.2002)
\bibitem[Riello \& Patat(2005)]{2005MNRAS.362..671R} Riello M.,
  Patat F.\ 2005, MNRAS, 362, 671
\bibitem[Richardson et al.(2002)]{2002AJ....123..745R} Richardson D.,
  Branch D., Casebeer D., Millard J., Thomas R.~C., Baron
  E.\ 2002, AJ, 123, 745
\bibitem[Richmond et al.(1994)]{1994AJ....107.1022R} Richmond M.~W.,
  Treffers R.~R., Filippenko A.~V., Paik Y., Leibundgut B.,
  Schulman E., Cox C.~V.\ 1994, AJ, 107, 1022
\bibitem[Richmond et al.(1996)]{1996AJ....111..327R} Richmond M.~W.,
  et al.\ 1996, AJ, 111, 327
\bibitem[Riess et al.(2009)]{} Riess A. G., et al.\ 2009, ApJ, 699,
  539
\bibitem[Sanders et al.(2003)]{sanders03} Sanders D.~B., Mazzarella
  J.~M., Kim D.-C., Surace J.~A., Soifer B.~T.\ 2003, AJ, 126,
  1607
\bibitem[Schlegel, Finkbeiner, \& Davis(1998)]{1998ApJ...500..525S} 
  Schlegel D.~J., Finkbeiner D.~P., Davis M.\ 1998, ApJ, 500, 525
\bibitem[Silverman et al.(2011)]{silver2011}Silverman J.~M., et
  al. 2011, in prep.
\bibitem[Singer et al.(2003)]{2003IAUC.8201....1S} Singer D.,
  Beutler B., Swift B., Li W., Yamaoka H., Itagaki K.\ 2003,
  IAU Circ., 8201, 1
\bibitem[Smartt et al.(2009)]{2009MNRAS.395.1409S} Smartt S.~J.,
  Eldridge J.~J., Crockett R.~M., Maund J.~R.\ 2009, MNRAS,
  395, 1409
\bibitem[Smith et al.(2007)]{2007ApJ...666.1116S} Smith N., et
  al.\ 2007, ApJ, 666, 1116
\bibitem[Smith et al.(2008)]{2008ApJ...686..467S} Smith N., Chornock
  R., Li W., Ganeshalingam M., Silverman J.~M., Foley R.~J.,
  Filippenko A.~V., Barth A.~J.\ 2008, ApJ, 686, 467
\bibitem[Smith et al.(2011a)]{smith2011a}Smith N., et al. 2011a, MNRAS,
  press (arXiv:1006.3899)
\bibitem[Smith et al.(2011b)]{smith2011b}Smith N., et al. 2011b, MNRAS,
  submitted (arXiv:1010.3718)
\bibitem[Soderberg et al.(2008)]{2008Natur.453..469S} Soderberg
  A.~M., et al.\ 2008, Nature, 453, 469
\bibitem[Sollerman, Cumming, \& Lundqvist(1998)]{1998ApJ...493..933S} 
  Sollerman J., Cumming R.~J., Lundqvist P.\ 1998, ApJ, 493, 933
\bibitem[Spergel et al.(2007)]{2007ApJS..170..377S} Spergel D.~N., et
  al.\ 2007, ApJS, 170, 377
\bibitem[Sullivan et al.(2006)]{2006ApJ...648..868S} Sullivan M., et
  al.\ 2006, ApJ, 648, 868
\bibitem[Tremonti et al.(2004)]{2004ApJ...613..898T} Tremonti C.~A.,
  et al.\ 2004, ApJ, 613, 898
\bibitem[Turatto et al.(1993)]{1993MNRAS.262..128T} Turatto M.,
  Cappellaro E., Danziger I.~J., Benetti S., Gouiffes C., Della
  Valle M.\ 1993, MNRAS, 262, 128
\bibitem[Valenti et al.(2009)]{2009Natur.459..674V} Valenti S., et
  al.\ 2009, Nature, 459, 674
\bibitem[van den Bergh, Li, \& Filippenko(2002)]{2002PASP..114..820V} 
   van den Bergh S., Li W., Filippenko A.~V.\ 2002, PASP, 114, 820
\bibitem[Van Dyk et al.(2000)]{2000PASP..112.1532V} Van Dyk S.~D.,
  Peng C.~Y., King J.~Y., Filippenko A.~V., Treffers R.~R., Li
  W., Richmond M.~W.\ 2000, PASP, 112, 1532
\bibitem[Vink, de Koter, \& Lamers(2001)]{2001A&A...369..574V} 
  Vink J.~S., de Koter A., Lamers H.~J.~G.~L.~M.\ 2001, A\&A, 369, 574
\bibitem[Vink \& de Koter(2005)]{2005A&A...442..587V} Vink J.~S.,
  de Koter A.\ 2005, A\&A, 442, 587
\bibitem[Wang et al.(2008)]{2008ApJ...675..626W} Wang X., et
  al.\ 2008, ApJ, 675, 626
\bibitem[Wang et al.(2009)]{2009ApJ...699L.139W} Wang X., et
  al.\ 2009, ApJ, 699, L139
\bibitem[Waxman et al.(2007)]{2007ApJ...667..351W} Waxman E., 
   M{\'e}sz{\'a}ros P., \& Campana S.\ 2007, ApJ, 667, 351
\bibitem[Wheeler et al.(1993)]{1993ApJ...417L..71W} Wheeler, J.~C., et al.\ 
   1993, ApJ, 417, L71
\bibitem[Young \& Branch(1989)]{1989ApJ...342L..79Y} Young T.~R.,
  Branch D.\ 1989, ApJ, 342, L79

\end{thebibliography}
\end{document}